\documentclass[iop]{emulateapj}

\slugcomment{temporal version}
\shorttitle{Suzaku Magnetar Catalog}
\shortauthors{ENOTO ET AL.}

\usepackage{color}

\newcommand{\suzaku}{{\it Suzaku} }

\renewcommand{\emph}[1]{\textcolor{red}{\bf #1}}

\usepackage[usenames,dvipsnames]{xcolor}
\usepackage{comment} 

\usepackage{subfigure}
\usepackage{here}
\usepackage{url}
\begin{document}

% ==================================================
\title{
Magnetar Broadband X-ray Spectra Correlated with Magnetic Fields: \\
Suzaku Archive of SGRs and AXPs Combined with NuSTAR, Swift, and RXTE\\
}
% ==================================================

% ----------------------------------------------------------------------------------------------------
\author{
%Teruaki Enoto and Suzaku Magnetar members
Teruaki Enoto\altaffilmark{1,2},
Shinpei Shibata\altaffilmark{3},
Takao Kitaguchi\altaffilmark{4}, 
\\
Yudai Suwa\altaffilmark{5}, 
Takahiko Uchide\altaffilmark{6},
Hiroyuki Nishioka\altaffilmark{7},
Shota Kisaka\altaffilmark{8},\\
Toshio Nakano\altaffilmark{9},
Hiroaki Murakami\altaffilmark{7},
and 
Kazuo Makishima\altaffilmark{7,10}
}
\altaffiltext{1}{The Hakubi Center for Advanced Research, Kyoto University, Kyoto 606-8302, Japan; teruaki\_enoto@10.alumni.u-tokyo.ac.jp}
\altaffiltext{2}{Department of Astronomy, Kyoto University, Kitashirakawa-Oiwake-cho, Sakyo-ku, Kyoto 606-8502, Japan}
\altaffiltext{3}{Department of Physics, Yamagata University, 
Kojirakawa 1-4-12, Yamagata, 990-8560 Japan}
\altaffiltext{4}{Department of Physical Science, Hiroshima University, 
	1-3-1 Kagamiyama, Higashi-Hiroshima, Hiroshima 739-8526, Japan}	
\altaffiltext{5}{Center for Gravitational Physics, Yukawa Institute for Theoretical
Physics, Kyoto University, Oiwake-cho, Kitashirakawa, Sakyo-ku, Kyoto
606-8502, Japan}
\altaffiltext{6}{Geological Survey of Japan, 
National Institute of Advanced Industrial Science and Technology (AIST), 
1-1-1 Higashi, AIST Central 7, Tsukuba, Ibaraki 305-8567, Japan}  
\altaffiltext{7}{Department of Physics, University of Tokyo,
    7-3-1 Hongo, Bunkyo-ku, Tokyo, 113-0033, Japan}
\altaffiltext{8}{Department of Physics and Mathematics, Aoyama Gakuin University, 5-10-1 Fuchinobe, Sagamihara, Kanagawa 252-5258, Japan}
\altaffiltext{9}{High Energy Astrophysics Laboratory, RIKEN Nishina Center, 2-1 Hirosawa, Wako, Saitama 351-0198, Japan}
\altaffiltext{10}{MAXI team, Institute of Physical and Chemical Research (RIKEN), 
	2-1 Hirosawa, Wako, Saitama 351-0198}
% ----------------------------------------------------------------------------------------------------

\begin{abstract}
Studies were made of the 1--70 keV persistent spectra 
	of fifteen magnetars
	as a complete sample observed 
	with {\it Suzaku}  from 2006 to 2013.
Combined with early {\it NuSTAR} observations of four hard X-ray emitters,
	nine objects showed 
	a hard power-law emission dominating at $\gtrsim$10 keV
	with the 15--60 keV flux of $\sim$1--$11\times 10^{-11}$\,ergs\,s$^{-1}$\,cm$^{-2}$.
The hard X-ray luminosity $L_{\rm h}$, 
	relative to that of a soft-thermal surface radiation $L_{\rm s}$, 
	tends to become higher  
	toward younger and strongly magnetized objects.
Updated from the previous study, 
	their hardness ratio, 
	defined as $\xi=L_{\rm h}/L_{\rm s}$, 
	is correlated with the measured spin-down rate $\dot{P}$ 
	as $\xi=0.62 \times (\dot{P}/10^{-11}\,{\rm s}\,{\rm s}^{-1})^{0.72}$,
	corresponding with positive and negative correlations of 
	the dipole field strength $B_{\rm d}$ ($\xi \propto B_{\rm d}^{1.41}$)
	and the characteristic age $\tau_{\rm c}$ ($\xi \propto \tau_{\rm c}^{-0.68}$), respectively.
Among our sample,
	five transients were observed 
	during X-ray outbursts,
	and the results are compared with their long-term 1--10 keV flux decays  
	monitored with {\it Swift}/XRT and {\it RXTE}/PCA.
Fading curves of three bright outbursts 
	are approximated by an empirical formula used in the seismology,
	showing a $\sim$10--40 d  plateau phase. 
Transients show the maximum luminosities of 
	$L_{\rm s}$$\sim$$10^{35}$\,erg\,s$^{-1}$,
	which is comparable to those of the persistently bright ones,
	and fade back to $\lesssim$$10^{32}$\,erg\,s$^{-1}$.	
Spectral properties are discussed in a framework of the magnetar hypothesis.
\end{abstract}

% =====================
\section{INTRODUCTION}
\label{section:introduction}
% =====================

%%%% SGR and AXP properties %%%%
Soft Gamma Repeaters (SGRs)
	and 
	Anomalous X-ray Pulsars (AXPs)
	are growing to
	a new population of young neutron stars.
These two subclasses 
	were historically discovered in different ways
	\citep{Kouveliotou1998Natur.393..235K, Mereghetti1995ApJ...442L..17M}, 
	but now are believed to have common observational properties
	(for recent reviews, see
	\citealt{Woods2006csxs.book..547W,Kaspi2007Ap&SS.308....1K,Mereghetti2008A&ARv..15..225M});
(a) a narrow range of slow spin periods, $P=$2--12\,s,
(b) high spin-down rates of $\dot{P}$=$10^{-12}$--$10^{-10}$\,s\,s$^{-1}$, 
(c) young characteristic ages as $\tau_{\rm c}=P/2\dot{P}\sim 1$--100\, kyr
(d) X-ray luminosities, 
	$L_{\rm x}$$\sim$$10^{34}$--$10^{35}$\,erg\,s$^{-1}$,
	that are brighter 
	than the spin-down power $L_{\rm sd}=3.9\times 10^{35}$\,erg\,s$^{-1}$$\dot{P}_{11}P_1^{-3}\sim 10^{32}$--$10^{34}$\,erg\,s$^{-1}$, 
	where period and its derivative are normalized as $P_1=P/(1\,\textrm{s})$ and $\dot{P}_{11}=\dot{P}/(10^{-11}\,\textrm{s\,s$^{-1}$})$, respectively,
(e) sporadic burst activities,
and 
(f) occasional associations with supernova remnants.

%%%% Theories %%%%
The magnetar hypothesis	for SGRs and AXPs 
	\citep{Duncan1992ApJ...392L...9D,
	Thompson1995MNRAS.275..255T,Thompson1996ApJ...473..322T} 
	has become the most popular model in the last decade.
This scenario describes 
	that 
	both SGRs and AXPs are 
	isolated ultra-strongly magnetized neutron stars 
	with their dipole magnetic field strength  
	reaching 
	$B_{\rm d}=1.0\times 10^{14}\,\textrm{G}(P_1\dot{P}_{11})^{1/2}$$\sim$$10^{14}$--$10^{15}$\,G.
In this model,
	high X-ray luminosity, $L_{\rm x}/L_{\rm sd}\sim 1$--$10^3$, 
	is interpreted as a release of magnetic energies 
	stored in the stellar interior. 
The ``magnetar" model has come to be widely recognized as a fascinating concept,
 	for example, in the context of supernova explosions \citep{Nicholl2013Natur.502..346N},
	and in the extreme fundamental physics exceeding 
	the quantum critical field 
	$B_{\rm QED}=m_{\rm e}^2c^3/(\hbar e)=4.414\times 10^{13}$\,G
	\citep{Harding2006RPPh...69.2631H}, 
	where $m_{\rm e}$, $e$, $c$, and $\hbar$ are
	the electron rest mass, electron charge, speed of light, 
	and Planck's constant, respectively.

%%%% Alternative models %%%%
Despite the accumulated evidence for the magnetar model,
	there are also alternative hypotheses proposed to explain 
	the observational features of SGRs and AXPs in terms of, 
	e.g., 
	accretion from a fossil disk 
	\citep{2001ApJ...554.1245A,Trumper2010A&A...518A..46T},
	invoking a quark star model 
	\citep{Ouyed2004A&A...420.1025O},
	or 
	as fast rotating massive white dwarves \citep{Malheiro2012PASJ...64...56M}.
Therefore,
	it is imperative at this stage 
	to observationally examine the radiation properties of this group 
	of objects,
	and understand their radiation properties in a unified way.

% ----------------------------------------------------
\begin{figure*}
\begin{center}
\includegraphics[scale=0.95]{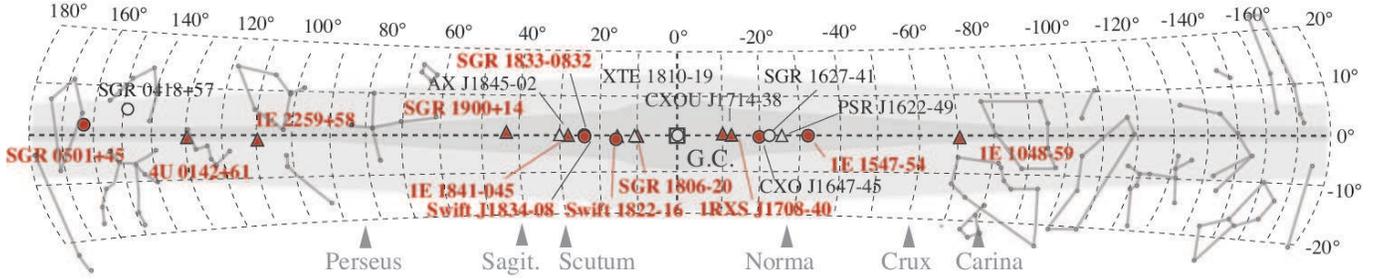}
\caption{
Known galactic SGRs and AXPs on the Galactic coordinate, 
	where 
	transient sources (circle) %\green{(change to star symblos)}
	and 
	persistently bright objects (triangle) %\green{(change to circle marks)}
	are classified as in Table~\ref{tab:list_suzaku_log}.
The red filled symbols 
	are sources observed with {\it Suzaku}. 
Spiral arms of our Galaxy are also indicated. 	
}
\label{fig:f_magnetar_location}
\end{center}
\end{figure*}
% ----------------------------------------------------

\begin{figure}
\begin{center}
\includegraphics[width=88mm]{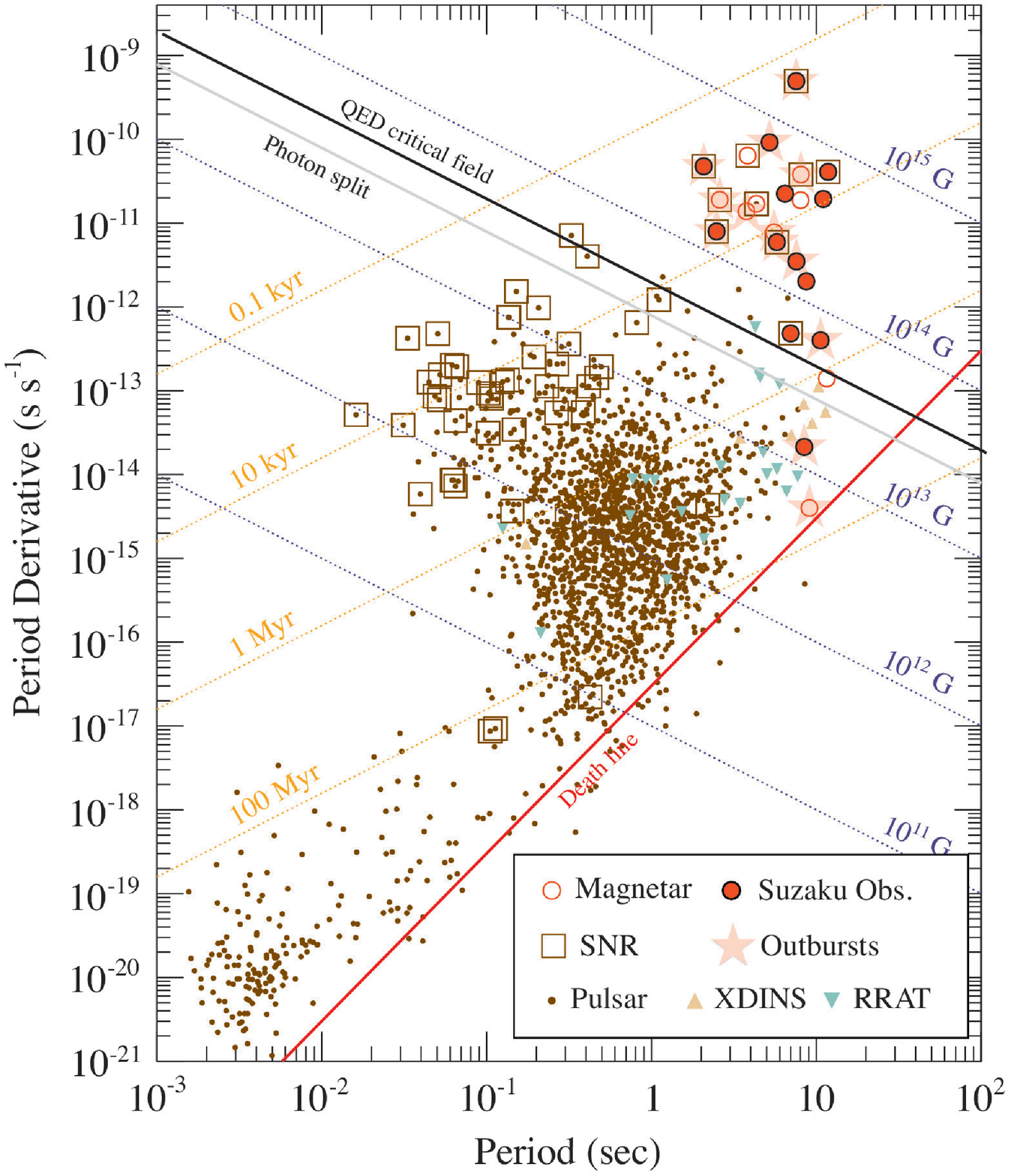}
\caption{
SGRs and AXPs on the $P$-$\dot{P}$ diagram,
	together with the grids of 
	$B_{\rm d}$, $\tau_{\rm c}$, and $L_{\rm sd}$.
The ATNF pulsar catalog is used for other pulsars 
	\citep{Manchester2005AJ....129.1993M}.
Star and square symbols are objects showing X-ray outbursts 
	and association with supernova remnants, respectively.
The pulsar death death 
	line \citep{Chen1993ApJ...402..264C,Zhang2000ApJ...531L.135Z},
	$B_{\rm QED}$, 
	and 
	photon splitting line \citep{Baring1998ApJ...507L..55B}
	are also indicated. 
}
\label{fig:p_pdot_diagram}
\end{center}
\end{figure}

Nearly 23 confirmed SGRs and AXPs are now known on the Galactic plane 
	as displayed in Figure \ref{fig:f_magnetar_location} 
	with their timing properties in Figure \ref{fig:p_pdot_diagram} 
	(see the detailed catalog, \citealt{Olausen2014ApJS..212....6O}). 
Some are persistently bright with stable $L_{\rm x}$, 
	intensively studied since the early days of X-ray astronomy: 
	e.g., 4U~0142+61 
	(e.g., \citealt{White1996ApJ...463L..83W, Enoto2011PASJ...63..387E, Dib2014ApJ...784...37D}).
The soft X-ray spectrum 
	is approximately an optically-thick radiation 
	with its blackbody temperature at $kT$$\sim$0.5 keV
	which is thought to originate 
	from the stellar surface or vicinity as a quasi-thermal emission
	\citep{Zane2009MNRAS.398.1403Z}.
In this paper, we call this soft radiation below 10 keV 
	``Soft X-ray Component (SXC)".	

Recent observations have revealed 
	a new distinctive ``Hard X-ray Component (HXC)"
	dominating above 10 keV.
First detected with INTEGRAL 
	from persistently bright sources  
	\citep{Kuiper2006ApJ...645..556K, Hartog2008A&A...489..263D, Hartog2008A&A...489..245D},
	the HXC was later reconfirmed by \suzaku and {\it NuSTAR} 
	\citep{Morii2010PASJ...62.1249M, Enoto2011PASJ...63..387E, An2013ApJ...779..163A}.
This HXC extends up to 100\,keV or more
	with a hard photon index $\Gamma_{\rm h}\sim 1$,
	but must cuts off 
	at some energies because of 
	an upper limit by the {\it CGRO}/COMPTEL at $\gtrsim$1\,MeV.
This power-law HXC is now believed to be an optically thin emission 
	presumably from a pulsar magnetosphere in the magnetar scheme (e.g., \citealt{Beloborodov2013ApJ...762...13B}).
	
There are also subsequent discoveries of transient objects,	
	mainly discovered by burst activities: 
	e.g., SGR~0501+4516 
	(e.g., \citealt{Enoto2009ApJ...693L.122E,Rea2009MNRAS.396.2419R}). 
Such transient sources 
	occasionally cause surges in persistent X-rays by a few orders of magnitude, 
	followed by a gradual decay \citep{Rea2011heep.conf..247R}.
During these ``outburst" states, 
	sporadic short bursts with short time-scale durations ($\sim$1 s) 
	have been detected
	\citep{Nakagawa2007PASJ...59..653N, Israel2008ApJ...685.1114I}.
Although a complete picture has yet to come, 
	bursts are thought to be originate 
	from magnetic reconnection \citep{Lyutikov2003MNRAS.346..540L}
	or 
	cracking of the crust with starquakes \citep{Thompson2002ApJ...574..332T}.

The SXC and HXC  
	match ideally with the simultaneous 0.2--600 keV broadband coverage of 
	the \suzaku satellite \citep{Mitsuda2007PASJ...59S...1M}.
Our previous study of 9 SGRs and AXPs utilizing \suzaku 
	(\citealt{Enoto2010ApJ...722L.162E}, hereafter Paper I)
	suggested that 
	1) phase-averaged X-ray radiation of SGRs/AXPs commonly consists 
	of the SXC below 10 keV and the HXC above 10 keV
	in both quiescent states and transient outbursts, 
	2) $\Gamma_{\rm h}$ depends on $B_{\rm d}$ and $\tau_{\rm ch}$,
	and 
	3) their wide-band spectral properties are tightly correlated 
	with $B_{\rm d}$ and $\tau_{\rm ch}$.

As the detailed description following Paper I,
	this paper provides
	a summary of \suzaku observations of SGRs and AXPs, 
	combining the systematic spectral study of all the \suzaku sources 
	and the X-ray decaying behavior of transient sources. 

% ----------------------------------------------------
% ----------------------------------------------------

% ---------------------------------------
% ~~~~~~~~~~~~~~~~~~~~~~~~~~~~~~~~~~~~~~~~~~~~~~~~~~~~
%\begin{table}
\begin{deluxetable*}{lccccrcclr}[h]
%\tabletypesize{\scriptsize}
\tablecaption{
Log of the \suzaku SGRs and AXPs observations.
\label{tab:list_suzaku_log}
}
\tablewidth{0pt}
\tablehead{
\colhead{Name} &
\colhead{$B_{\rm d}$} &
\colhead{ObsID} &
\colhead{Start Time} &
\colhead{Epoch} &
\colhead{Exp.} &
%\colhead{Epoch} &
\colhead{Nominal} &
\colhead{XIS Mode} & 
\colhead{Process} 
\\
&
\colhead{($10^{14}$\,G)} &
& & & \colhead{(ks)} & 
\colhead{Pointing}& 
\colhead{(XIS0, 1, 3)} & 
\colhead{Version} 
%(erg s$^{-1}$ cm$^{-2}$) & ($10^{22}$ cm$^{-2}$) & (erg s$^{-1}$ cm$^{-2}$) 
}
% -------------------
\startdata
% -------------------
\multicolumn{9}{c}{Persistently Bright Sources \rule[0mm]{0mm}{3mm}} \\ 
\hline
SGR~1806$-$20 & 24 & 401092010 & 2006-09-09 22:13:43 & AO1 & 48.9  & HXD & (full, full, full) & 2.0.6.13  \\ 
            & -- & 401021010 & 2007-03-30 15:08:00 & AO1 & 19.6  & HXD & (1/8, 1/8, 1/8) & 2.0.6.13  \\ 
            & -- & 402094010 & 2007-10-14 05:35:49 & AO2 & 42.7  & HXD & (full, full, full) & 2.1.6.15  \\ 
1E 1841$-$045 (SNR Kes 73) & 7.1  & 401100010 & 2006-04-19 10:51:40 & PV & 97.0  & HXD & (1/8, 1/8, 1/8) & 2.0.6.13  \\ 
SGR~1900$+$14 & 7.0 & 401022010 & 2006-04-01 08:42:57 & AO1 & 0.9  & HXD & (1/4, 1/8, 1/4) & 2.0.6.13  \\ 
            & -- & 404077010 & 2009-04-26 18:23:44 & Key & 40.6  & HXD & (1/4, 1/4, 1/4) & 2.3.12.25  \\ 
CXOU~J171405.7$-$381031 & 5.0 & 501007010 & 2006-08-27 01:27:07 & AO1 & 75.6  & XIS & (full ,full ,full ) & 2.0.6.13 \\
1RXS~J170849.0$-$400910 & 4.7 & 404080010 & 2009-08-23 16:25:08 & Key & 50.9  & HXD & (1/4, 1/4, 1/4) & 2.4.12.26  \\ 
                      & --  & 405076010 & 2010-09-27 14:41:52 & Key & 47.0  & HXD & (1/4, 1/4, 1/4) & 2.5.16.28  \\ 
1E~1048.1$-$5937 & 4.2 & 403005010 & 2008-11-30 23:02:01 & AO3 & 85.0  & HXD & (full, full, full) & 2.2.11.22  \\ 
4U~0142$+$61 & 1.3 & 402013010 & 2007-08-13 04:04:13 & AO2 & 71.9  & HXD & (1/4, 1/4, 1/4) & 2.1.6.15  \\ 
	       &  & 404079010 & 2009-08-12 01:41:15 & Key & 82.7  & HXD & (1/4, 1/4, 1/4) & 2.4.12.26  \\ 
		   &  & 406031010 & 2011-09-07 15:43:32 & ToO & 36.7  & XIS & (full, full, 1/4) & 2.7.16.30  \\ 
           &  & 408011010 & 2013-07-31 10:05:39 & AO8 & 79.8  & XIS & (1/8, 1/4, 1/4) & 2.8.20.35  \\ 
1E~2259$+$586 & 0.59 & 404076010 & 2009-05-25 20:00:17 & Key & 89.2  & HXD & (1/4, 1/4, 1/4) & 2.4.12.26  \\ 
%CXOU J171405.7-381031 (SNR CTB37B) \\
AX~J1818.8$-$1559 & $\cdots$ & 406074010 & 2011-10-15 13:17:36  & AO6 & 88.5  & XIS & (1/8 ,1/8 ,1/8 ) & 2.7.16.30\\
\hline
\multicolumn{9}{c}{Transient Sources \rule[0mm]{0mm}{4mm}} \\ 
\hline
1E~1547.0$-$5408    & 3.2 & 903006010 & 2009-01-28 21:34:12 & ToO & 10.6  & HXD & (psum, 1/4, 1/4) & 2.3.12.25  \\ 
                  & -- & 405024010 & 2010-08-07 03:52:07 & AO5 & 34.1  & HXD & (1/4, 1/4, psum) & 2.5.16.28  \\ 
SGR~0501$+$4516 & 1.9 & 903002010 & 2008-08-26 00:24:42 & ToO & 43.3  & XIS & (1/4, 1/4, 1/4) & 2.2.8.20  \\ 
			  & -- & 404078010 & 2009-08-17 20:21:51 & Key & 29.5  & HXD & (1/4, 1/4, 1/4) & 2.4.12.26  \\ 
			  & -- & 405075010 & 2010-09-20 17:27:15 & Key & 49.1  & HXD & (1/4, 1/4, 1/4) & 2.5.16.28  \\ 
			  & -- & 408013010 & 2013-08-31 23:25:40 & AO8 & 41.2  & XIS & (1/8, 1/4, full) & 2.8.20.35  \\ 
SGR~1833$-$0832 & 1.8 & 904006010 & 2010-03-27 09:03:32 & ToO & 35.7  & HXD & (1/8, full, full) & 2.5.16.28  \\ 
CXOU~J164710.2$-$455216 & 1.6 & 901002010 & 2006-09-23 06:59:17 & ToO & 38.7  & XIS & (1/8, 1/8, 1/8) & 2.0.6.13  \\ 
Swift~J1834.9$-$0846 & 1.4 & 408015010 & 2013-10-17 07:17:57 & AO8 & 30.4  & XIS & (full, full, full) & 2.8.20.35   \\ 
Swift~J1822.3$-$1606 & 0.14 & 906002010 & 2011-09-13 09:59:07 & ToO & 36.1  & HXD & (1/8, full, full) & 2.7.16.30
% -------------------
 \enddata
  \tablenotetext{a}{
 Although there is accumulated evidence 
	that SGRs and AXPs are intrinsically the same class of magnetars 
	(\citealt{Gavriil2002Natur.419..142G}, \citealt{Mereghetti2008A&ARv..15..225M}, Paper I),
	let us retain,
	in this paper, the historical terminology of 
	``SGR" (usually discovered from burst activities) 
	and ``AXP" (identified as bright X-ray sources). 
%	as an observationally-based terminology.
Conventionally,
	some sources have a duplicated labeling both as the SGR and AXP; 
	for example 1E~1547.0$-$5408 is also called SGR~1550$-$5418 or PSR~J1550$-$5418.
For such an object, 
	we employ ether of the conventional names.}
 \tablenotetext{b}{Preceding \suzaku studies: 
 	SGR~1806$-$20 \citep{Esposito2007A&A...476..321E,Nakagawa2009PASJ...61..109N},
 	1E~1841$-$45 \citep{Morii2010PASJ...62.1249M},
 	SGR~1900$+$14 \citep{Nakagawa2009PASJ...61..109N}
 	4U~0142$+$61 \citep{Enoto2011PASJ...63..387E,Makishima2014PhRvL.112q1102M},
 	AX~J1818.8$-$586 \citep{Mereghetti2012A&A...546A..30M},
 	1E~1547.0$-$5408 \citep{Enoto2010PASJ...62..475E,Iwahashi2013PASJ...65...52I,Enoto2012MNRAS.427.2824E,2015PASJ..tmp..263M},
 	SGR~0501$+$4516 \citep{Enoto2009ApJ...693L.122E,Enoto2010ApJ...715..665E,Nakagawa2011PASJ...63S.813N}
 	CXOU~J164710.2$-$455216 \citep{Naik2008PASJ...60..237N},
 	SGR~1833$-$0832 \citep{Esposito2011MNRAS.416..205E},
 	Swift~J1822.3$-$1606 \citep{Rea2012ApJ...754...27R},
	and Paper~I.
 	}
\tablenotetext{c}{Objects are sorted by the dipole field $B_{\rm d}$
	calculated from the period $P$ and its derivative $\dot{P}$
	assuming the magnetic dipole radiation.}
\tablenotetext{d}{Type of observations:
	PV (Performance Verification phase, i.e., first 9 month after the launch), 
	AO (Announcement of Opportunity observations), 
	ToO (Target of Opportunity observations), 
	and 
	Key (Key Project observations).
	}
\tablenotetext{e}{Exposure time (ks) of one XIS instrument, the maximum value among the three cameras is shown.}
\tablenotetext{f}{Nominal pointing position (XIS or HXD) at the observations,
	see \cite{Mitsuda2007PASJ...59S...1M}. 
	}
\tablenotetext{g}{XIS Observation modes:
	The full window reads out every 8 s,
	while 1/4 and 1/8 window modes read out every 2 and 1 s, respectively.
	The timing P-sum mode (psum) provids a $\sim$7.8 ms time resolution
	 together with one-dimentionally projected position information. 
	}
\tablenotetext{g}{Processing version of the data set.}
\end{deluxetable*}
% ~~~~~~~~~~~~~~~~~~~~~~~~~~~~~~~~~~~~~~~~~~~~~~~~~~~~
% ---------------------------------------

% ==========================================
\section{Observation and Data Reduction}
% ==========================================

% ----------------------------------------------------
\begin{figure*}[h]
\begin{center}
\includegraphics[height=57mm]{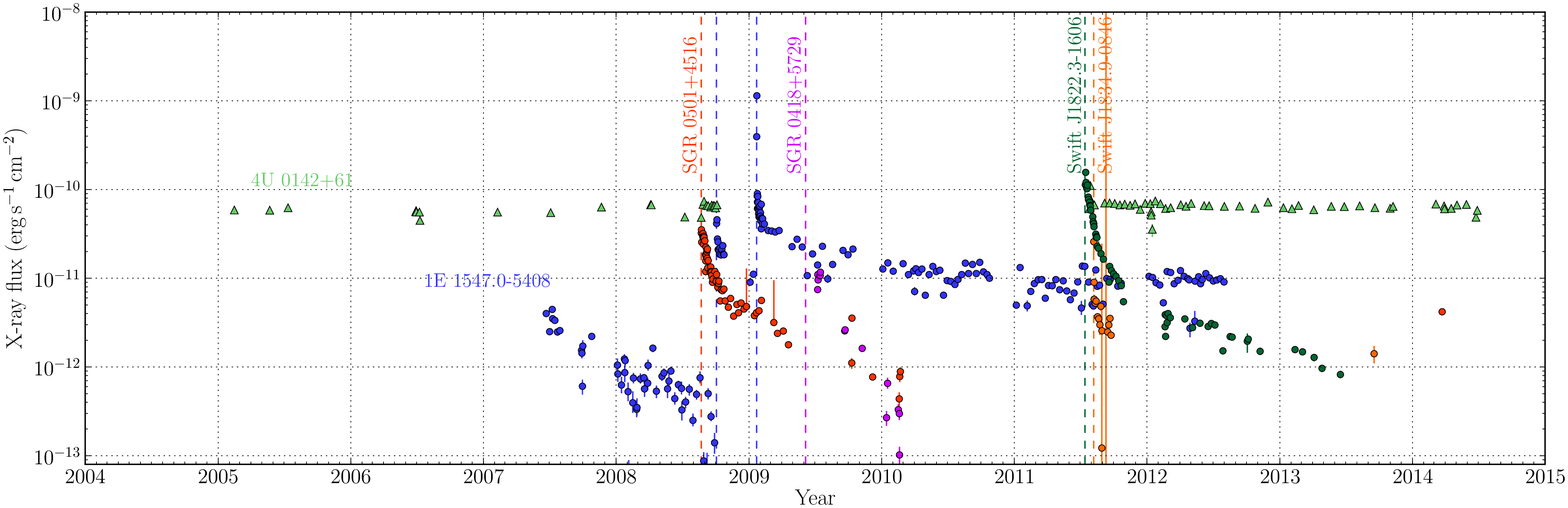}
\caption{
Examples of long-term histories in soft X-rays,
	during the {\it Suzaku} operation period,  %2?--10 keV X-ray flux 
	of persistently bright source (filled triangles) %, e.g., 4U~0142+61, SGR~1806-20)
	and transient objects (filled circles) % , e.g., SGR~0501+4516, Swift~J1822.3$-$1606) %in Table~\ref{tab:list_suzaku_log} 
	monitored with {\it Swift}/XRT and {\it RXTE}/PCA. 
First short bursts (usually the discovery of transient sources) 
	are indicated by the vertical dashed line for each outburst.
The data process is described in \S3.2.
}
\label{fig:transient_vs_persisten}
\end{center}
\end{figure*}
% ----------------------------------------------------	

% ==========================================
\subsection{Suzaku Observations}
% ==========================================
\label{section:suzaku}

% ==========================================
\subsubsection{Persistently bright or transient sources}
% ==========================================

Table~\ref{tab:list_suzaku_log}
	summarizes 
	all SGRs and AXPs 
	which {\it Suzaku} has observed as of 2013 December.
In this table, 
	the ``transient sources"
	exhibit prominent soft X-ray increases 
 	by 2--3 orders of magnitudes 
	and subsequent decays on timescales of months to years, 
	while 
	the ``persistently bright ones" 
	are relatively stable with their X-ray luminosities around $10^{35}$\,erg\,s$^{-1}$.
This is illustrated in Figure~\ref{fig:transient_vs_persisten} as 
	long-term X-ray flux records of representative sources. 
Since this ``persistent" and ``transient" classification is somehow phenomenological 
	without a clear consensus (e.g., \citealt{Pons2012ApJ...750L...6P}),
	we classified in this paper relatively variable sources as ``transients".

Our \suzaku sample in Table~\ref{tab:list_suzaku_log}
	includes 
	16 pointings of 9 persistently bright sources 
	and 
	10 target of opportunity (ToO) observations    
	to follow-up 6 transient objects.
This covers 15 objects of all the $\sim$29 sources or candidates 
	known to date\footnote{The latest magnetars and candidates 
	are listed in \url{http://www.physics.mcgill.ca/~pulsar/magnetar/main.html}
	(see also., \citealt{Olausen2014ApJS..212....6O}}.
Due to operational constraints,
	we were unable to observe recent transients 
	SGR~0418$+$5729 and SGR~1745$-$29.	
In the following analyses,
	we reprocessed all the published data while adding newly observed objects,
	and performed comprehensive phase-averaged spectroscopic analyses.

% ==========================================
\subsubsection{Reduction of broad-band Suzaku spectra}
% ==========================================
\label{Reduction of Broad-band X-ray Spectra}

% ----------------------------------------------------
%\begin{figure*}[!h]
\begin{figure*}[!t]
\begin{center}
\includegraphics[width=180mm]{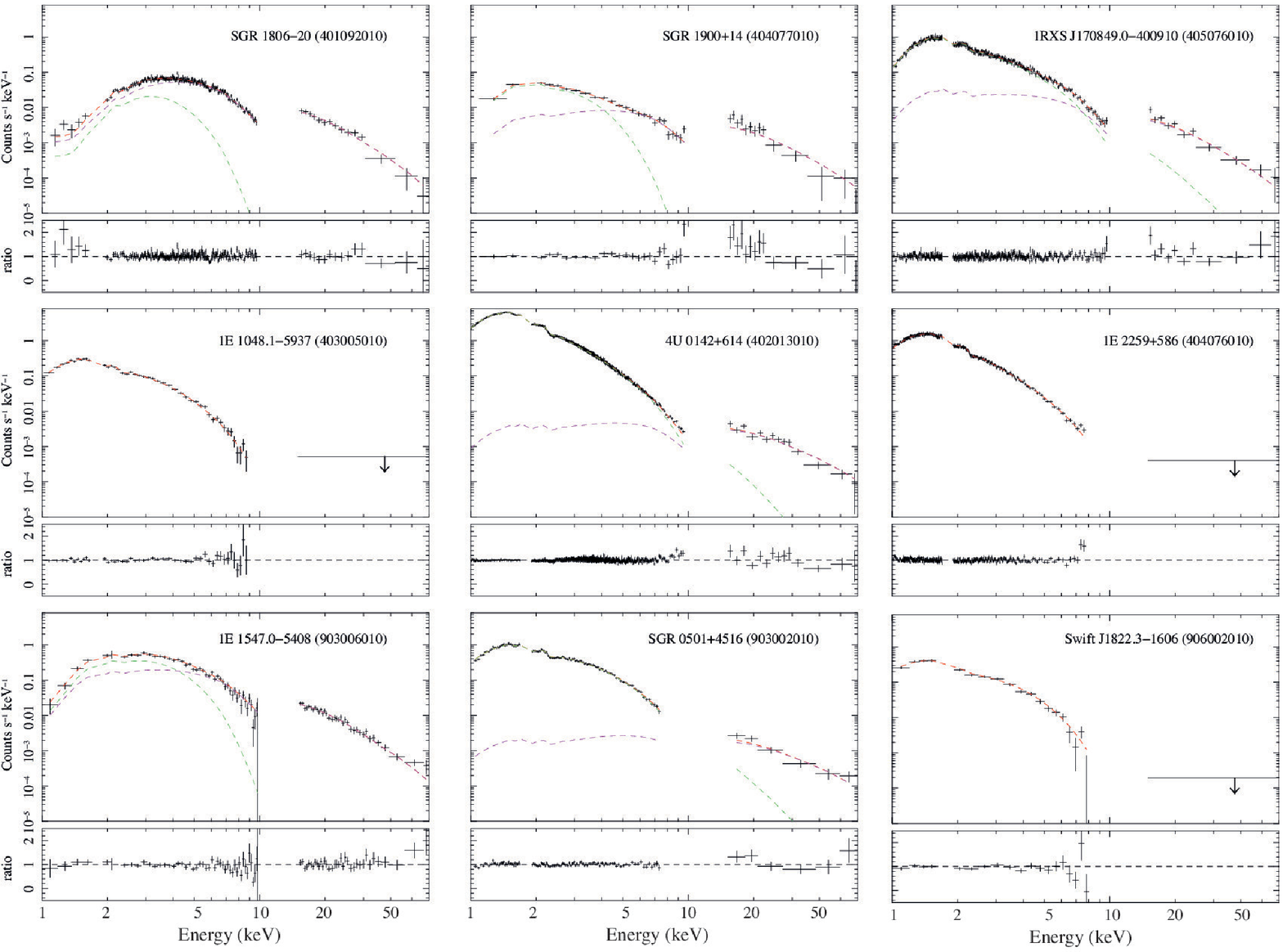}
\caption{
Examples of the XIS0+3 and HXD-PIN spectra of nine observations,
	peresented
	after subtracting the background, 
	but not removing the instrumental responses.
When the HXC is not detected, 
	its 3$\sigma$ upper limits are shown.
The XIS and HXD-PIN spectra are simultaneously fitted 
	by the CBB model (green) plus the additional power-law for the HXC if detected.
}
\label{fig:suzaku_raw_spec}
\end{center}
\end{figure*}
% ----------------------------------------------------	
We reprocessed the X-ray Imaging Spectrometer 
	(XIS, 0.2--12 keV; \citealt{Koyama2007PASJ...59S..23K}) 
	and 
	the Hard X-ray Detector 
	(HXD, 10--600 keV; 
	\citealt{Takahashi2007PASJ...59S..35T,Kokubun2007PASJ...59S..53K})
	data
	using FTOOLS ``\texttt{aepipeline}" of the HEADAS version 6.14 or later
	with latest calibration database (CALDB) and the standard screening criteria.	
Only the full window, 1/4, or 1/8 window modes 
	of XIS0, 1, and 3 (Table~\ref{tab:list_suzaku_log})
	are analyzed,	
	since XIS2 has been out of operation 
	due to damage by a micro meteorite in 2006 November.
As for the HXD,
	only the HXD-PIN data were utilized 
	since the typical SGR/AXP intensity in hard X-rays  
	($\lesssim$1 mCrab; Paper~I) % \citealt{Enoto2010ApJ...722L.162E})
	is below the detection limit of HXD-GSO. 

% === XIS spectra ===
The on-source and background XIS spectra 	
	were extracted 
	from a circular region of $2'.5$ radius 
	and 
	an annulus with the inner radius of $4'.0$
	and outer radius $7'.5$ centered on the source, respectively.
The XIS spectra of our magnetar sample are free from 
	the pile-up effect, 
	less than $\sim$1\% \citep{Yamada2012PASJ...64...53Y}.
The response matrix file (rmf) and auxiliary response file (arf)
	were produced using the FTOOLS 
	\texttt{xisrmfgen} and \texttt{xissimarfgen}
	\citep{Ishisaki2007PASJ...59S.113I}.	
Two XIS0 and XIS3 spectra were 
	summed up, with the rmf and arf also combined.

% === HXD === 
From the deadtime-corrected HXD-PIN spectrum of each source, 
	we subtracted the Non X-ray Background (NXB), 
	created 
	with the LCFITDT method \citep{Fukazawa2009PASJ...61S..17F},
	and filtered 
	with the same criteria as those used in the observed on-source data. 
The Cosmic X-ray Background (CXB) 
	was modeled as described by \cite{Enoto2010PASJ...62..475E}
	utilizing the refined spectral model as given by \cite{Moretti2009A&A...493..501M}.
Corresponding standard response files are used for this HXD-PIN spectrum.
Thanks to the collimated field of view (34'$\times$34' FWHM), 
	the spectra are free from source contamination 
	except for three sources:
	CXOU~J164710.2-45516 (a nearby bright X-ray source GX~340+0, \citealt{Naik2008PASJ...60..237N}), 
	AX~J1818.8$-1559$ (a nearby hard source AX~J1819.2$-$1601),	
	and CXOU~J171405.7$-$381031 (a potential contamination 
	from a surrounding supernova remnant CTB~37B, \citealt{Nakamura2009PASJ...61S.197N}). 

% === GRXE ===
The Galactic Ridge X-ray emission (GRXE; e.g., 
	\citealt{Krivonos2007A&A...463..957K})
	was further subtracted from the HXD-PIN spectrum
	when targets are close to the Galactic center\footnote{
	The GRXE are subtracted from SGR~1806$-$20, 
	1RXS~J170849.0$-$400910,
	SGR~1833$-$0832,
	and 1E~1547.0$-$5408.
	For example,
	a blank-sky data (ObsID 500008010) was used for SGR 1806$-$20,
	while an observation of supernova remnant G25.5+0.0 (ObsID 504099010), 
	which emits negligible signals in the HXD band,
	was utilized for SGR~1833$-$0832.}.
We fixed the GRXE photon index at 2.1 \citep{Valinia1998ApJ...505..134V}
	and employed normalization adjusted to reproduce 
	near-by blank sky observations.
The GRXE contribution is typically $\lesssim$3\% of the NXB.

In the following analyses,
	all uncertainties quoted are given at the 68\% (1$\sigma$) confidence level
	for one parameter of interest unless stated otherwise.
%	all errors quoted here are statistical uncertainties at the 68\% confidence level
%	for one parameter of interest unless stated otherwise.

% ==========================================
\subsubsection{Detections of the soft and hard X-rays}
% ==========================================

% ----------------------------------------------------
\begin{figure}[h]
\begin{center}
\includegraphics[width=70mm]{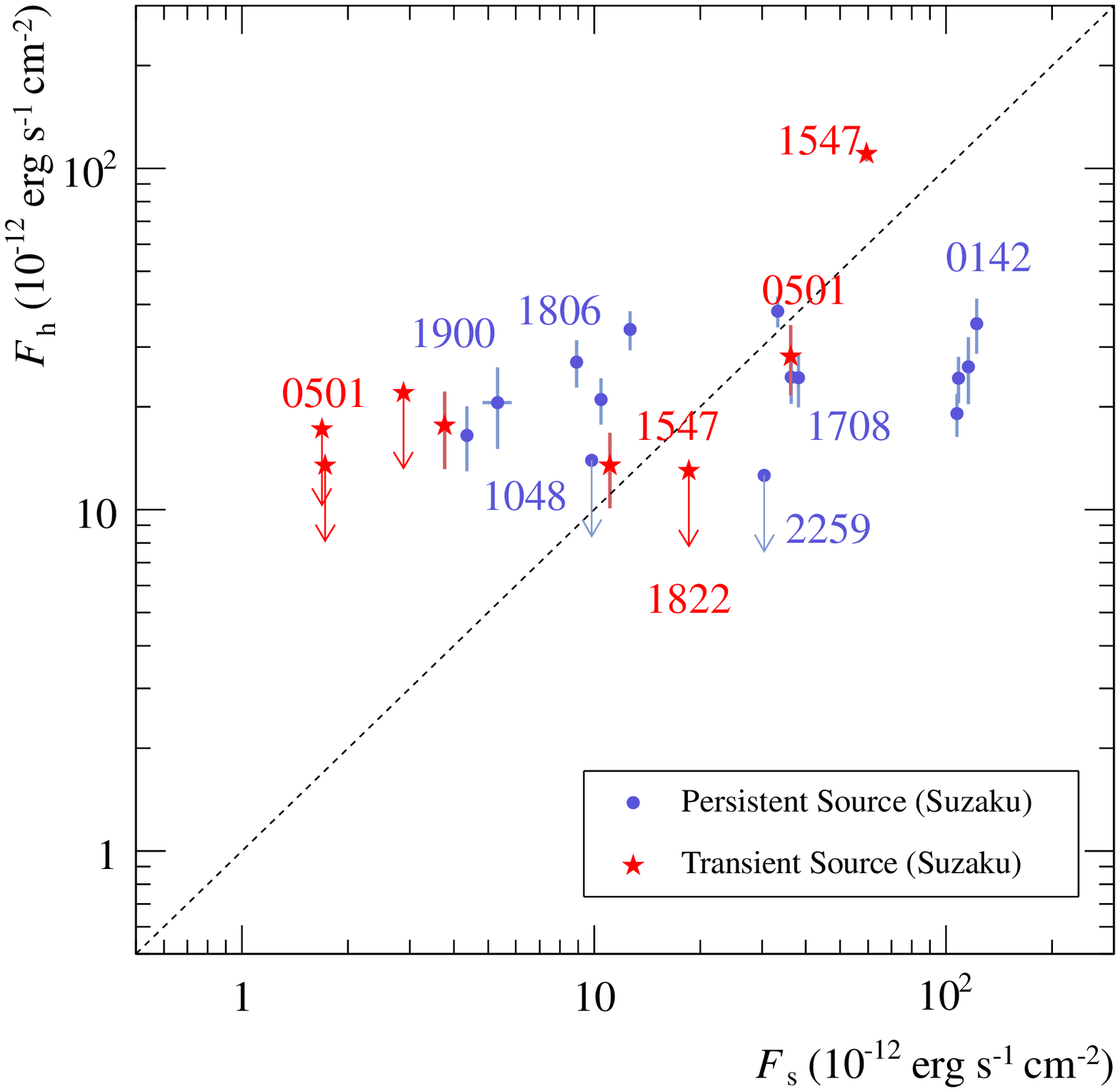}
\caption{
Comparison of the 1--10\,keV (XIS) and 15--60\,keV (HXD-PIN)
	absorbed soft and hard fluxes, $F_{\rm s}$ and $F_{\rm h}$,
	respectively. 
The first four digits of the source names are indicated. 
}
\label{fig:Fs_vs_Fh}
\end{center}
\end{figure}
% ----------------------------------------------------	

% ---------------------------------------
% ~~~~~~~~~~~~~~~~~~~~~~~~~~~~~~~~~~~~~~~~~~~~~~~~~~~~
\begin{deluxetable}{lrrrr}
\renewcommand{\arraystretch}{1.5}
%\tablewidth{290pt}
%\tabletypesize{\scriptsize}
\tablecaption{HXD-PIN source rate and fluxes of the HXC.
	\label{tab:suzaku_detections}
	}
\tablewidth{0pt}
\tablehead{
\colhead{Name}  &
\colhead{ObsID} &
\colhead{$T_{\rm pin}^{\rm a}$} &
\colhead{$R_{\rm pin}^{\rm b}$} &
\colhead{$F_{\rm h}^{\rm c}$} 
\\
 &
 & 
(ks) & 
 &
}
% -------------------
\startdata
% -------------------
% \multicolumn{3}{c}{Persistent Sources \rule[0mm]{0mm}{3mm}} \\ 
% \hline
1806$-$20 & 401092010 & 51.9 & $6.0\pm 0.3\pm 0.3$ & $33.7\pm 4.4$  \\
1806$-$20 & 401021010 & 15.6 & $5.8\pm 0.5\pm 0.5$ & $21.1\pm 3.3$  \\
1806$-$20 & 402094010 & 46.6 & $4.7\pm 0.3\pm 0.3$ & $27.1\pm 4.3$  \\
1841$-$04 & 401100010 & 59.8 & $6.7\pm 0.3\pm 0.3$ & $48.9\pm 0.3$  \\
1900$+$14 & 401022010 & 13.5 & $3.4\pm 0.6\pm 0.6$ & $20.6\pm 5.5$  \\
1900$+$14 & 404077010 & 39.1 & $3.5\pm 0.3\pm 0.3$ & $16.5\pm 3.5$  \\
%1714$-$38 & 501007010 & 64.6 & $\cdots$ & $\cdots$  \\
1708$-$40 & 404080010 & 47.9 & $4.0\pm 0.3\pm 0.3$ & $24.4\pm 4.4$  \\
1708$-$40 & 405076010 & 55.4 & $4.5\pm 0.3\pm 0.3$ & $24.4\pm 4.0$  \\
1048$-$59 & 403005010 & 63.3 & $<2.3$ & $<13.2$   \\
0142$+$61 & 402013010 & 94.7 & $4.1\pm 0.2\pm 0.2$ & $35.1\pm 6.4$  \\
0142$+$61 & 404079010 & 92.5 & $3.1\pm 0.2\pm 0.2$ & $26.2\pm 5.8$  \\
0142$+$61 & 406031010 & 39.3 & $4.7\pm 0.3\pm 0.3$ & $24.3\pm 3.7$  \\
0142$+$61 & 408011010 & 96.2 & $3.8\pm 0.2\pm 0.2$ & $19.1\pm 2.8$  \\
2259$+$58 & 404076010 & 96.0 & $<1.8$ & $<10.1$   \\
%1818$-$15 & 406074010 & 77.8 & $\cdots$ & $\cdots$  \\
1547$-$54 & 903006010 & 31.0 & $17.4\pm 0.4\pm 0.4$ & $110.2\pm 5.2$  \\
1547$-$54 & 405024010 & 39.6 & $3.2\pm 0.3\pm 0.3$ & $13.5\pm 3.3$  \\
0501$+$45 & 903002010 & 50.7 & $3.0\pm 0.3\pm 0.3$ & $28.1\pm 6.5$  \\
0501$+$45 & 404078010 & 25.5 & $<3.7$ & $<20.9$   \\
0501$+$45 & 405075010 & 48.9 & $<2.9$ & $<16.2$   \\
0501$+$45 & 408013010 & 33.4 & $<2.3$ & $<12.7$   \\
1833$-$08 & 904006010 & 10.0 & $3.9\pm 0.7\pm 0.7$ & $17.6\pm 4.5$  \\
%1647$-$45 & 901002010 & 32.9 & $\cdots$ & $\cdots$  \\
1822$-$16 & 906002010 & 33.7 & $<0.9$ & $<4.9$   
\enddata
\tablenotetext{a}{$T_{\rm pin}$: The effective HXD exposure.}
\tablenotetext{b}{The 15--60 keV HXD-PIN count rates with $1\sigma$ statistical and systematic errors. 
If not detected, the $3\sigma$ upper-limits are shown. 
}
\tablenotetext{c}{The 15--60 keV absorbed flux 
	in an unit of $10^{-12}$\,ergs\,s$^{-1}$\,cm$^{-2}$ with 1$\sigma$ statistical and systematic errors. 
If not detected, the $3\sigma$ upper-limits are shown converted from the count rate $R_{\rm pin}$
	with a conversion factor of $F_{\rm x}/R_{\rm pin}=6\times 10^{-12}$\,erg\,s$^{-1}$\,cm$^{-2}$/(0.01\,cnt\,s$^{-1})$. 
	}	
\tablenotetext{d}{The \suzaku detected HXC above were also reported with other satellite:
	4U~0142+61, 1RXS~J170849.0$-$400910 
	\citep{Kuiper2006ApJ...645..556K, Hartog2008A&A...489..245D, 
	Hartog2008A&A...489..263D}; 
	1E~1841$-$045 \citep{Hongjun2013ApJ...779..163A};
	SGR~1806$-$20 
	SGR~1900$+$14 \citep{Gotz2006A&A...449L..31G}; 
	1E~1547.0$-$5408 \citep{	Kuiper2012ApJ...748..133K};
	SGR~0501+4516 \citep{Rea2009MNRAS.396.2419R}.
	}		
\end{deluxetable}
% ~~~~~~~~~~~~~~~~~~~~~~~~~~~~~~~~~~~~~~~~~~~~~~~~~~~~	 

% ---------------------------------------

% === fitting information ===
Figure~\ref{fig:suzaku_raw_spec} 
	illustrates nine examples of the 1--10 keV XIS and 15--60 keV HXD-PIN spectra
	derived by the procedures in \S\ref{Reduction of Broad-band X-ray Spectra}.
The SXC below 10 keV was clearly detected with the XIS 
	from all the observations in Table~\ref{tab:list_suzaku_log}
	except 
	Swift\,J1834.9$-$0846
	which had already been faded 
	to become undetectable. 
Hereafter, we analyze the other sources. 
	
% ~~~~~~~~~~~~~~~~~~~~~~~~~~~~~~~~~~~~~~~~~~~~~~~~~~~~
\begin{deluxetable*}{lcccccccc}[H]
%\tabletypesize{\scriptsize}
%\tabletypesize{\small}
\tablecaption{
List of archived {\it NuSTAR} and {\it Swift} observations of SGRs and AXPs used in this work.
\label{tab:list_of_nustar}
}
\tablewidth{0pt}
% --------------------
\tablehead{
\colhead{Name} &
\multicolumn{3}{c}{{\it NuSTAR}} &
& 
\multicolumn{3}{c}{{\it Swift/XRT}} &
\colhead{Ref.}
\\
\cline{2-4} 
\cline{6-8} 
\\
&
\colhead{ObsID} &
\colhead{Obs. Date} &
\colhead{(ks)} & 
& 
\colhead{ObsID} &
\colhead{Obs. Date} &
\colhead{(ks)} & 
%\\
}
\startdata
%1E~1841$-$045 &
%	30001025004,6,8,10,12 & 
%	2012-11-09, 2013-09-05,07,12,14,21 & 
%	35, 66, 38, 33, 101, 37, 16, 16, 88 &
%	\cite{An2013ApJ...779..163A}
%\\
%1E~1841$-$045 &
%	30001025004,6, & 
%	2012-11-09, 2013-09-05,07, & 
%	35, 66, 38, 33, 101, 37, &
%	\cite{An2013ApJ...779..163A}
%\\
%&
%	8,10,12 & 
%	2013-09-12,14,21 & 
%	16, 16, 88 &
%	-- 
%\\
%4U~0142+61 & 
%	30001023002, 3 & 
%	2014-03-27,28 & 
%	24, 144 & 
%	-- 
1E~1841$-$045 & 
	30001025[04,06,08,10,12] & 
	13 Sep 5-23 & 
	273 & 
	& 
%	000318630[50,51],00080220004 &
	00080220004 &
	13 Sep 21 & 
	1.8 & 
	[1]
\\	
%1E~1841$-$045 & 
%	- & 
%	- & 
%	- & 
%	& 
%	- &
%	- &
%	- & 
%	--
%\\	
%1E~1048.1$-$5937 & 
%	300010240[02,03,05,07] & 
%	13 July 17--27 & 
%	316 & 
%	& 
%	000312202[43,44,45] &
%	13 July 15--28 &
%	4(w),17(p) & 
%	[2]
%\\	
4U~0142+61 & 
	300010230[02,03] &
	14 March 27--30 & 
	168 &
	& 
	000800260[01--03] & 
	14 March 27--30 & 
	24 & 
%	24(w) & 	
	[2]
\\	
1E~2259$+$586 & 
	300010260[02,03,05] &
	13 April 24--27 &
	157 &
	&
	000802920[02,03,04] &
	13 April 25--28 &
	30 &
%	30(p) &
	[3]
\\	
1E~2259$+$586 & 
	30001026007 &
	13 May 16--18 &
	88 &
	&
	000802920[05,07] &
	13 May 16--18 &
	8.7 &
	--- 
%1E~1048.1$-$5937 & 
%	30001024002,3,5,7 & 
%	2013-07-17,19,25 & 
%	25, 24, 157, 111   &
%	\cite{An2014ApJ...790...60A}
%\\
%1E~2259$+$586 & 
%	30001026002,3,5,7 & 
%	2013-03-24,25,26, 05-16 & 
%	-- & 
%	\cite{Vogel2014ApJ...789...75V}
%\\
%SGR~1745$-$29 & 80002013002--26 & --
\enddata
\tablecomments{
	Data of the Galactic center soft gamma repeater SGR~1745$-$29 (80002013002--26)
	is not yet available. 
}
\tablecomments{
	[1] \cite{An2013ApJ...779..163A}; %[2] \citealt{An2014ApJ...790...60A}; 
	[2] \cite{2015ApJ...808...32T}
	[3] \cite{Vogel2014ApJ...789...75V}
}
%\tablenotetext{a}{
%Time onset of the outbursts defined at the first short burst detected by {\it Swift}/BAT or {\it Fermi}/GBM.
%See the above references. 
%}
\end{deluxetable*}
% ~~~~~~~~~~~~~~~~~~~~~~~~~~~~~~~~~~~~~~~~~~~~~~~~~~~~
	
After the NXB and CXB subtractions,
	 we tabulate in Table~\ref{tab:suzaku_detections}
	 the 15--60 keV HXD-PIN source count rates $R_{\rm pin}$ 
	 together with 1$\sigma$ statistical and systematic uncertainties.
The systematic uncertainty of the HXD-PIN background  
	is a sum of 1\% level of the NXB 
	(reproducibility of the LCFITDT model; \citealt{Fukazawa2009PASJ...61S..17F})
	and 10\% of the CXB (1$\sigma$ sky-to-sky fluctuation).
The associated 15--60 keV flux $F_{\rm x}$ was calculated via 
	power-law fitting of the HXD data.
If the HXC is undetectable, 
	we set 3$\sigma$ upper-limits on the count rates,
	which were converted to those on $F_{\rm x}$ 
	assuming spectral shapes of detected sources 
	(caption of Table.\ref{tab:suzaku_detections}).
We assigned 3$\sigma$ upper limits on 1E~2259$+$586, 1E~1048.1$-$537,
	Swift~J1822.3$-$1606, and latter observations of SGR~0501$+$4516.
The upper limit on  1E~2259$+$586 is consistent with 
	a recent detection by NuSTAR \citep{Vogel2014ApJ...789...75V}.
		
As reported in Paper I and references therein,
	we have detected the HXC from 7 sources 
	with $>$3$\sigma$ significance
	(see the caption in Table~\ref{tab:list_suzaku_log}).
Out of 10 new observations of 6 sources added to Paper~I,
	the HXC was reconfirmed from 4U~0142+61 in 2011, 2013 
	and RXS~J170849.0$-$400910 in 2010.
A signature of the HXC was suggested in SGR~1833$-$0832,
	but $F_{\rm h}$ is rather weak compared with other sources,
	and the detection is marginal. 
Thus, we do not use this source in the correlation fittings in \S\ref{sec:Analysis_and_Results}. 
Figure~\ref{fig:Fs_vs_Fh} illustrates a comparison of 
	$F_{\rm s}$ and $F_{\rm h}$.
\suzaku has detected the HXC
	down to $\sim 10^{-11}$\,ergs\,cm$^{-2}$\,s$^{-1}$ in the 15--60 keV band. 

% ===========================
\subsection{NuSTAR Observations}
\label{sec:nustar}
% ===========================

The Nuclear Spectroscopic Telescope Array ({\it NuSTAR},
	\citealt{Harrison2013ApJ...770..103H}) 
	provides a high spectral sensitivity in the 3--79~keV band,
	and has already observed bright AXPs
	\citep{An2014AN....335..280A,
	An2013ApJ...779..163A,
	An2014ApJ...790...60A,
	Vogel2014ApJ...789...75V,
	2015ApJ...808...32T,
	2015arXiv150908115Y}.
In order to verify our results  
	performed by the non-imaging instrument HXD,
	here we analyze initial {\it NuSTAR} data sets available in the archive
	listed in Table~\ref{tab:list_of_nustar}.

% ----------------------------------------------------
\begin{figure}[h]
\begin{center}
\includegraphics[height=57mm]{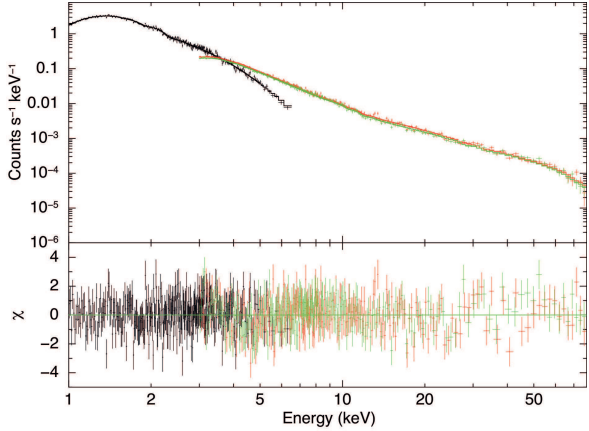}
\caption{
Background-subtracted {\it NuSTAR} ($>$2.5 keV)
	and {\it Swift} X-ray spectra 
	of 4U~0142$+$61 in 2014 March. 
}
\label{fig:nustar_spec}
\end{center}
\end{figure}
% ----------------------------------------------------  

% ----------------------------------------------------
% ~~~~~~~~~~~~~~~~~~~~~~~~~~~~~~~~~~~~~~~~~~~~~~~~~~~~
%\begin{deluxetable*}{p{7.2pc}cp{1.5pc}p{1.4pc}p{1.9pc}p{2.2pc}ccc}[h]
\begin{deluxetable*}{lrcrrrrrrrr}
%\tabletypesize{\scriptsize}
%\tabletypesize{\small}
\tablecaption{List of X-ray outbursts from transient sources. 
        \label{tab:list_of_outbursts}}
\tablewidth{0pt}
% --------------------
\tablehead{
\colhead{Name} &
\colhead{$B_{\rm d}$} &
\colhead{Time Origin$^{\rm a}$} &
\colhead{Ref.} & 
\colhead{Model$^{\rm b}$}       & 
%\multicolumn{6}{c}{PD function$^{c}$} 
\colhead{$L_0$$^{\rm c}$}       & 
\colhead{$\tau_0$$^{\rm c}$}    &
\colhead{$p$$^{\rm c}$}    &
\colhead{$E_{\rm tot}$$^{\rm c}$}    
\\
&
\colhead{($10^{14}$\,G)} &
(UT) & & 
& 
\colhead{($10^{34}$\,erg\,s$^{-1}$)} &
\colhead{(d)} & 
\colhead{} & 
\colhead{(erg)} 
}
\startdata
CXOUJ164710$-$455216 & 1.6      & 2006-09-21 01:34:52 & 1a-d & PL    & 8.2$\pm$0.3 & --           & 0.23$\pm$0.02 & $3.1\times 10^{41}$   \\
SGR~0501$+$4516          & 1.9      & 2008-08-22 12:41:59 & 2a-d & PD    & 5.9$\pm$0.2 & 15.9$\pm$2.9 & 0.76$\pm$0.05 & $1.5\times 10^{43}$  \\
1E~1547.0$-$5408           & 3.2      & 2008-10-03 09:28:08 & 3a-d &  PL    & 9.1$\pm$0.4 & --           & 0.13$\pm$0.02 & $4.8\times 10^{41}$ \\
1E~1547.0$-$5408           & 3.2      & 2009-01-22 01:32:41 & 4a-e & PL     & 25.7$\pm$0.6 & --           & 0.29$\pm$0.01 & $7.9\times 10^{41}$ \\
SGR~0418$+$5729          &  0.061 & 2009-06-05 20:40:48 & 5a-e &  PD   & 6.1$\pm$0.4 & 42.9$\pm$11.6 & 1.99$\pm$0.30 & $2.3\times 10^{41}$ \\
SGR~1833$-$0832           & 1.8      & 2010-03-19 18:34:50 & 6a-b &  PL   & 8.8$\pm$1.2 & --           & 0.07$\pm$0.06 & $5.8\times 10^{41}$   \\
Swift~J1822$-$16069       & 0.14    & 2011-07-14 12:47:47 & 7a-c &   PD  & 7.2$\pm$0.3 & 11.2$\pm$0.9 & 1.25$\pm$0.02 & $2.9\times 10^{41}$  \\
Swift~J1834.9$-$0846      & 1.4      & 2011-08-07 19:57:46 & 8a-c &   PL   & 1.8$\pm$0.4 & --           & 0.28$\pm$0.09 & $5.7\times 10^{40}$ 
\enddata
\tablenotetext{a}{
Time onset of the outbursts defined at the first short burst detected by {\it Swift}/BAT or {\it Fermi}/GBM.
See the above references. 
}
\tablenotetext{b}{
Applied fitting model used to unabsorbed 2--10 keV X-ray light curves; 
A single power-law model (PL, $L_{\rm x}(t)=L_0(t/1\,{\rm d})^{-p}$) 
and 
a plateau decaying function (PD, $L_{\rm x}(t)=L_0(1+t/\tau_0)^{-p}$).
}
\tablenotetext{c}{
Best-fit parameters of unabsorbed 2--10 keV X-ray light curves (see details in \S\ref{section:outburst_lightcurve}).
%	A plateau flux $K_{\rm plt}$ ($10^{-11}$\,erg\,s$^{-1}$\,cm$^{-2}$), 
%	its duration $\tau_{\rm plt}$ (d), and decaying power-law slope $p$.
%Corresponding total fluence $S_{\rm tot}$ ($10^{-9}$\,erg\,cm$^{-2}$) 
%	and emitted energy $E_{\rm tot}$ ($10^{41}$\,erg) 
%	of individual outbursts are also listed.
$L_{\rm 0}$ is an initial X-ray luminosity at 1 day after the onset or during the plateau phase 
for the PL and PD models, respectively. 
$\tau_{\rm 0}$ and $p$ are a duration of the plateau and the slope, respectively. 
The total emitted energy $E_{\rm total}$ is evaluated as 
	an integration up to 100\,d if $p<1$.	
}
\tablecomments{
X-ray outbursts after the {\it Suzaku} and {\it Swift} launches, 
i.e., outbursts from 1E~2259+586 in 2002, XTE~J1810-197 in 2003, and 1E 1048.1-5937 in 2007 are not included. 
Further details of individual outbursts are listed in Ref., 
(1a) \citet{Krimm2006GCN..5581....1K}; 
(1b) \citet{Israel2007ApJ...664..448I};
(1c) \citet{Naik2008PASJ...60..237N};
(1d) \citet{Clark2005A&A...434..949C};
%(2a) \citet{Palmer2008ATel.1548....1P};
%(2b) \citet{Esposito2008MNRAS.390L..34E};
%(2c) \citet{Corbel1999ApJ...526L..29C};
(2a) \citet{Holland2008GCN..8112....1H};
(2b) \citet{Enoto2009ApJ...693L.122E};
(2c) \citet{Rea2009MNRAS.396.2419R};
(2d) \citet{Aptekar2009ApJ...698L..82A};
(3a) \citet{Krimm2008GCN..8311....1K}; 
(3b) \citet{Israel2010MNRAS.408.1387I}; 
(3c) \citet{Tiengo2010ApJ...710..227T};
(4a) \citet{Gronwall2009GCN..8833....1G};
(4b) \citet{Enoto2010PASJ...62..475E};
(4c) \citet{Ng2011ApJ...729..131N};
(5a) \citet{Horst2009ATel.2077....1V};
(5b) \citet{Horst2010ApJ...711L...1V};
(5c) \citet{Rea2010Sci...330..944R};
(5d) \citet{Esposito2010MNRAS.405.1787E};
(5e) \citet{Guver2011MNRAS.418.2773G};
(5d) \citet{Rea2013ApJ...770...65R};
(6a) \citet{Gelbord2010GCN..10526...1G};
(6b) \citet{Gogus2010ApJ...718..331G};
(6c) \citet{Esposito2011MNRAS.416..205E};
(7a) \citet{Cummings2011ATel.3488....1C};
(7b) \citet{Rea2012ApJ...754...27R};
(7c) \citet{Scholz2012ApJ}; % read 
(8a) \citet{DElia2011GCN..12253...1D};
(8b) \citet{Kargaltsev2012ApJ...748...26K};
(8c) \citet{Esposito2013MNRAS.429.3123E};
%(9a) \citet{Kennea2013ApJ...770L..24K};
%(10a) \citet{Stamatikos2014GCN..16520...1S};
}
\end{deluxetable*}
% ~~~~~~~~~~~~~~~~~~~~~~~~~~~~~~~~~~~~~~~~~~~~~~~~~~~~

% ----------------------------------------------------

The data were processed and filtered 
	with the standard \texttt{nupipeline} and \texttt{nuproducts} softwares
	of \texttt{HEASOFT} version 6.16 
	and the {\it NuSTAR} CALDB version 20141020. 
The on-source spectra were extracted 
	from a circular region of 1$'$.0 radius centered on the target objects 
	within which nearly 90\% photons are collected.
Since some AXPs are faint hard X-ray sources,
	we used the background modeling software
	\texttt{nuskybkg} \citep{Wik2014ApJ...792...48W}
	for accurate background subtraction.
This tool generates a simulated background spectrum 
	expected on the selected source region
	by fitting blank-sky spectra from the same focal plane.
We selected,
	for the background spectral modeling,
	three annual regions	
	with radii 2$'$.0--5$'$.0, 5$'$.0--8$'$.0, and 8$'$.0--12$'$.3 
	centered on the target sources for each telescope,
	and adjusted model parameters
	to explain the actual blank-sky data.
The background spectrum was simulated 
	from the best fit modeling parameters
	with a 100 times longer exposure to reduce statistical uncertainties. 
The accuracy of the simulated background spectrum 
	was verified using blank sky data as described in \cite{Kitaguchi2014ApJ...782....3K}. 	

If there are simultaneous {\it Swift} coverage of 
	 during the {\it NuSTAR} observations,
	 we also utilized the archived {\it Swift}/XRT data
	 listed in Table~\ref{tab:list_of_nustar},
	after the data processing as described in \S\ref{subsec:swift_rxte_observations}.
If the observations were carried out in a sequential way during a month,
	we merged continuous data with different observation ID 
	into one spectrum and response 
	for our long-term and phase averaged analyses. 
Two series of observations of 1E~2259$+$586 
	were performed in 2013 April and March, 
	so we derived two spectra (Table~\ref{tab:list_of_nustar}). 
As an example,	
	we show the background-subtracted X-ray spectra of 
	4U~0142+61 in Figure~\ref{fig:nustar_spec}.
The derived {\it NuSTAR} spectra are consistent with previous works by
	\citealt{An2013ApJ...779..163A},	
	\citealt{Vogel2014ApJ...789...75V},
	and \citealt{2015ApJ...808...32T}.

% ==========================================
\subsection{Swift and RXTE Observations of Outbursts}
\label{subsec:swift_rxte_observations}
% ==========================================

% ----------------------------------------------------
\begin{figure}[!t]
\begin{center}
\includegraphics[width=88mm]{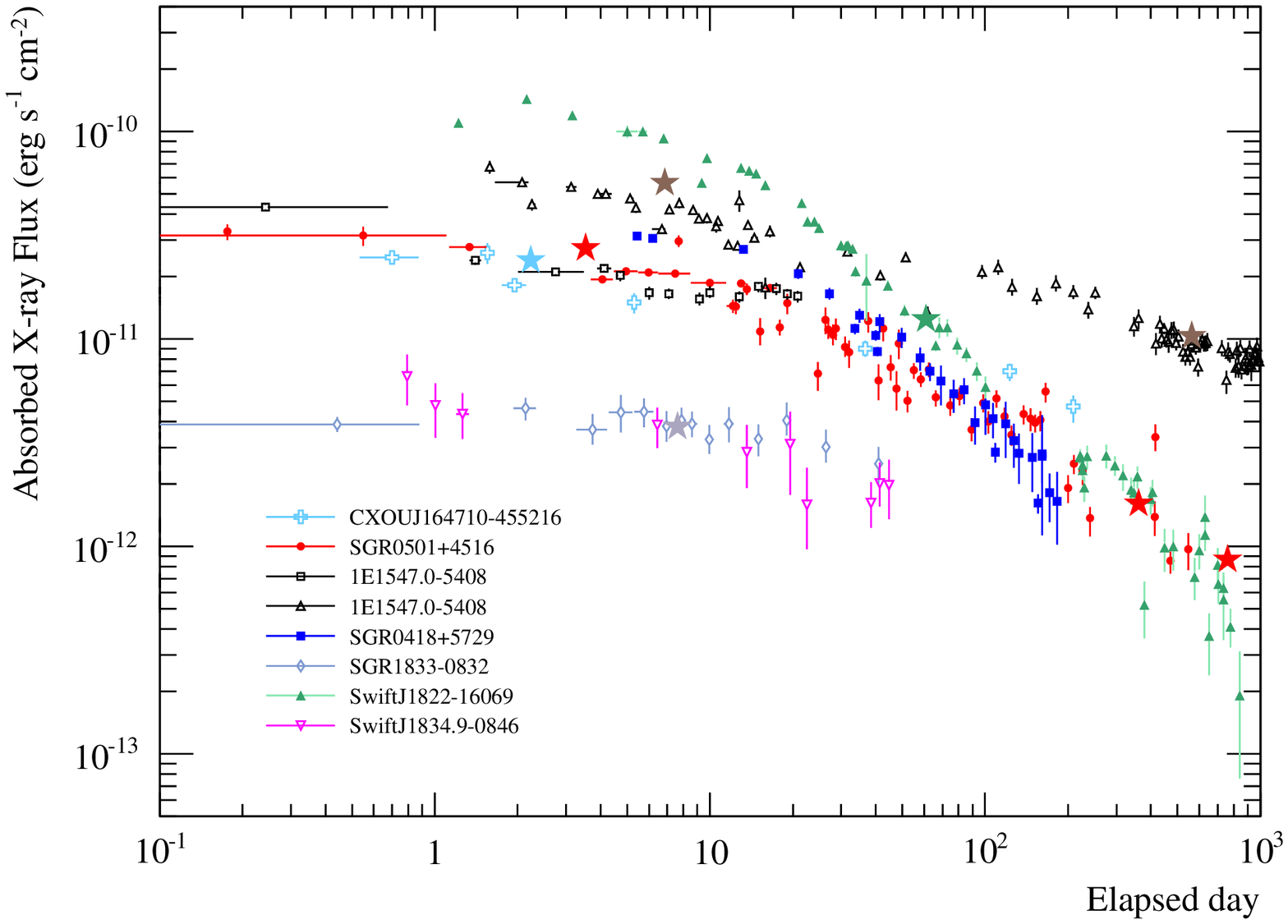}
\caption{
Absorbed 2--10\,keV flux decay 
	of known 8 outbursts of transient SGRs and AXPs
	with time onsets defined at the first short burst detected by {\it Swift}/BAT or {\it Fermi}/GBM
	listed in Table~\ref{tab:list_of_outbursts}.
Star symbols (same color as the legend) represent {\it Suzaku}/XIS observations.
The 2--10 keV flux of $2\times 10^{-11}$\,erg\,cm$^{-2}$\,s$^{-1}$
	corresponds to 1\,mCrab intensity. 
}
\label{fig:outburst_flux_decay}
\end{center}
\end{figure}
% ----------------------------------------------------	

Transient magnetars are 
	characterized by sporadic X-ray outbursts,
	namely sudden increases of the persistent emission,  
	which have recently provided a drastic increase of the number of this class 
	(e.g., as early studies, 
	\citealt{Gavriil2002Natur.419..142G,Kouveliotou2003ApJ...596L..79K,Kaspi2003ApJ...588L..93K}
	and for a recent review, \citealt{Rea2011heep.conf..247R}).	
The onset of an outburst is usually 
	noticed by a detection of short bursts 
	by the Swift Burst Alert Telescope (BAT;
	\citealt{Gehrels2004ApJ...611.1005G,Barthelmy2005SSRv..120..143B}),
	and then monitored 
	with the Swift X-ray Telescope (XRT; \citealt{Burrows2005SSRv..120..165B})	
	or
	with 
	{\it Rossi X-Ray Timing Explorer} 
	({\it RXTE}; \citealt{Bradt1993A&AS...97..355B}), 
	as shown in Figure~\ref{fig:outburst_flux_decay}.

% Suzaku observations
We conducted \suzaku ToO observations of some transients 
	usually within a week after.
The obtained \suzaku snap shots 
	are also presented in Figure~\ref{fig:outburst_flux_decay}.
We studied characteristics of outbursts 
	of latest known 8 outbursts of 7 sources 
	which occurred after the \suzaku launch,
	as listed in Table~\ref{tab:list_of_outbursts}
	(see also Figure~\ref{fig:transient_vs_persisten}).
The onset of an outburst is defined 
	as the first short bursts reported by {\it Swift}/BAT 
	or {\it Frermi}/Gamma-ray Burst Monitor (GBM). 

% Swift/XRT 
We uniformly processed  
	all the public available XRT data of 8 outbursts 
	via the standard procedure FTOOLS {\tt xrtpipeline} 	with default filtering criteria.
We used the latest available RMF matrix in CALDB v20140610,
	while generated the ARF files with the {\tt xrtmkarf} tool. 
For the imaging Photon Counting (PC) mode, 
	we extracted source photons from 
	a circular region with a 48\arcsec (20 pixels) radius centered on the target,
	while collected background spectra from annular regions 
	with the inner and outer radii of 167\arcsec (70 pixels) and 286\arcsec (120 pixels), respectively.
When the PC-mode count rates exceed $\sim$0.5 count s$^{-1}$, 
	we excluded a central 8.0\arcsec (3.4 pixels) region 
	following a standard procedure\footnote{http://www.swift.ac.uk/analysis/xrt/pileup.php}.
For the Windowed Timing (WT) mode with 
	an one-dimensional information and 1.76\,ms time resolution,
	we extracted source and background spectra 
	from a strip of 94\arcsec  width around the source
	and surrounding regions by 140\arcsec  away from the target,
	respectively. 
When the count rate exceeds 100\,cnt\,s$^{-1}$,
	we excluded the central 14\arcsec strip to avoid the pile-up. 
After the above standard pipelines, 
	we discarded observations with poor photon statistics
	if the total source count is smaller than 100 cnts per observation.

% RXTE/PCA 	
The non-imaging {\it RXTE} Proportional Counter Array (PCA; \citealt{Jahoda2006ApJS..163..401J})
	operates in the 2--60 keV energy band 
	with a full width at half-maximum field of view of $\sim$1$^{\circ}$.
Due to nearby sources, 
	we only used the PCA data for 
	SGR~0501$+$4516
	and SGR~0418$+$5729.
The data were processed via the standard procedure 
	using FTOOLS {\tt rex}, {\tt pcarsp}, and {\tt recofmi} tasks.

% ==========================================
\section{Analysis and Results} 
\label{sec:Analysis_and_Results}
% ==========================================

% ==========================================
\subsection{Spectral modeling of two X-ray components}
\label{SpectralModeling}
% ==========================================

% ----------------------------------------------------
\begin{figure*}[!t]
\begin{center}
\includegraphics[width=182mm]{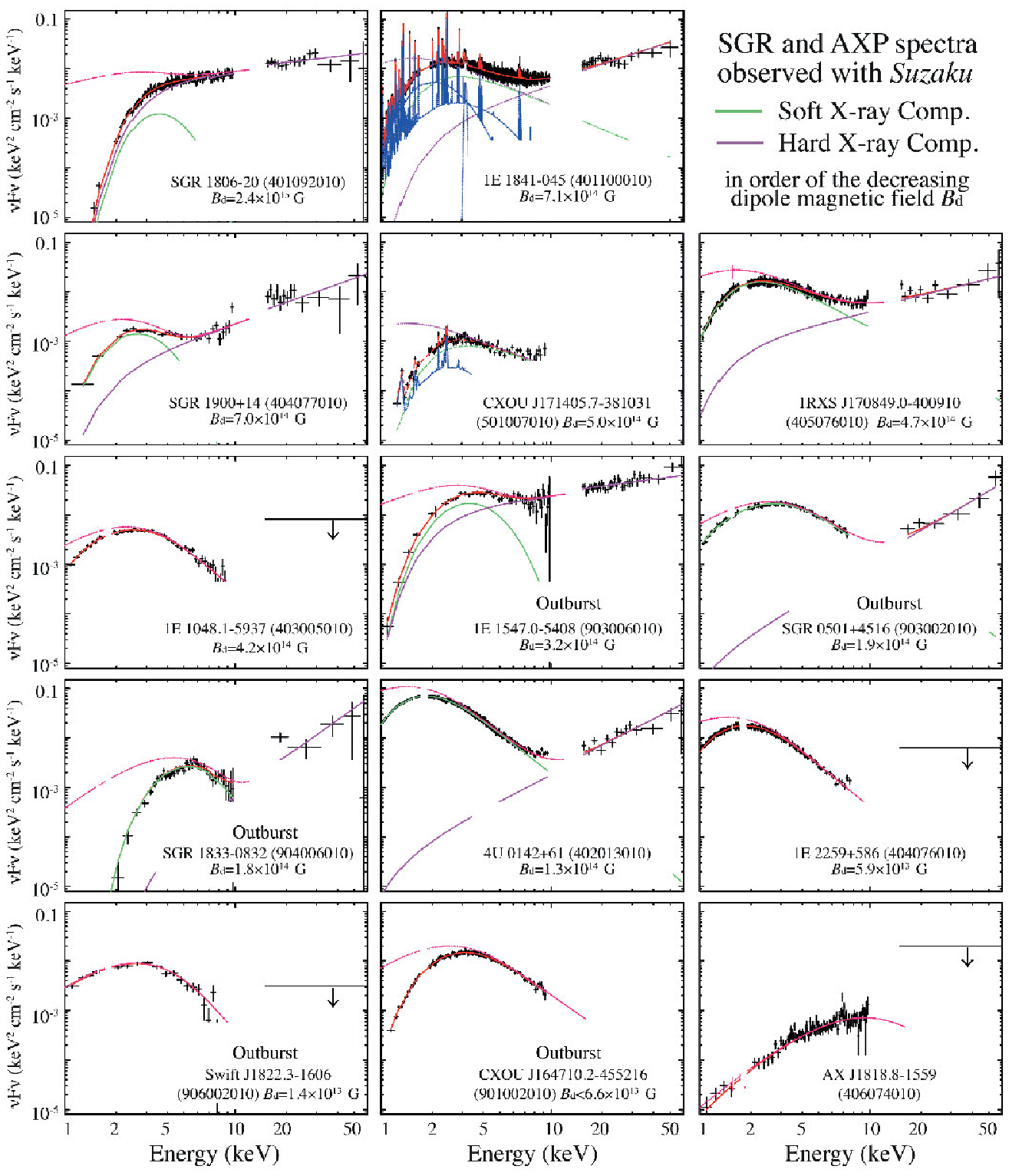}
\vspace{2mm}
\caption{
\suzaku $\nu F \nu$ spectra (black points)
	of persistent X-ray emission from fourteen SGRs and AXPs,
	deconvolved using the quasi-thermal CBB (green)
	and the hard PL (purple) model.
The photoelectric absorption is included. 	
Corresponding best-fit parameters are listed in Table~\ref{tab:suzaku_fit_summary_cbb}.
Absorption-corrected best-fit model is shown in pink. 
The spectra are shown in the order of decreasing dipole field $B_{\rm d}$ except for CXOU~J164710.2$-$455216 (only the upper limit) and a magnetar candidate AX~J1818.8$-$1559.
The SNR components are shown in blue  
	for 1E~1841-045 (Kes~73) 
	and 
	CXOU~J171405.7$-$381031 (CTB~37B).
}
\label{fig:suzaku_spec_eeuf}
\end{center}
\end{figure*}
% ----------------------------------------------------	

We carried out 
	unified spectral fitting of 
	the \suzaku phase-average broad-band spectra.
Phase-resolved spectroscopy, 
	available only for bright and slowly rotating objects (e.g., 1RXS J18049.0$-$400910),
	is beyond the scope of this paper.
In order to avoid instrumental calibration uncertainties of the XIS,
	the 1.7--1.9 keV data were discarded, 
	and 
	a 2\% systematic error was assigned to the XIS spectral bins. 
The XIS and HXD-PIN spectra were binned 
	so that each bin has either $>$5$\sigma$ significance or $>$30 counts.
The cross-normalization factor of HXD-PIN was fixed at 1.164 and 1.181 
	relative to XIS-FI in cases of the XIS- and HXD-nominal pointing positions
	\citep{MaedaSuzakuMemo2008}, respectively.
The normalization of XIS-BI to XIS-FI was allowed to differ by up to 5\%.
For the simultaneous {\it Swift} and {\it NuSTAR} spectra (\S\ref{sec:nustar})
	the normalization was fixed to that of the {\it Swift}/XRT 
	and the cross-normalization between the telescopes (FPMA and FPMB)
	was allowed by up to 5\%.

\begin{deluxetable*}{llllccccccccccccccc}[h]
%\tabletypesize{\scriptsize}
\tablecaption{
Best-fit parameters of SGR/AXP observations using the CBB+PL model with {\it Suzaku} and {\it NuSTAR}. 
\label{tab:suzaku_fit_summary_cbb}
}
\tablewidth{0pt}
\tablehead{
\colhead{Name} &
\colhead{ObsID} &
\colhead{$F_{1-10}$} &
\colhead{$F_{15-60}$} &
\multicolumn{2}{c}{Unabsorbed $F_{\rm s}$,$F_{\rm h}$} &
\colhead{$N_{\rm H}$}  &
\colhead{$kT$} & 
\colhead{$R$} &
\colhead{$\Gamma_{\rm s}$} &
\colhead{$\Gamma_{\rm h}$} &
\colhead{$\chi^2_{\nu}$ (dof)} &
\\
\cline{5-6}
 &
\colhead{Month} &
 &
 & 
\colhead{SXC} &
\colhead{HXC} &
 {\scriptsize ($10^{22}$\,cm$^{-2}$)} &	
(keV) & 
(km) & 
 & 
 & 
 &
}
% -------------------
\startdata
% -------------------
% ~~~~~~~~~~~~~~~~~~~~~~~~~~~~~~~~~~~~~~~~~~~~~~~~~~~~
\multicolumn{12}{c}{{\it Suzaku} observations} \\
1806$-$20 & 401092010 & $12.6_{-0.1}^{+0.1}$ & $33.7$(4.4) & $6.3_{-0.8}^{+0.9}$ & $58.4_{-1.4}^{+1.4}$ & $6.7(3)$ & $0.61(4)$ & 1.8 & $\cdots$ & $1.62$(5) & 0.97 (377)\\
1806$-$20 & 401021010 & $10.5_{-0.2}^{+0.1}$ & $21.1$(3.3) & $3.4_{-0.8}^{+0.9}$ & $50.1_{-2.6}^{+2.9}$ & $5.5(5)$ & $0.68(9)$ & 1.1 & $\cdots$ & $1.51$(9) & 1.18 (171)\\
1806$-$20 & 402094010 & $8.9_{-0.2}^{+0.1}$ & $27.1$(4.3) & $5.9_{-0.8}^{+1.0}$ & $42.2_{-1.7}^{+1.8}$ & $6.5(5)$ & $0.65(6)$ & 1.5 & $\cdots$ & $1.50$(7) & 1.22 (193)\\
1841$-$04 & 401100010 & $19.4_{-0.1}^{+0.1}$ & $48.9$(0.3) & $35.1_{-1.4}^{+1.4}$ & $50.9_{-2.1}^{+2.2}$ & $2.5(1)$ & $0.27(1)$ & 21.1 & 3.41(10) & $0.87$(8) & 1.22 (2087)\\
1900$+$14 & 401022010 & $5.3_{-0.5}^{+0.5}$ & $20.6$(5.5) & $4.6_{-0.6}^{+0.1}$ & $25.0_{-3.4}^{+3.2}$ & $1.8(3)$ & $0.57(2)$ & 2.5 & $\cdots$ & $0.96$(14) & 1.13 (44)\\
1900$+$14 & 404077010 & $4.3_{-0.1}^{+0.1}$ & $16.5$(3.5) & $4.5_{-0.3}^{+0.3}$ & $26.3_{-2.5}^{+2.9}$ & $1.9(1)$ & $0.52(2)$ & 3.0 & $\cdots$ & $0.78$(9) & 1.34 (57)\\
1714$-$38 & 501007010 & $1.7_{-0.1}^{+0.1}$ & $\cdots$ & $5.0_{-0.3}^{+0.5}$ & $\cdots$ & $3.5(1)$ & $0.24(4)$ & 15.7 & 3.25(8) & $\cdots$ & 1.28 (179)\\
1708$-$40 & 404080010 & $38.1_{-0.2}^{+0.5}$ & $24.4$(4.4) & $69.1_{-3.3}^{+3.5}$ & $28.4_{-2.7}^{+2.8}$ & $1.3(1)$ & $0.26(2)$ & 14.2 & 3.48(9) & $0.67$(24) & 0.98 (394)\\
1708$-$40 & 405076010 & $35.8_{-0.1}^{+0.4}$ & $24.4$(4.0) & $55.0_{-2.7}^{+4.0}$ & $33.4_{-2.2}^{+2.1}$ & $1.2(1)$ & $0.30(1)$ & 9.5 & 3.80(21) & $1.14$(16) & 1.05 (540)\\
1048$-$59 & 403005010 & $9.9_{-0.1}^{+0.1}$ & $<13.2$ & $12.3_{-0.2}^{+0.2}$ & $<19.5$ & $0.47(3)$ & $0.45(1)$ & 4.7 & 4.88(20) & $\cdots$ & 1.32 (78)\\
0142$+$61 & 402013010 & $122.0_{-0.1}^{+0.1}$ & $35.1$(6.4) & $185.1_{-0.7}^{+0.7}$ & $38.3_{-2.3}^{+2.3}$ & $0.61(1)$ & $0.28(1)$ & 19.0 & 4.68(2) & $0.24$(7) & 1.43 (1725)\\
0142$+$61 & 404079010 & $115.6_{-0.2}^{+0.2}$ & $26.2$(5.8) & $176.8_{-1.5}^{+1.5}$ & $27.1_{-2.2}^{+2.3}$ & $0.62(1)$ & $0.28(1)$ & 18.6 & 4.71(4) & $0.39$(14) & 1.15 (1401)\\
0142$+$61 & 406031010 & $108.3_{-0.9}^{+1.1}$ & $24.3$(3.7) & $164.8_{-4.4}^{+4.6}$ & $41.3_{-4.0}^{+4.1}$ & $0.63(3)$ & $0.28(1)$ & 18.0 & 4.65(10) & $0.32$(21) & 1.08 (1062)\\
0142$+$61 & 408011010 & $107.8_{-0.4}^{+0.4}$ & $19.1$(2.8) & $162.7_{-1.4}^{+1.4}$ & $32.9_{-2.2}^{+2.3}$ & $0.60(1)$ & $0.28(1)$ & 17.9 & 4.81(4) & $0.26$(11) & 1.37 (838)\\
2259$+$58 & 404076010 & $30.4_{-0.1}^{+0.1}$ & $<10.1$ & $44.8_{-0.5}^{+0.5}$ & $<14.9$ & $0.55(1)$ & $0.29(1)$ & 7.8 & 4.85(4) & $\cdots$ & 1.20 (510)\\
1818$-$15 & 406074010 & $1.0_{-0.1}^{+0.1}$ & $\cdots$ & $1.4_{-0.1}^{+0.1}$ & $\cdots$ & $2.1(21)$ & $1.61(6)$ & 0.1 & $\cdots$ & $\cdots$ & 1.31 (83)\\
1547$-$54 & 903006010 & $59.7_{-0.8}^{+0.9}$ & $110.2$(5.2) & $50.6_{-1.7}^{+1.7}$ & $158.7_{-3.3}^{+3.3}$ & $2.8(1)$ & $0.67(2)$ & 1.9 & $\cdots$ & $1.53$(4) & 1.29 (140)\\
1547$-$54 & 405024010 & $11.1_{-0.2}^{+0.1}$ & $13.5$(3.3) & $17.5_{-0.6}^{+0.6}$ & $23.4_{-2.4}^{+2.5}$ & $2.8(1)$ & $0.62(1)$ & 1.3 & $\cdots$ & $1.15$(12) & 1.23 (69)\\
0501$+$45 & 903002010 & $36.2_{-0.2}^{+0.3}$ & $28.1$(6.5) & $42.2_{-0.8}^{+0.6}$ & $30.4_{-3.7}^{+4.0}$ & $0.40(2)$ & $0.49(1)$ & 2.7 & 4.35(17) & $0.10$(35) & 1.23 (206)\\
0501$+$45 & 404078010 & $2.9_{-0.1}^{+0.1}$ & $<20.9$ & $3.8_{-0.2}^{+0.3}$ & $<30.8$ & $0.44(10)$ & $0.30(3)$ & 2.2 & 4.16(27) & $\cdots$ & 0.92 (77)\\
0501$+$45 & 405075010 & $1.7_{-0.1}^{+0.1}$ & $<16.2$ & $2.0_{-0.2}^{+0.2}$ & $<23.9$ & $0.24(13)$ & $0.30(4)$ & 1.6 & 4.20(45) & $\cdots$ & 0.80 (34)\\
0501$+$45 & 408013010 & $1.7_{-0.1}^{+0.1}$ & $<12.7$ & $2.3_{-0.1}^{+0.1}$ & $<18.8$ & $0.41(7)$ & $0.26(2)$ & 2.3 & 3.79(15) & $\cdots$ & 0.96 (367)\\
1833$-$08 & 904006010 & $3.8_{-0.1}^{+0.1}$ & $17.6$(4.5) & $7.9_{-0.2}^{+0.3}$ & $35.6_{-8.8}^{+9.3}$ & $9.6(5)$ & $1.08(4)$ & 0.7 & $\cdots$ & $-0.38$(40) & 1.37 (167)\\
1647$-$45 & 901002010 & $27.2_{-0.1}^{+0.1}$ & $\cdots$ & $45.9_{-0.5}^{+0.5}$ & $\cdots$ & $1.7(1)$ & $0.49(1)$ & 3.4 & 4.39(6) & $\cdots$ & 1.26 (248)\\
1822$-$16 & 906002010 & $18.2_{-0.3}^{+0.4}$ & $<4.9$ & $18.5_{-0.4}^{+0.5}$ & $<7.2$ & $0.02(4)$ & $0.54(2)$ & 0.7 & 5.86(71) & $\cdots$ & 1.41 (59)\\
\hline
\multicolumn{12}{c}{{\it NuSTAR} observations} 
\\
% ~~~~~~~~~~~~~~~~~~~~~~~~~~~~~~~~~~~~~~~~~~~~~~~~~~~~
% Swift XRT normalization is fixed at 1.0 
% 4u0142_sim_both_compbbint_v150128a_swiftbase.xcm
1841$-$04 & 30001025004 & 
	$23.6_{-0.1}^{+0.1}$ & 
	$35.2$(0.6) & 
	$29.9_{-0.1}^{+0.1}$ & 
	$48.6_{-0.5}^{+0.5}$ & 
	$2.3(1)$ & $0.28(3)$ & 18.1 & 3.45(6) & 1.13(2) & 1.15 (1493)
\\
%0142$+$61 & 2014-04 & 
%	$114.6_{-0.1}^{+0.3}$ & $21.7_{-0.3}^{+0.3}$ & 
%	$167.1_{-0.6}^{+0.6}$ & $27.4_{-0.4}^{+0.4}$ & 
%	$0.58(2)$ & 
%	$0.30(1)$ & 
%	$15.7$ & 
%	$4.84(2)$ & 
%	$0.61(2)$ & 
%	$1.11$ (765)
%\\
0142$+$61 & 30001023002 & $114.6_{-0.1}^{+0.3}$ & $21.7$(0.3) & $167.1_{-0.1}^{+0.1}$ & $27.4_{-0.4}^{+0.4}$ & $0.58(2)$ & $0.30(1)$ & 15.8 & 4.84(2) & 0.61(2) & 1.11 (765)
\\
%2259$+$58 & 2013-04 &
%	$28.9_{-0.9}^{+0.1}$ & $3.1_{-0.2}^{+0.3}$ & 
%	$51.1_{-2.3}^{+2.8}$ & $3.7_{-0.3}^{+0.3}$ & 
%	0.84(4) & 
%	0.28(1) & 
%	8.9 & 
%	5.00(5) & 
%	0.58(13) & 
%	1.07 (404)
%\\	
2259$+$58 & 30001026002 & $28.9_{-0.9}^{+0.1}$ & $3.1$(0.3) & $51.1_{-2.3}^{+2.8}$ & $3.7_{-0.3}^{+0.3}$ & $0.84(4)$ & $0.28(1)$ & 8.9 & 4.95(6) & 0.58(13) & 1.07 (404)
\\
%2259$+$58 & 2013-05 &
%	$24.9_{-2.3}^{+0.1}$ & $3.1_{-0.2}^{+0.3}$ & 
%	$42.0_{-3.7}^{+4.9}$ & $3.4_{-0.2}^{+0.2}$ & 
%	0.81(7) & 
%	0.30(1) & 
%	7.0 & 
%	5.06(7) & 
%	0.58(9) & 
%	0.95 (223)
%\\	
2259$+$58 & 30001026007 & $24.9_{-2.3}^{+0.1}$ & $3.0$(0.3) & $42.0_{-3.7}^{+4.9}$ & $3.4_{-0.2}^{+0.2}$ & $0.78(7)$ & $0.30(1)$ & 7.0 & 5.06(8) & 0.66(14) & 0.94 (222)
\\
1048$-$59 & 2013-07 &
	$\sim8.9$ & $\lesssim6.7$ & 
	$\sim18.3$ & N/A &  %$\lesssim6.3$ &
	\multicolumn{6}{c}{from \cite{2015arXiv150908115Y}} 	
\enddata
\tablenotetext{a}{$F_{1-10}$ and $F_{15-60}$: Absorbed 1-10 keV and 15-60 keV fluxes ($10^{-12}$\,erg\,s$^{-1}$\,cm$^{-2}$)}
\tablenotetext{b}{Unabsorbed flx $F_{\rm s}$ and $F_{\rm h}$ of the SXC and HXC in the 1--60 keV band ($10^{-12}$\,erg\,s$^{-1}$\,cm$^{-2}$).}
\tablenotetext{c}{Radius is evaluated from $R=0.09643\times (d/{\rm kpc}) (T/{\rm keV})^{-2} (F_{\rm s}/10^{-11}\textrm{\,erg\,s$^{-1}$\,cm$^{-1}$})^{0.5}$}
\end{deluxetable*}

A photo-absorption factor was multiplied to 
	an intrinsic continuum spectral model of the SXC.
We employed 
	the {\tt phabs} model \citep{Balucinska-Church1992ApJ} in {\tt XSPEC}
	with the solar metallicity abundance {\bf angr} \citep{Anders1989GeCoA}
	and cross-section {\bf bcmc};
	this combination has been widely used in the literature
	and allows us to compare with previous results. 
When the HXC is detected in the 15--60\,keV band,
	it was expressed by an additional single power-law.
We further added a plasma emission model 
	when analyzing the 1E~1841$-$045 
	and 
	CXU~J171405.74$-$381031
	data,  
	to describe 
	the surrounding supernova remnant (SNR) 
	Kes~73 \citep{Kumar2014ApJ...781...41K}
	and 
	CTB~37B \citep{Sato2010PASJ...62L..33S}, 
	respectively,
	as described in Appendix \ref{section:suzaku:individual_sources}.
	
The intrinsic SXC spectrum is conventionally fitted by 
	a model comprising  
	two blackbody components (hereafter 2BB model),
	or 
	a combination of a blackbody plus 
	an additional soft power-law (BB+PL model), 
The second high-energy component of both models 
	is considered to represent either  
	a temperature anisotropy over the stellar surface,
	an up-scattering of soft photons in the magnetosphere, 
	or effects of a magnetised neutron star atmosphere
	(e.g., \citealt{Lyutikov2006MNRAS.368..690L, Rea2008ApJ...686.1245R,Guver2011MNRAS.418.2773G}).
These empirical 2BB and BB+PL models roughly explain the data,
	though only approximately sometimes. 
The low and high blackbody temperatures in the 2BB model, 
	$kT_{\rm L}$ and $kT_{\rm H}$,
	are known to follow a relation of $kT_{\rm L}/kT_{\rm H}\sim 0.4$
	 \citep{Nakagawa2009PASJ...61..109N},
	thus the number of free spectral parameters 
	is expected to be three rather than four of the 2BB or BB+PL models. 
	
Since the SXC spectral modeling has not 
	reached a consensus, 
	we utilize an empirical blackbody shape 
	with a Comptonization-like power-law tail
	(hereafter CBB model, \citealt{Tiengo2005A&A...437..997T,Halpern2008ApJ...676.1178H,Enoto2010PASJ...62..475E},
	and Paper I).
This CBB model is mathematically described by three parameters; 
	the temperature $kT$,
	soft-tail power-law photon index $\Gamma_{\rm s}$,
	and normalization corresponding to the emission radius $R$,
	and the model reproduces 
	the soft-tail at $\gtrsim$5\, keV of the BB+PL model 
	without a large $N_{\rm H}$ that is needed by the 2BB. 
	
\begin{figure*}[t]
\begin{center}
\includegraphics[width=180mm]{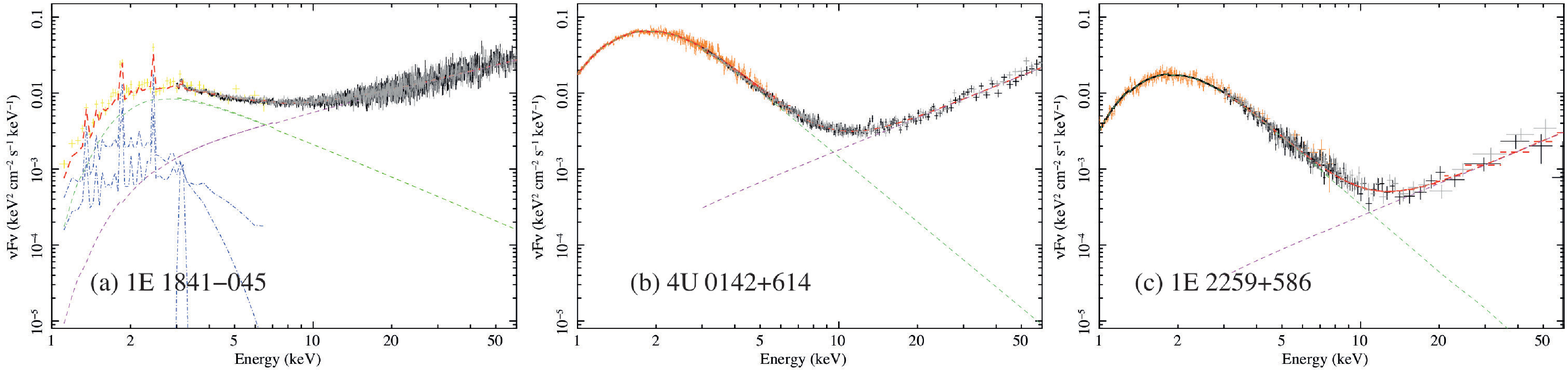}
\caption{
Same as Figure~\ref{fig:suzaku_spec_eeuf}, 
	but for {\it NuSTAR} observations
	of 
	(a) 1E~1841$-$045 in 2013 September,
	(b) 4U~0142$+$61 in 2014 March,
	and (c) 1E~2259$+$586.
}
\label{fig:nustar_eeuf}
\end{center}
\end{figure*}

We present  
	the \suzaku CBB best-fit parameters 
	in Table~\ref{tab:suzaku_fit_summary_cbb}
	and corresponding $\nu F_{\nu}$ spectra
	in Figure~\ref{fig:suzaku_spec_eeuf}.
The {\it NuSTAR} fit results are given in Table~\ref{tab:suzaku_fit_summary_cbb}
	together with some $\nu F_{\nu}$ examples 
	in Figure~\ref{fig:nustar_eeuf}.
The {\it NuSTAR} $\nu F_{\nu}$ shapes of bright AXPs,
	4U~0142$+$61 and 1E~1841$-$045, 
	are consistent with those of {\it Suzaku}.
The HXC of 1E~2259$+$586 was detected with {\it NuSTAR}  \citep{Vogel2014ApJ...789...75V},
	but not with {\it Suzkau}.
	
% ----------------------------------------------------
% ----------------------------------------------------  

% ~~~~~~~~~~~~~~~~~~~~~~~~~~~~~~~~~~~~~~~~~~~~~~~~~~~~
\begin{deluxetable*}{lrrrrrrrrrrrrr}[h]
\renewcommand{\arraystretch}{1.5}
%\tablewidth{290pt}
%\tabletypesize{\scriptsize}
\tablecaption{X-ray luminosities of the soft and hard components
	from magnetars measured with {\it Suzaku}.
	\label{tab:suzaku_obs_Lx}
	}
\tablewidth{0pt}
\tablehead{
\colhead{Name}  &
\colhead{ObsID} &
\colhead{Month} & 
\colhead{$P$} &
\colhead{$\dot{P}_{11}$} &
\colhead{$B_{14}$} &
\colhead{$\tau$} &
\colhead{$L_{\rm sd}$} &
\colhead{$L_{\rm s}$} &
\colhead{$L_{\rm h}$} &
\colhead{$L_{\rm x}$} & 
\colhead{Abs. HR} &
\colhead{HR} 
\\
\colhead{}  &
\colhead{} &
\colhead{yyyy-mm} &
\colhead{(s)} &
\colhead{} &
%\colhead{($10^{-11}$\,s\,s$^{-1}$)} &
%\colhead{($10^{14}$\,G)} &
\colhead{} &
\colhead{(kyr)} & 
\colhead{} & 
\colhead{} & 
\colhead{} & 
\colhead{(total)} & 
%\colhead{$F_{15-60}/F_{1-10}$} & 
\scriptsize{($\frac{F_{15-60}}{F_{1-10}}$)} & 
\colhead{($\xi$=$L_{\rm h}/L_{\rm s}$)} 
%\colhead{$=L_{\rm hxc}/L_{\rm sxc}$} 
}
% -------------------
\startdata
% -------------------
\multicolumn{13}{c}{Suzaku \rule[0mm]{0mm}{3mm}} \\ 
 \hline 
 1806$-$20 & 401092010 & 2006-09 & 7.548 & 49.5 & 19.6 & 0.2 & 4.6 & $5.7_{-0.1}^{+0.8}$ & $53.0_{-0.1}^{+1.2}$ & $58.8_{1.5}^{+1.5}$ & $2.7_{-0.4}^{+0.4}$ & $9.3_{-1.2}^{+1.4}$ & \\ 

1806$-$20 & 401021010 & 2007-03 & $\cdots$ & $\cdots$ & $\cdots$ & $\cdots$ & $\cdots$ & $3.1_{-0.1}^{+0.9}$ & $45.5_{-0.1}^{+2.6}$ & $48.6_{2.5}^{+2.8}$ & $2.0_{-0.3}^{+0.3}$ & $14.7_{-3.7}^{+4.2}$ & \\ 

1806$-$20 & 402094010 & 2007-10 & $\cdots$ & $\cdots$ & $\cdots$ & $\cdots$ & $\cdots$ & $5.3_{-0.1}^{+0.9}$ & $38.4_{-0.1}^{+1.7}$ & $43.7_{1.7}^{+1.9}$ & $3.0_{-0.5}^{+0.5}$ & $7.2_{-1.0}^{+1.3}$ & \\ 

1841$-$04 & 401100010 & 2006-04 & 11.789 & 4.1 & 7.0 & 4.6 & 0.100 & $30.5_{-0.1}^{+1.2}$ & $44.1_{-0.1}^{+1.9}$ & $74.6_{2.2}^{+2.2}$ & $2.5_{-0.1}^{+0.1}$ & $1.4_{-0.1}^{+0.1}$ & \\ 

1900$+$14 & 401022010 & 2006-04 & 5.200 & 9.2 & 7.0 & 0.9 & 2.6 & $8.7_{-0.1}^{+0.2}$ & $46.9_{-0.1}^{+6.0}$ & $55.5_{6.5}^{+6.0}$ & $3.9_{-1.1}^{+1.1}$ & $5.4_{-1.0}^{+0.7}$ & \\ 

1900$+$14 & 404077010 & 2009-04 & $\cdots$ & $\cdots$ & $\cdots$ & $\cdots$ & $\cdots$ & $8.4_{-0.1}^{+0.5}$ & $49.3_{-0.1}^{+5.5}$ & $57.7_{4.8}^{+5.5}$ & $3.8_{-0.8}^{+0.8}$ & $5.9_{-0.7}^{+0.7}$ & \\ 

1714$-$38 & 501007010 & 2006-08 & 3.825 & 6.4 & 5.0 & 0.9 & 4.6 & $10.5_{-0.1}^{+0.9}$ & $\cdots$ & (=$L_{\rm s}$) &$\cdots$ & $\cdots$ & \\ 

1708$-$40 & 404080010 & 2009-08 & 11.005 & 1.9 & 4.7 & 9.0 & 0.058 & $12.0_{-0.1}^{+0.6}$ & $4.9_{-0.1}^{+0.5}$ & $16.9_{0.7}^{+0.8}$ & $0.64_{-0.12}^{+0.12}$ & $0.41_{-0.04}^{+0.05}$ & \\ 

1708$-$40 & 405076010 & 2010-09 & $\cdots$ & $\cdots$ & $\cdots$ & $\cdots$ & $\cdots$ & $9.5_{-0.1}^{+0.7}$ & $5.8_{-0.1}^{+0.4}$ & $15.3_{0.6}^{+0.8}$ & $0.68_{-0.11}^{+0.11}$ & $0.61_{-0.05}^{+0.06}$ & \\ 

1048$-$59 & 403005010 & 2008-11 & 6.458 & 2.3 & 3.9 & 4.6 & 0.33 & $11.9_{-0.1}^{+0.2}$ & $<$$19.0$ & (=$L_{\rm s}$) &$<$$1.3$ & $<$$2.0$ & \\ 

0142$+$61 & 402013010 & 2007-08 & 8.689 & 0.20 & 1.3 & 68.1 & 0.012 & $28.8_{-0.1}^{+0.1}$ & $6.0_{-0.1}^{+0.4}$ & $34.7_{0.4}^{+0.4}$ & $0.29_{-0.05}^{+0.05}$ & $0.21_{-0.01}^{+0.01}$ & \\ 

0142$+$61 & 404079010 & 2009-08 & $\cdots$ & $\cdots$ & $\cdots$ & $\cdots$ & $\cdots$ & $27.5_{-0.1}^{+0.2}$ & $4.2_{-0.1}^{+0.3}$ & $31.7_{0.4}^{+0.4}$ & $0.23_{-0.05}^{+0.05}$ & $0.15_{-0.01}^{+0.01}$ & \\ 

0142$+$61 & 406031010 & 2011-09 & $\cdots$ & $\cdots$ & $\cdots$ & $\cdots$ & $\cdots$ & $25.6_{-0.1}^{+0.7}$ & $6.4_{-0.1}^{+0.6}$ & $32.1_{0.9}^{+1.0}$ & $0.22_{-0.03}^{+0.03}$ & $0.25_{-0.03}^{+0.03}$ & \\ 

0142$+$61 & 408011010 & 2013-07 & $\cdots$ & $\cdots$ & $\cdots$ & $\cdots$ & $\cdots$ & $25.3_{-0.1}^{+0.2}$ & $5.1_{-0.1}^{+0.4}$ & $30.4_{0.4}^{+0.4}$ & $0.18_{-0.03}^{+0.03}$ & $0.20_{-0.01}^{+0.01}$ & \\ 

2259$+$58 & 404076010 & 2009-05 & 6.979 & 0.048 & 0.59 & 229.0 & 0.0057 & $5.5_{-0.1}^{+0.1}$ & $<$$1.8$ & (=$L_{\rm s}$) &$<$$0.3$ & $<$$0.5$ & \\ 

1818$-$15 & 406074010 & 2011-10 & 2.482 & 0.80 & 1.4 & 4.9 & 2.1 & $0.30_{-0.01}^{+0.02}$ & $\cdots$ & (=$L_{\rm s}$) &$\cdots$ & $\cdots$ & \\ 

1547$-$54 & 903006010 & 2009-01 & 2.072 & 4.8 & 3.2 & 0.7 & 21.4 & $9.3_{-0.1}^{+0.3}$ & $29.1_{-0.1}^{+0.6}$ & $38.4_{0.7}^{+0.7}$ & $1.8_{-0.1}^{+0.1}$ & $3.1_{-0.1}^{+0.1}$ & \\ 

1547$-$54 & 405024010 & 2010-08 & $\cdots$ & $\cdots$ & $\cdots$ & $\cdots$ & $\cdots$ & $3.2_{-0.1}^{+0.1}$ & $4.3_{-0.1}^{+0.5}$ & $7.5_{0.4}^{+0.5}$ & $1.2_{-0.3}^{+0.3}$ & $1.3_{-0.1}^{+0.2}$ & \\ 

0501$+$45 & 903002010 & 2008-08 & 5.762 & 0.59 & 1.9 & 15.4 & 0.12 & $5.5_{-0.1}^{+0.1}$ & $4.0_{-0.1}^{+0.5}$ & $9.5_{0.5}^{+0.5}$ & $0.78_{-0.18}^{+0.18}$ & $0.72_{-0.09}^{+0.10}$ & \\ 

0501$+$45 & 404078010 & 2009-08 & $\cdots$ & $\cdots$ & $\cdots$ & $\cdots$ & $\cdots$ & $0.50_{-0.01}^{+0.04}$ & $<$$4.0$ & (=$L_{\rm s}$) &$<$$7.3$ & $<$$10.7$ & \\ 

0501$+$45 & 405075010 & 2010-09 & $\cdots$ & $\cdots$ & $\cdots$ & $\cdots$ & $\cdots$ & $0.26_{-0.01}^{+0.03}$ & $<$$3.1$ & (=$L_{\rm s}$) &$<$$9.6$ & $<$$14.2$ & \\ 

0501$+$45 & 408013010 & 2013-08 & $\cdots$ & $\cdots$ & $\cdots$ & $\cdots$ & $\cdots$ & $0.30_{-0.01}^{+0.02}$ & $<$$2.5$ & (=$L_{\rm s}$) &$<$$7.4$ & $<$$10.9$ & \\ 

1833$-$08 & 904006010 & 2010-03 & 7.565 & 0.35 & 1.6 & 34.3 & 0.032 & $7.6_{-0.1}^{+0.3}$ & $34.6_{-0.1}^{+9.0}$ & $42.3_{8.6}^{+9.0}$ & $4.7_{-1.2}^{+1.2}$ & $4.5_{-1.1}^{+1.2}$ & \\ 

1647$-$45 & 901002010 & 2006-09 & $10.61$ & $\cdots$ & $\cdots$ & $\cdots$ & $\cdots$ & $8.4_{-0.1}^{+0.1}$ & $\cdots$ & (=$L_{\rm s}$) &$\cdots$ & $\cdots$ & \\ 

1822$-$16 & 906002010 & 2011-09 & 8.438 & 0.0021 & 0.14 & 6250.0 & 0.00014 & $0.57_{-0.01}^{+0.01}$ & $<$$0.2$ & (=$L_{\rm s}$) &$<$$0.3$ & $<$$0.4$ & \\ 

\hline \hline \multicolumn{13}{c}{NuSTAR \rule[0mm]{0mm}{3mm}} \\ 
 \hline 
 1841$-$04 & 30001025004 & 2013-09 & 11.789 & 4.1 & 7.0 & 4.6 & 0.100 & $25.9_{-0.1}^{+0.1}$ & $42.1_{-0.1}^{+0.4}$ & $68.1_{0.4}^{+0.4}$ & $1.5_{-0.1}^{+0.1}$ & $1.6_{-0.1}^{+0.1}$ & \\ 
0142$+$61 & 30001023002 & 2014-01 & 8.689 & 0.20 & 1.3 & 68.1 & 0.012 & $26.0_{-0.1}^{+0.1}$ & $4.3_{-0.1}^{+0.1}$ & $30.2_{0.1}^{+0.1}$ & $0.19_{-0.01}^{+0.91}$ & $0.16_{-0.01}^{+0.01}$ & \\ 
2259$+$58 & 30001026002 & 2013-04 & 6.979 & 0.048 & 0.59 & 229.0 & 0.0057 & $6.3_{-0.1}^{+0.3}$ & $0.5_{-0.1}^{+0.1}$ & $6.7_{0.3}^{+0.3}$ & $0.11_{-0.01}^{+0.01}$ & $0.07_{-0.01}^{+0.01}$ & \\ 
2259$+$58 & 30001026007 & 2013-05 & $\cdots$ & $\cdots$ & $\cdots$ & $\cdots$ & $\cdots$ & $5.2_{-0.1}^{+0.6}$ & $0.4_{-0.1}^{+0.1}$ & $5.6_{0.5}^{+0.6}$ & $0.12_{-0.02}^{+0.01}$ & $0.08_{-0.01}^{+0.01}$ & \\
1048$-$59 & 30001024002 & 2013-07 & 6.458 & 2.3 & 3.9 & 4.6 & 0.33 & $17.8_{-0.1}^{+0.1}$ & $<6.1_{-0.1}^{+0.1}$ & $23.9_{0.0}^{+0.0}$ & $<0.74$ & $<0.34$ & 

\enddata
\tablecomments{{\bf Pulsar timing information:} spin period $P$ (s), period derivative $\dot{P}_{11}=\dot{P}/(10^{-11}$s\,s$^{-1}$), surface magnetic field ${B}_{14}=B/(10^{14}$\,G), and spin-down luminosity $L_{\rm sd}=4\pi I^2\dot{P}/P^3=3.2\times 10^{33}\dot{P}_{11} (P/5\,{\rm s})^{-3}$\,erg\,s$^{-1}$, where $I=10^{45}$\,g\,cm$^2$ is the neutron star momentum of inertia.\\
{\bf Spectral information:} The 1--60\, keV luminosity of the SXC and HXC, $L_{\rm s}$, $L_{\rm h}$, and their toral $L_{\rm x}=L_{\rm s}+L_{\rm h}$. All the luminosities, $L_{\rm sd}$, $L_{\rm s}$, $L_{\rm h}$, and $L_{\rm x}$ are shown in an unit of $10^{34}$\,erg\,s$^{-1}$. \\
{\bf Hardness ratio (HR):} Evaluated from absorbed fluxes of 15--60 keV and 1-10 keV ($\eta=F_{15-60}/F_{1-10}$) or luminosities ($\xi=L_{\rm h}/L_{\rm s}$) after correcting absorption. 
}
%\tablenotetext{a}{Period derivative  $\dot{P}_{11}=\dot{P}/(10^{-11}$s\,s$^{-1}$).}
%\tablenotetext{b}{Surface magnetic field ${B}_{14}=B/(10^{14}$\,G)
%	evaluated from $B=3.2\times 10^{19}\sqrt{P \dot{P}}$\,G.}
%\tablenotetext{c}{Spin-down luminosity $L_{\rm sd}=4\pi I^2\dot{P}/P^3=3.6\times 10^{35}\dot{P}_{11} (P/1\,{\rm s})^{-3}$\,erg\,s$^{-1}$, where $I=10^{45}$\,g\,cm$^2$ is the neutron star momentum of inertia.
%}
%\tablenotetext{d}{$L_{\rm sd}$, $L_{\rm sxc}$, and $L_{\rm hxc}$ are unit of $10^{34}$\,erg\,s$^{-1}$.}
%\tablenotetext{d}{$\xi=L_{\rm hxc}/L_{\rm sxc}$.}
%\tablenotetext{b}{The 15--60 keV HXD-PIN count rates with $1\sigma$ statistical and systematic errors. 
%If not detected, the $3\sigma$ upper-limits are shown. 
%}
%\tablenotetext{c}{The 15--60 keV absorbed flux 
%	in an unit of $10^{-12}$\,ergs\,s$^{-1}$\,cm$^{-2}$ with 1$\sigma$ statistical and systematic errors. 
%If not detected, the $3\sigma$ upper-limits are shown converted from the count rate $R_{\rm pin}$.
%	}	
\end{deluxetable*}
% ~~~~~~~~~~~~~~~~~~~~~~~~~~~~~~~~~~~~~~~~~~~~~~~~~~~~	 

% ==========================================
\subsection{Correlations among spectral parameters}
% ==========================================

% --------------------------------------------------------------------------
\subsubsection{Ratio of HXC to SXC vs. magnetic field}
\label{Ratio of HXC to SXC vs. magnetic field}
% --------------------------------------------------------------------------

% === Explanation of the correlation === 
In Paper I,	
	we proposed a broad-band spectral evolution of this class:
	i.e.,
	i) the hardness ratio of the HXC to SXC 
	is positively (or negatively) 
	correlated to their $B_{\rm d}$ (or $\tau_{\rm c}$),
	and
	ii) 		
	the HXC photon index $\Gamma_{\rm h}$ becomes harder 
	toward the weaker $B_{\rm d}$ 
	sources (see also \citealt{Kaspi2010ApJ...710L.115K}).

\begin{figure*}
\begin{center}
\includegraphics[width=88mm]{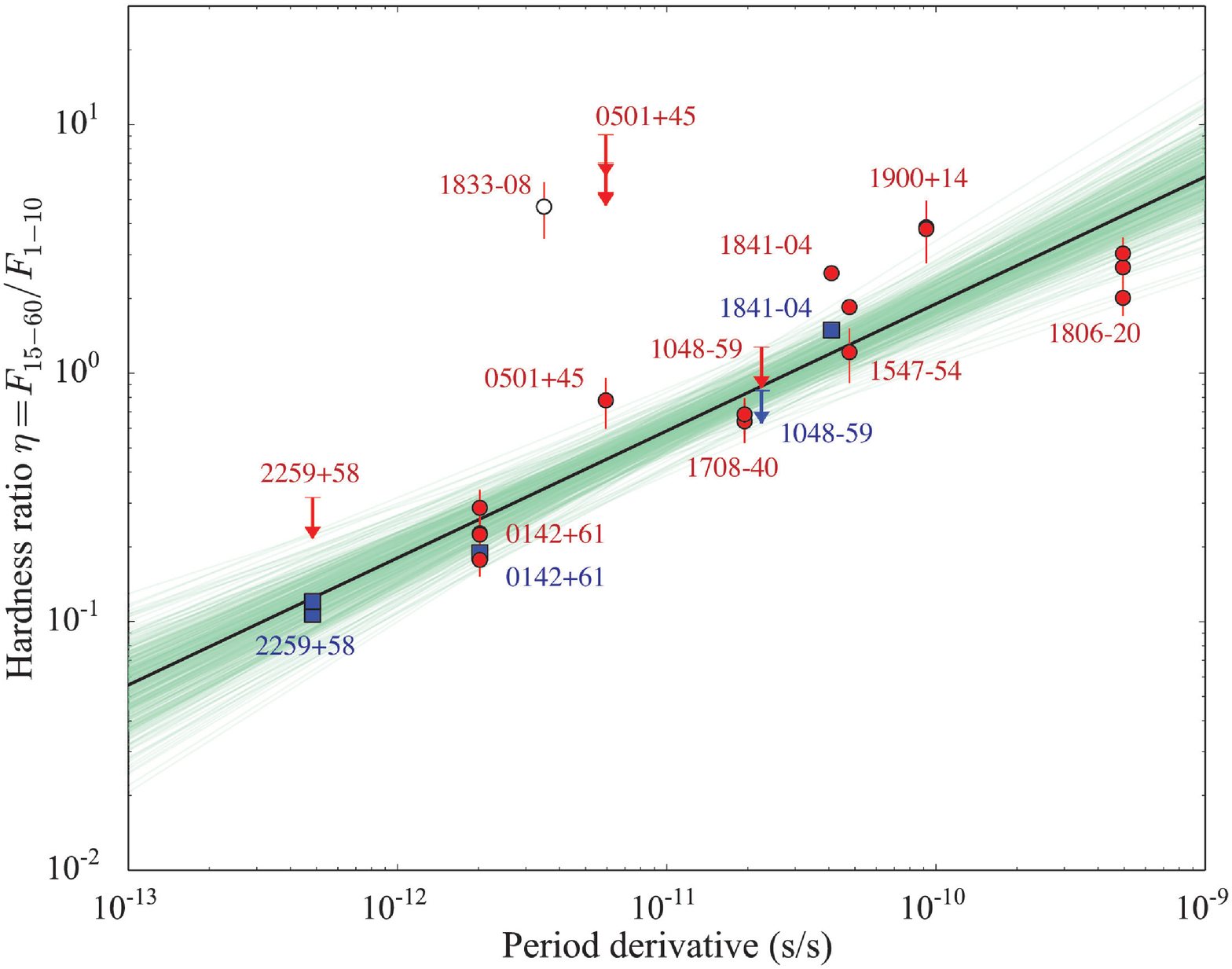}
\includegraphics[width=88mm]{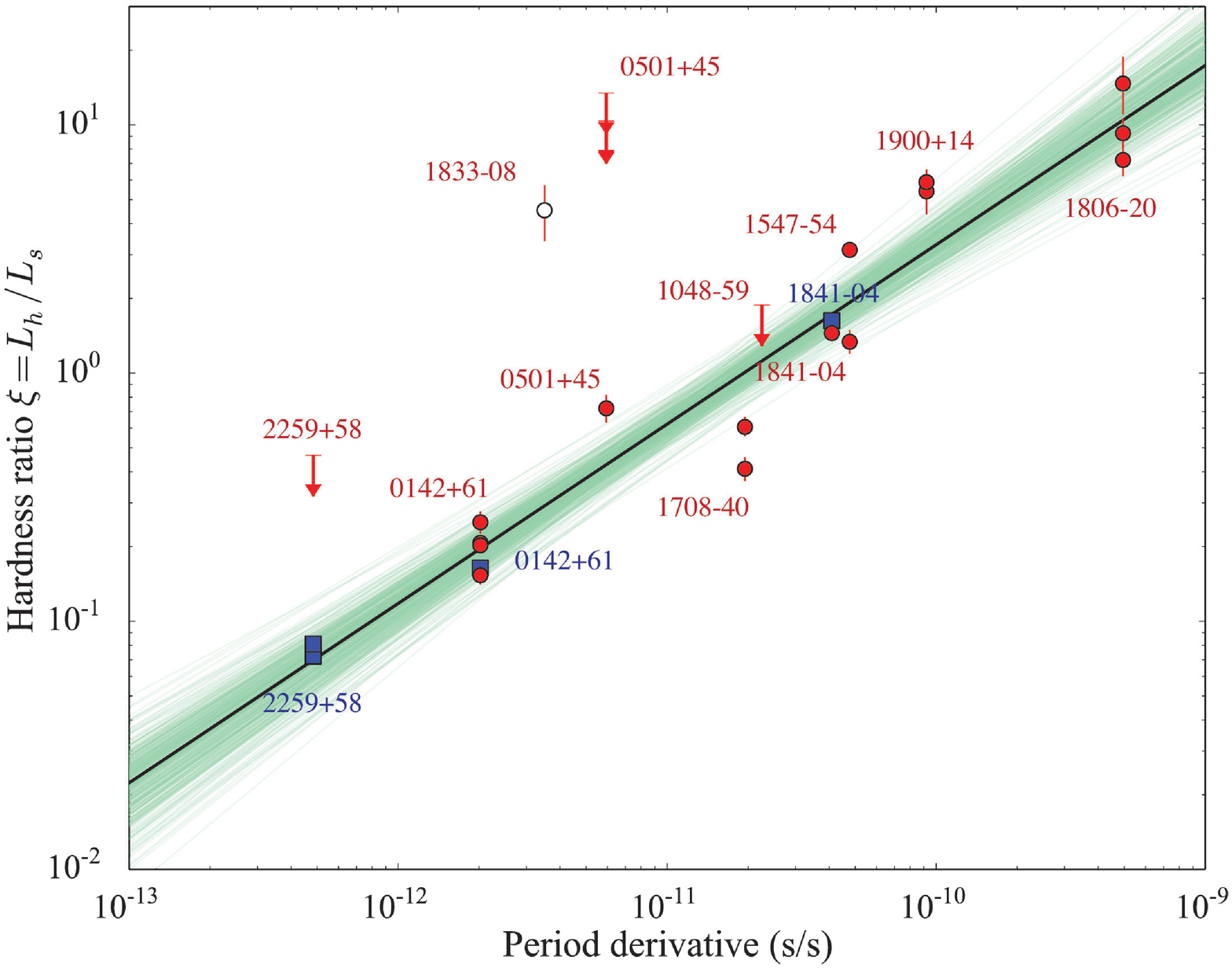}
\includegraphics[width=88mm]{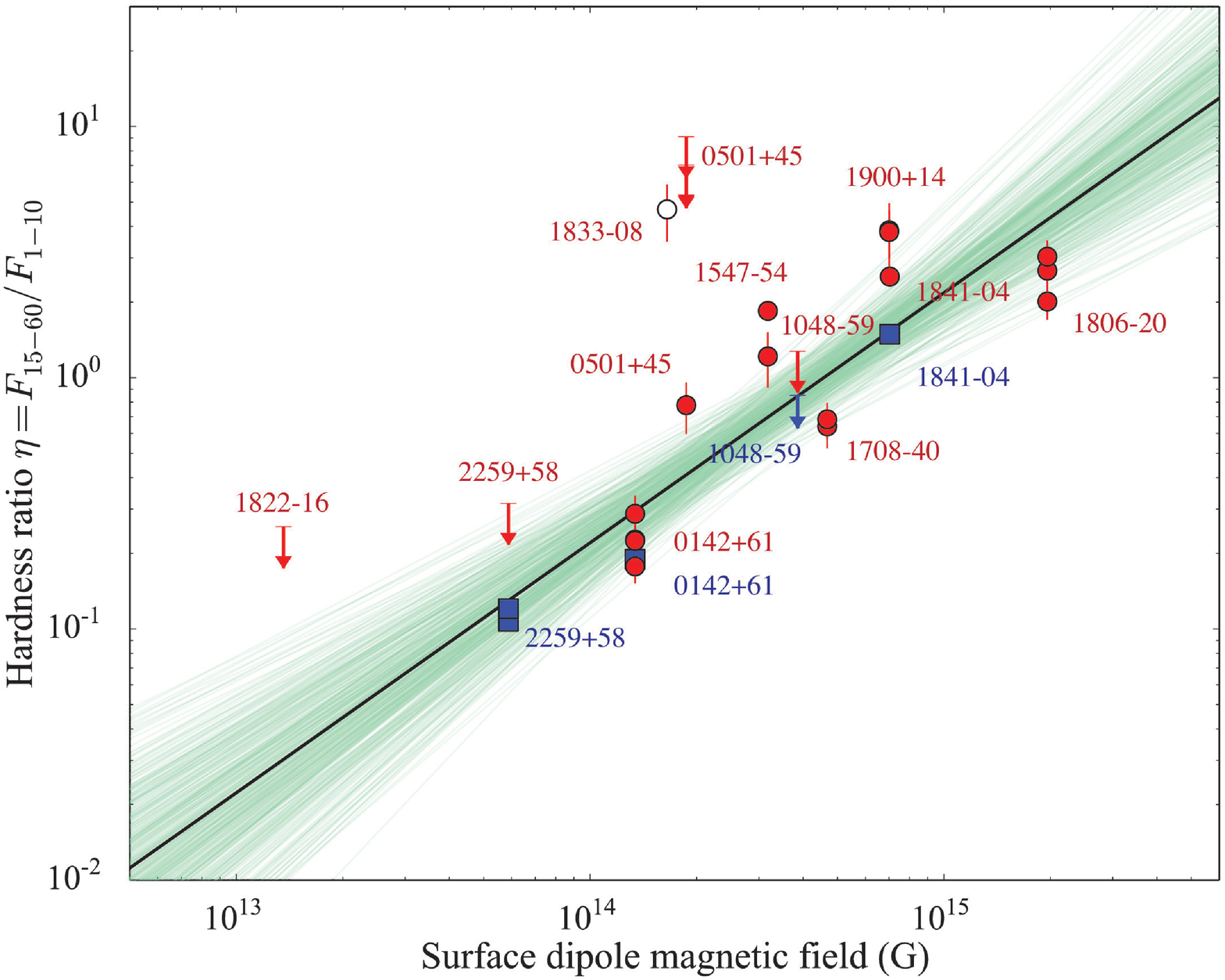}
\includegraphics[width=88mm]{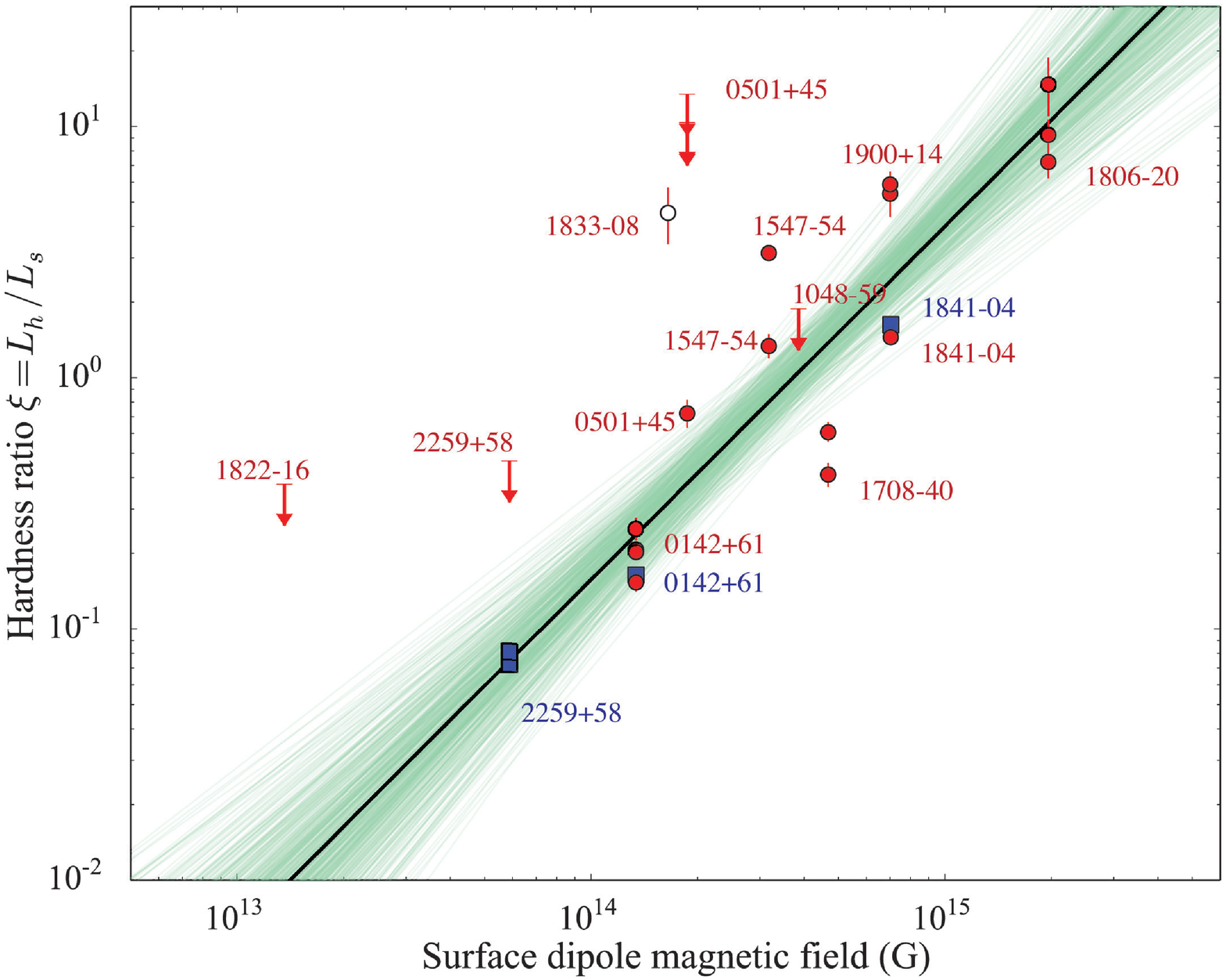}
\includegraphics[width=88mm]{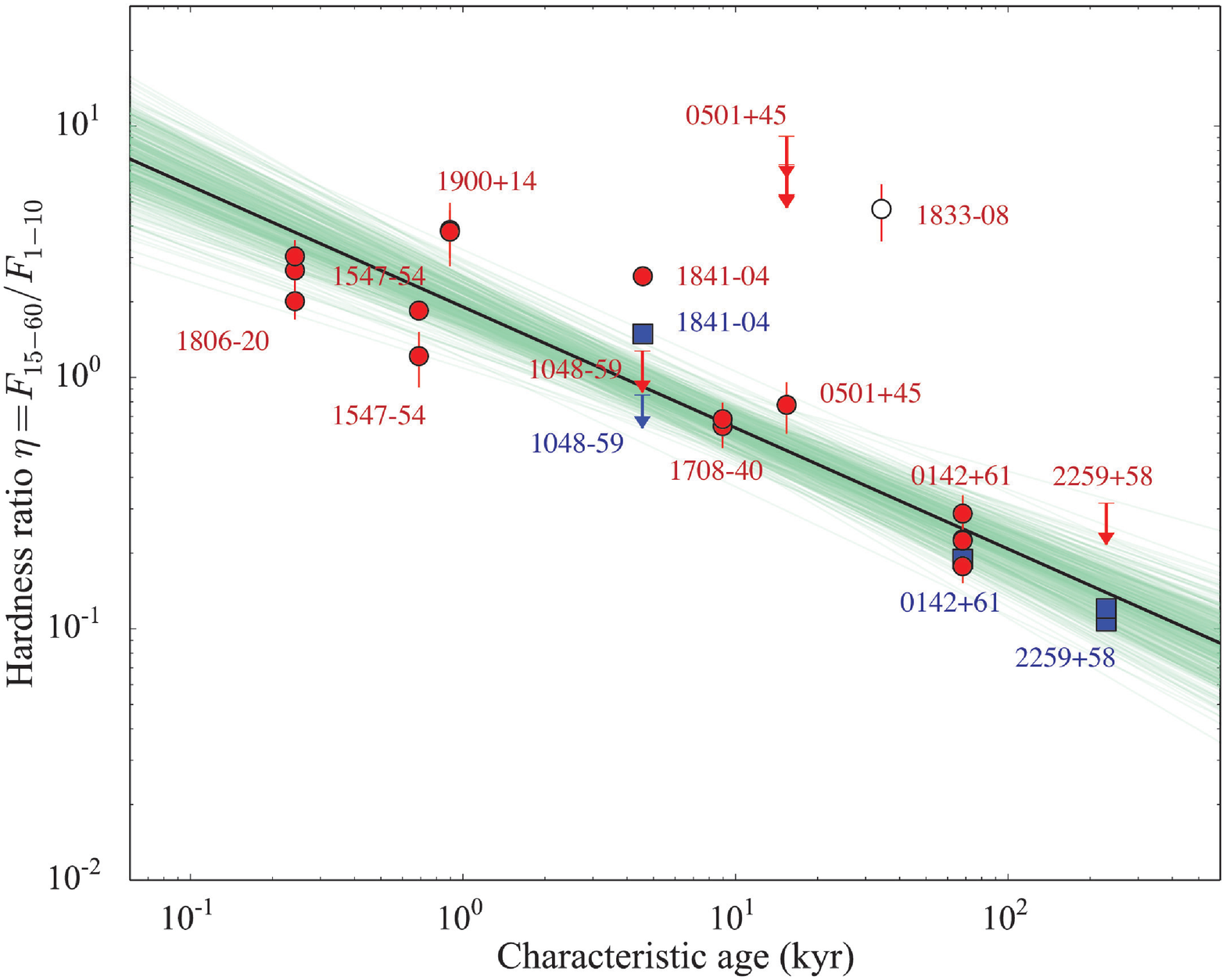}
\includegraphics[width=88mm]{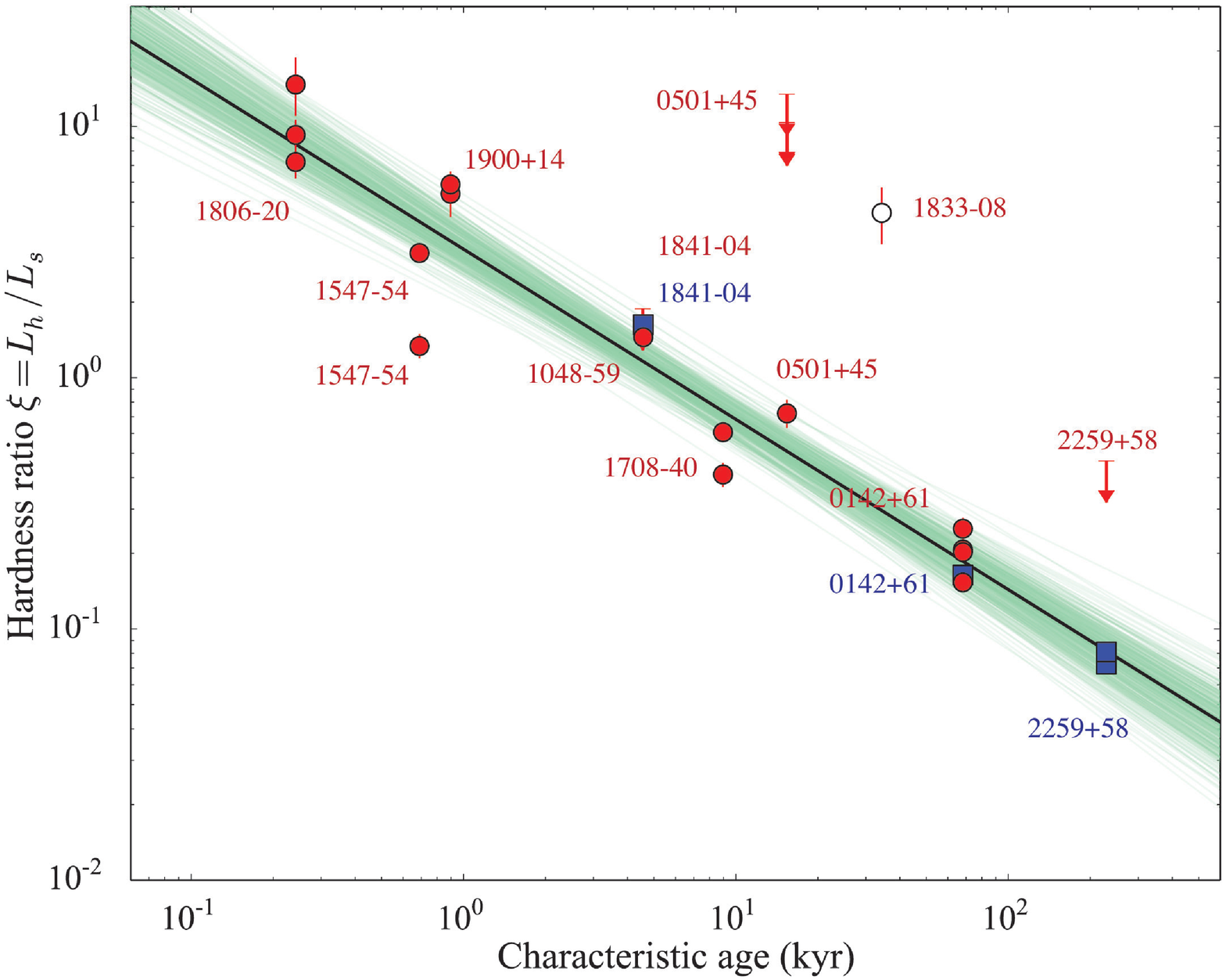}
\caption{
Hardness ratios of the HXC to the SXC (Table~\ref{tab:suzaku_obs_Lx}), 
	defined 
	using the absorbed fluxes $\eta=F_{15-60}/F_{1-10}$ (left panels) 
	or 
	using the unabsorbed luminosities $\eta=L_{\rm h}/L_{\rm s}$ (right panels),
	as a function 
	of $\dot{P}$ (top panels).
	$B_{\rm d}$ (middle),  
	and $\tau_{\rm c}$ (bottom).
Filled circles and triangles are the data from {\it Suzaku} and {\it NuSTAR}, respectively.
All the upper limits (arrows) are also indicated.
The solid black line is the best fit regression model
	calculated  in the linmix package \citep{2007ApJ...665.1489K}.
Green lines are samples from the corresponding posterior distribution 
	of the model parameters. 
}
\label{fig:correlations}
\end{center}
\end{figure*}

Based on our updated sample 
	shown in Figure~\ref{fig:suzaku_spec_eeuf},
	we revised these correlations.
Figure~\ref{fig:correlations} top-left panel shows 
	the ratio of absorbed fluxes, $\eta=F_{15-60}/F_{1-10}$, 
	as a function of $\dot{P}$, 
	which is derived from the pulsar timing information
	independently from the spectroscopy.  
The Spearman's rank-order test
	of this correlation
	gives a significantly high value, $r_s=0.94$.
The correlation is fitted as,
\begin{eqnarray}
\eta &=& F_{15-60}/F_{1-10} \nonumber \\
&=& (0.59\pm 0.07) \times (\dot{P}/10^{-11}\,{\rm s\, s^{-1}})^{0.51\pm 0.05},
\end{eqnarray}
using the Bayesian method in \cite{2007ApJ...665.1489K} (linmix package) 
	to account for measurement errors and intrinsic scatter of the data. 
While the correlations is derived from the spectral analyses below 70\,keV,
	the potential HXC cutoff does not strongly affect the correlation (Paper~I).
Due to limited photon statistics of several {\it Suzaku} sources,
	we can only study total emission.
Deconvolution into pulsed and un-pulsed components \citep{Kuiper2012ApJ...748..133K}
	will be reported in future publications.
	
As shown in $\nu F_{\nu}$ plots (Figure~\ref{fig:suzaku_spec_eeuf}),
	the HXC largely contributed to the SXC band below $\lesssim$10~keV
	in stronger $B_{\rm d}$ objects (e.g., SGR~1806$-$20, SGR~1900$+$14, and 1E~1547.0$-$5408).
To remove this mixing, 
	and to eliminate the effects of the photo-absorption, 	
	we also define
	the absorption-corrected luminosity ratio between the two components, 
	$\xi=L_{\rm h}/L_{\rm s}$ (listed in Table~\ref{tab:suzaku_obs_Lx}), 
	in the same way as Paper I. 
The Spearman's rank-order significance becomes $r_s=0.97$.
The correlation slope of $\xi$ becomes steeper than those of $\eta$ as,
\begin{eqnarray}
\label{xi-correlation:1}
\xi &=& L_{\rm h}/L_{\rm s} \nonumber \\
\label{eq:xi_vs_P}
&=& (0.62\pm 0.07) \times (\dot{P}/10^{-11}\,{\rm s\, s^{-1}})^{0.72\pm 0.05}.
\end{eqnarray}
This is shown in the top-right panel of Figure~\ref{fig:correlations}. 
On these correlations, 
	we revised timing information ($P$ and $\dot{P}$) from Paper~I, 
	referring to the McGill catalog 	
	\citep{Olausen2013arXiv1309.4167O},
	and further added {\it NuSTAR} observations.
Data points of canonical AXPs 4U~0142$+$61 and 1E~1841$-$045 
	are consistent with those with {\it Suzaku},
	and the new HXC detection from 1E~2259$+$586 \citep{Vogel2014ApJ...789...75V}
	falls on the correlation. 
The Galactic center source SGR~J1745$-$29,
	though not shown in Figure~\ref{fig:correlations},
	is also expected to follow the relation  
	since its wide-band spectrum resembles that of 1E~1547.0$-$5408
	\citep{Mori2013ApJ...770L..23M}.

The above correlation to the directly measured quantity
$\dot{P}$ can be converted to correlations to $B_{\rm d}$ and $\tau_{\rm d}$,
	although the additional two relations are not independent to Eq. (\ref{eq:xi_vs_P})
	since $B_{\rm d}$ and $\tau_{\rm c}$ are estimated using combinations of the same $P$ and $\dot{P}$;
	i.e., 	$B_{\rm d} \propto P^{1/2} \dot{P}^{1/2}$ and $\tau_{\rm c}\propto P\dot{P}^{-1}$. 
Furthermore, considering 
	the clustering of rotational periods of magnetars in a narrow range ($P=2$--11\,s), 
	Correlation of  Eq. (\ref{eq:xi_vs_P}), $\xi \propto \dot{P}^{-k}$ (k$\sim$0.72), gives 
	$\xi \propto B_{\rm d}^{2k} \sim B_{\rm d}^{1.4}$ and 
	$\xi \propto \tau_{\rm c}^{-k} \sim \tau_{\rm c}^{-0.72}$.
These relations are shown in the middle and bottom 	panels in Figure~\ref{fig:correlations},
	and same fitting procedures give,
\begin{eqnarray}
\eta &=& F_{15-60}/F_{1-10} \nonumber \\
&=& (0.097\pm 0.036)\times (B_{\rm d}/B_{\rm QED})^{1.00\pm 0.14} \\
&=& (1.91\pm 0.33) \times (\tau_{\rm c}/1\, {\rm kyr})^{-0.48\pm 0.05}  
\end{eqnarray}
and 
\begin{eqnarray}
\xi &=& L_{\rm h}/L_{\rm s} \nonumber   \\
\label{eq:xi_vs_B}
&=& (0.050\pm 0.022) \times (B_{\rm d}/B_{\rm QED})^{1.41\pm 0.16} \\
\label{eq:xi_vs_tau}
&=& (3.25\pm 0.50) \times (\tau_{\rm c}/1\, {\rm kyr})^{-0.68\pm 0.05}. 
\end{eqnarray}
The slopes of Eq. (\ref{eq:xi_vs_B}) and (\ref{eq:xi_vs_tau})
	are consistent with those from Paper I within error bars. 
The {\it Suzaku} upper limit for the second lowest $B_{\rm d}$-field source Swift~J1822.3$-$1606 
	($B_{\rm d}=1.4\times 10^{13}$\,G, i.e., $B_{\rm d}/B_{\rm QED}\sim 0.32$)
	is also consistent with this picture.	
Thus we reconfirm, and reinforce, 
	the evolution in $\eta$ and $\xi$ as reported in Paper~I.

% ----------------------------------------------------
% ----------------------------------------------------

% --------------------------------------------------------------------------
\subsubsection{Photon index $\Gamma_{\rm h}$ of HXC vs. magnetic field}
\label{PhotonIndex}
\begin{figure}
\begin{center}
\includegraphics[width=85mm]{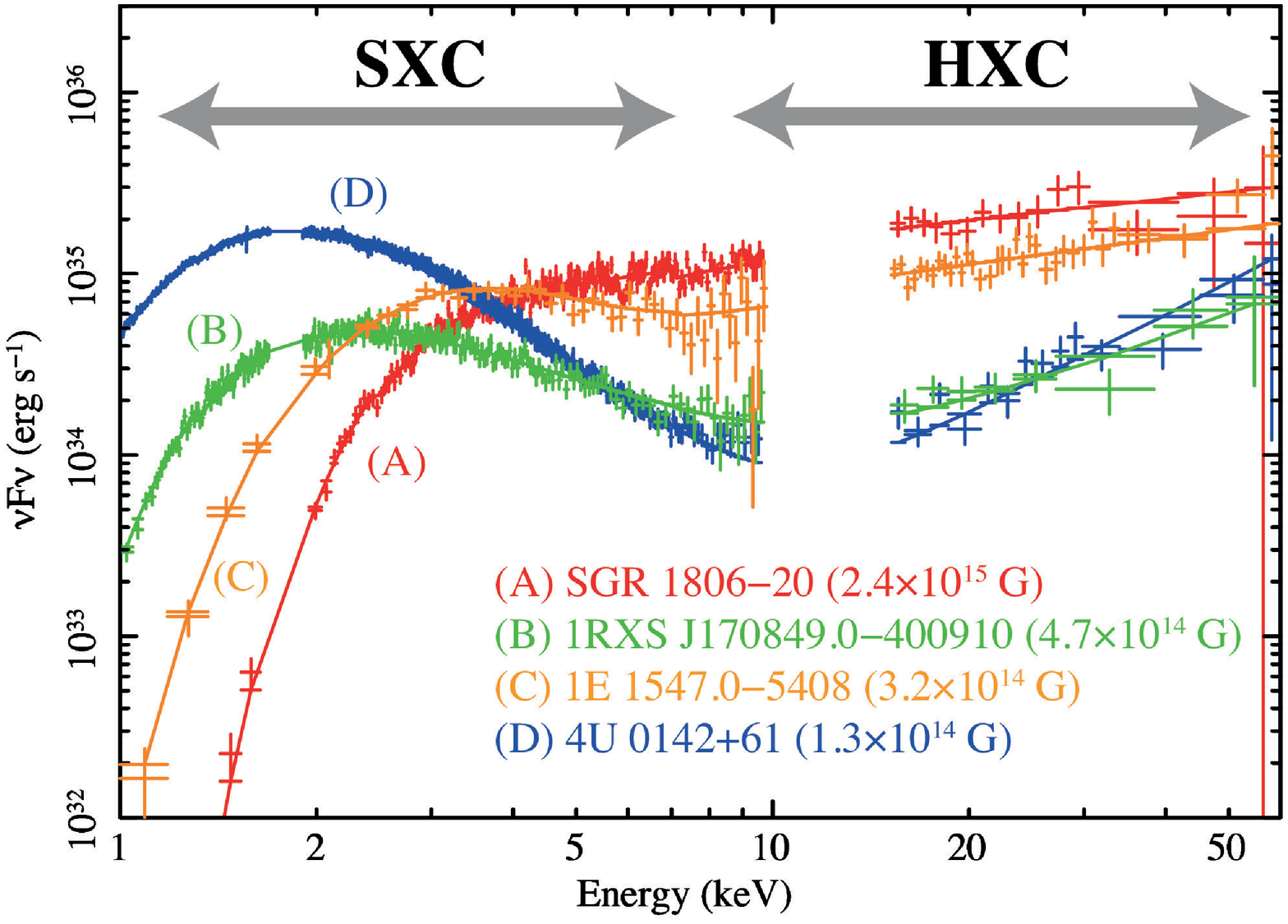}
\includegraphics[width=85mm]{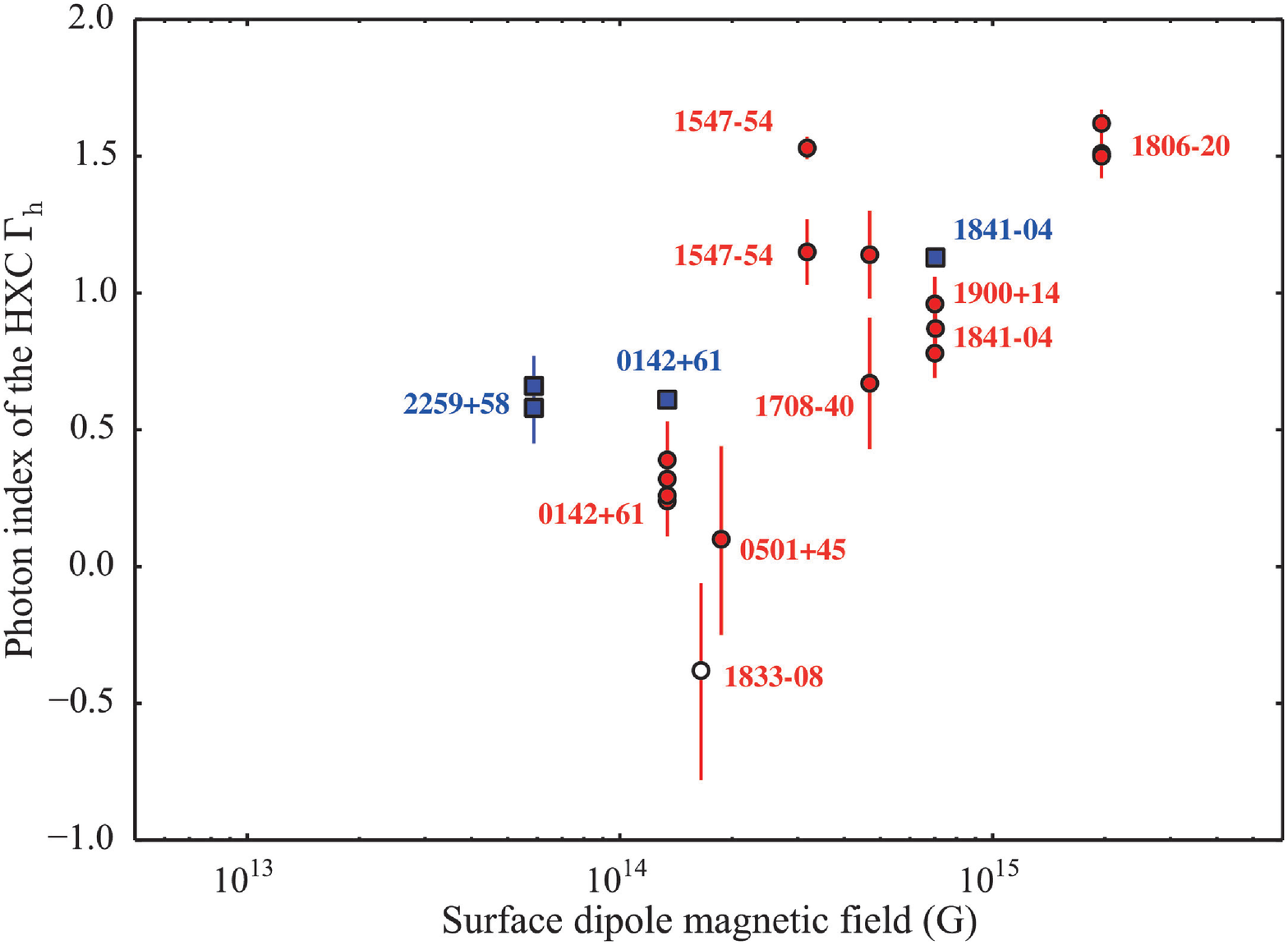}
\caption{
(top)
Comparison of representative X-ray spectra of magnetars 
	after correcting the data for their distances.
(bottom)
The photon index $\Gamma_{\rm h}$ of the HXC
	as a function of the magnetic field $B_{\rm d}$.
Symbols are the same as in Figure~\ref{fig:correlations}.
}
\label{fig:Gh_vs_B}
\end{center}
\end{figure}

% --------------------------------------------------------------------------
The second prediction of Paper I
	is the HXC spectral hardening 
	toward weaker-$B_{\rm d}$ objects,
	as seen in representative $\nu F_{\nu}$ spectra in Figure~\ref{fig:Gh_vs_B} (top).
Figure~\ref{fig:Gh_vs_B} (bottom) also shows 
	the HXC photon index $\Gamma_{\rm h}$
	as a function of $B_{\rm d}$, which is fitted as 
\begin{equation}
\label{eq:Gh_vs_B}
\Gamma_{\rm h} = (0.40\pm 0.11)  \times (B_{\rm d}/B_{\rm QED})^{0.35\pm 0.09}.
\end{equation}
The $\Gamma_{\rm h}$ values are stable on a long-time scale
	for persistently bright sources 
	such as 4U~0142+61, SGR~1806$-$20, and 1E~1841$-$045,
	while some transients show slope change during the outbursts.
This was also reported 
	during the $\sim$400 days {\it INTEGRAL} monitoring of 1E~1547.0$-$5408
	changing from $\Gamma \sim 1.4$ to $\sim 0.9$ \citep{Kuiper2012ApJ...748..133K}.
This 	figure clearly indicates a peculiar trend that 
	relatively weaker HXC intensity sources show harder spectral slope of the HXC.
	
% ----------------------------------------------------
% ----------------------------------------------------

% ----------------------------------------------------------------------
\subsubsection{Surface temperature $kT$ of SXC vs. $B_{\rm d}$}
% ----------------------------------------------------------------------

\begin{figure}
\begin{center}
\includegraphics[width=85mm]{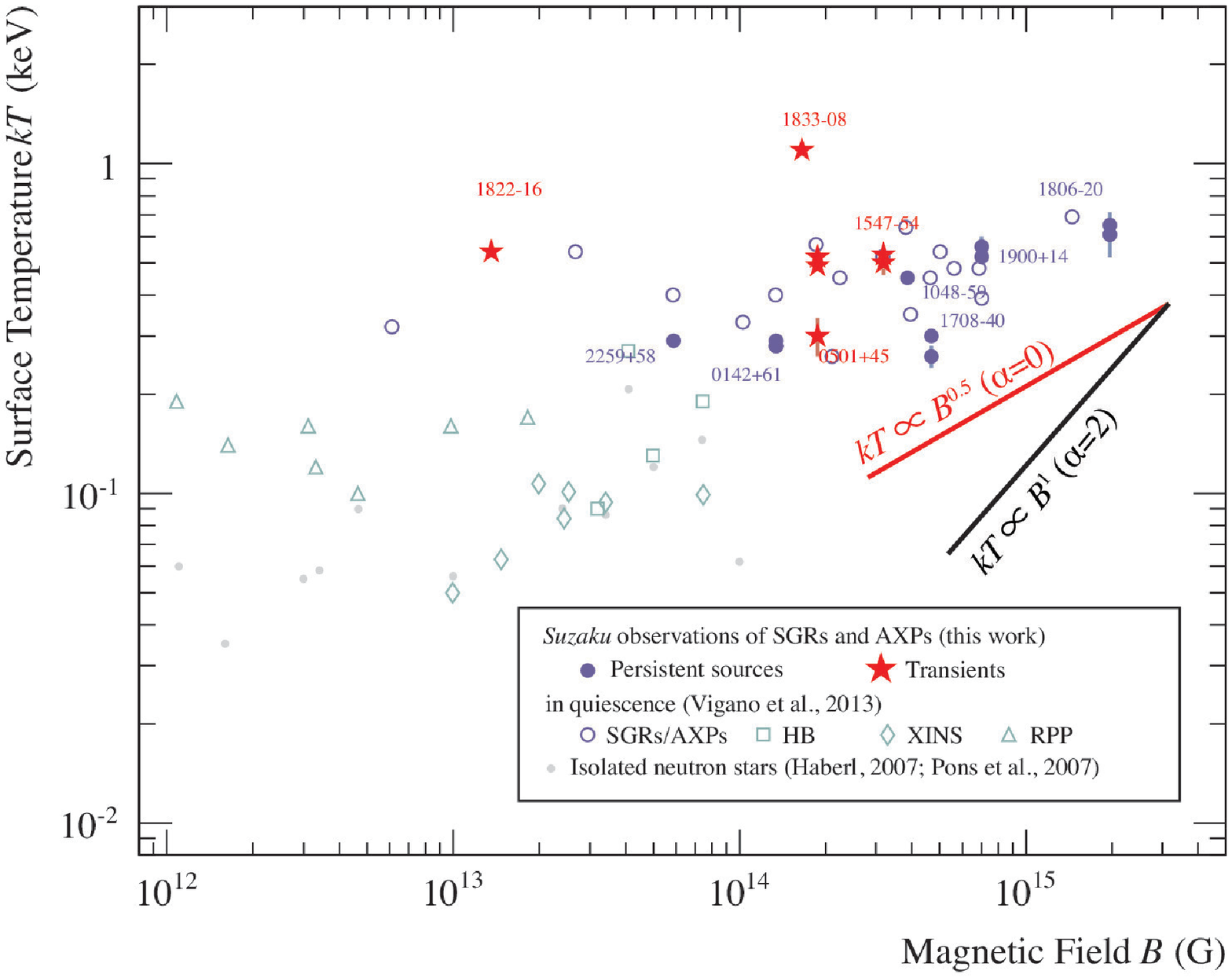}
\caption{
Surface temperature of magnetars measured with \suzaku in the CBB model (filled symbols)
	as a function of the surface dipole magnetic field $B_{\rm d}$.
For comparison,
	isolated neutron stars in quiescence are shown (open symbols)
	from \cite{2013MNRAS.434..123V}; 
	SGRs/AXPs (open circles),
	high-B pulsars (open squares),
	X-ray isolated neutron stars (open diamonds),
	and rotation powered pulsars (open triangles). 
Data from \cite{2007Ap&SS.308..181H,Pons2007PhRvL..98g1101P}	 are also compared. 
Two slopes are indicated with $T\propto B^1$ ($\alpha=2$) 
	and $kT\propto B^{0.5}$ ($\alpha=0$, exponential decay of magnetic field) 
}
\label{fig:kT_vs_B}
\end{center}
\end{figure}

The surface temperature $T_{\rm s}$ of the SXC
	is plotted as a function of $B_{\rm d}$ in Figure~\ref{fig:kT_vs_B}
	where we also added $T_{\rm s}$ of other isolated neutron stars 
	from previous studies.
This plot indicates 
	1) higher $T_{\rm s}$ of magnetars than that of other isolated neutron stars,
	2) a tendency of the positive correlation between $T_{\rm s}$ and $B_{\rm d}$ 
	in the quiescent neutron star sample,
	and 
	3) increase of $T_{\rm s}$ during transient magnetar outbursts.
All these properties suggest that the values of $T_{\rm s}$
	reflect the effects of magnetic energy dissipation,
	which is an implicit but direct consequence of the magnetar hypothesis. 

% === Magnetic Field Decay ====
The magnetic field decay in the high field regime $B_{\rm d}\gtrsim 10^{13}$\,G 
	\citep{Goldreich1992ApJ...395..250G},
	postulated in the magnetar hypothesis,
	may be formulated as \citep{Colpi2000ApJ...529L..29C,DallOsso2012MNRAS.422.2878D}
\begin{equation}
\frac{dB}{dt}=-aB^{1+\alpha},
\label{eq:colpi}
\end{equation}	
	where $a$ and $\alpha$ are the normalization and decay index, respectively.
The solution is obtained as 
\begin{equation}
B(t)=\frac{B_0}{(1+\alpha t/\tau_B)^{1/\alpha}}
\label{eq:B-decay}
\label{eq:omori}
\end{equation}
	with $\tau_B=(aB_0^{\alpha})^{-1}$. 
This formula successfully explains 
	the clustering of rotational period of SGRs and AXPs 
	around 2--12\,s \citep{Colpi2000ApJ...529L..29C},
	and resolves the overestimation of $\tau_{\rm ch}$ in 1E~2259+586 
	when compared with the age derived from plasma diagnostics of 
	the surrounding SNR CTB~109 
	(\citealt{2015PASJ...67....9N}, but see also \citealt{2014MNRAS.443.3586S}).
Then, let us here assume that 
	the surface temperature $T_{\rm s}$ 
	is determined by a balance between the radiative cooling and heating 
	by the magnetic energy dissipation of $d(B^2/8\pi)/dt$ in the crust.
Following \cite{Pons2007PhRvL..98g1101P}, this is described as 
\begin{equation}
S\sigma T^4 = -S\triangle R \frac{d}{dt}\left(\frac{B^2}{8\pi} \right),
\label{eq:surface}
\end{equation}
	where $S$, $\triangle R$, and $\sigma$ 
	are the surface area, crust thickness, and the Stefan-Boltzmann constant, respectively.
Combining this with Eq. (\ref{eq:colpi}),
	we derive 
\begin{equation}
4\pi \sigma T^4 = a\triangle R B^{2+\alpha},
\end{equation}
	or the relation of $T\propto B^{1/2+\alpha/4}$. 
The slope of Figure~\ref{fig:kT_vs_B} 
	is close to a range, $\alpha \sim$1--2,
	estimated from comparisons between pulsar and SNR ages \citep{2015PASJ...67....9N}.
This is another support for the magnetar hypothesis. 

% ----------------------------------------------------
% ----------------------------------------------------

% ~~~~~~~~~~~~~~~~~~~~~~~~~~~~~~~~
\subsection{Empirical modeling of X-ray outbursts}
% ~~~~~~~~~~~~~~~~~~~~~~~~~~~~~~~~
\label{section:outburst_lightcurve}

In our sample, transients were observed basically during X-ray outbursts.  
The decay of the SXC is often composed of 
	two distinct components 
	(e.g., \citealt{Woods2004ApJ...605..378W}): 
	an initial quickly fading emission within $\sim$1\,d
	and 
	a longer time-scale one decaying on $\sim$month.
The former component, 
	possibly related with burst activities, 
	is beyond our present analyses.
Below, we focus on the longer-decay component.

The absorbed SXC fluxes
	were already shown in Figure~\ref{fig:outburst_flux_decay},
	which are in agreement with previous studies 
	listed in Table~\ref{tab:list_of_outbursts}.
The light curves were derived 
	in the following way over individual outbursts
	from the data described in \S\ref{subsec:swift_rxte_observations}.
Individual {\it Swift} and {\it RXTE} spectra 
	were at first fitted by a single blackbody model (\S\ref{subsec:swift_rxte_observations}),
	while	
	the CBB model (\S\ref{section:suzaku}) was further employed
	to explain 	residuals in the higher energy band 
	if the resultant reduced chi-square is not accepted ($\chi^2_{\nu}>1.4$).
All the spectral parameters were allowed to vary in each observation
	except the absorption column density $N_{\rm H}$ 
	which is fixed at the value determined by high statistics observations
	from early phase observations with  {\it Suzaku}/XIS or {\it Swift}/XRT. 
Below, 
	we use flux measurements of acceptable fits, %($0.02<\chi^2_{\nu}<1.40$),
	and detailed spectral characterization will be discussed elsewhere.

\begin{figure}[!t]
\begin{center}
\includegraphics[width=80mm]{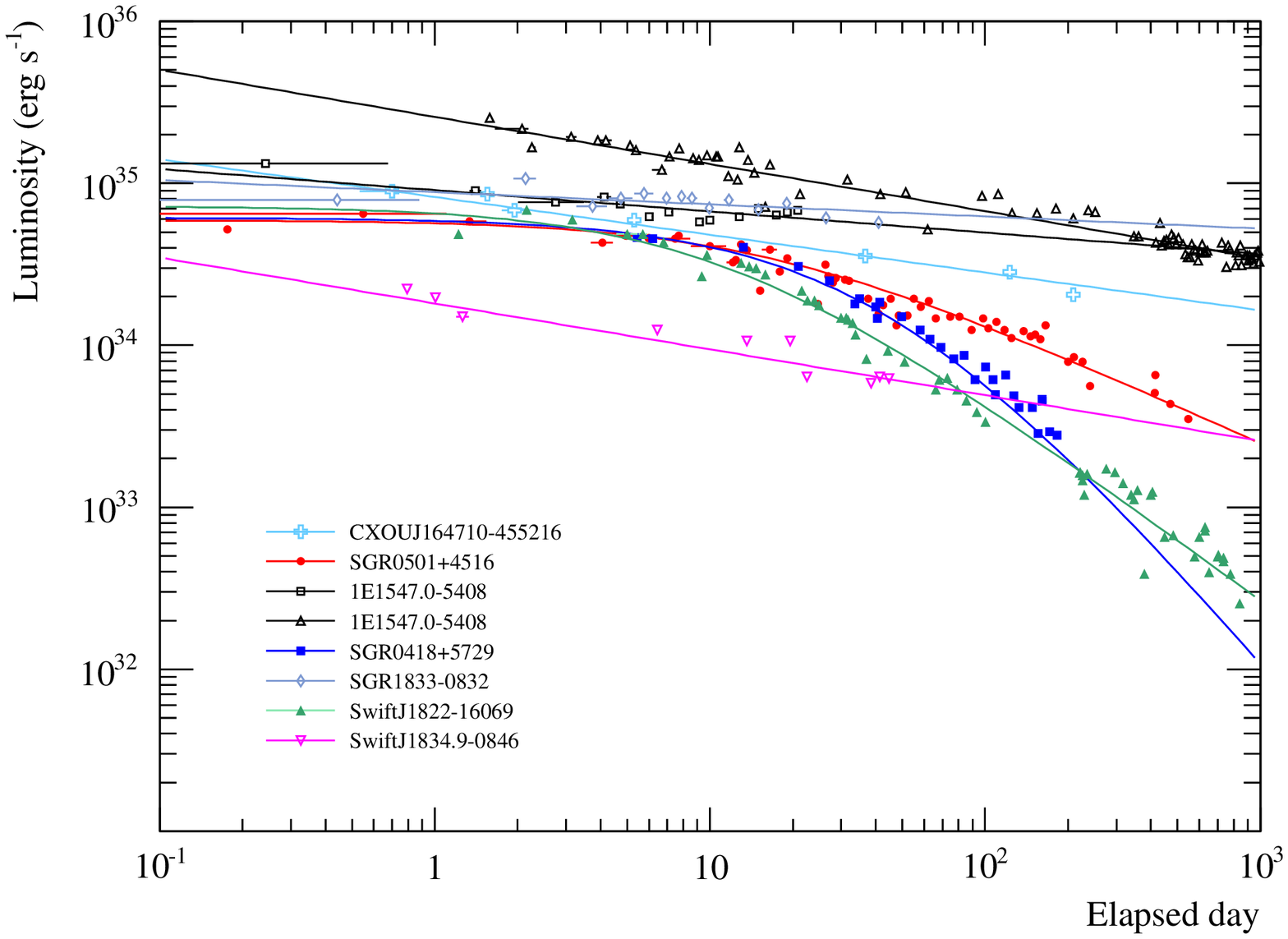}
\caption{
The same as Figure~\ref{fig:outburst_flux_decay},
	but the observed flux was converted to the unabsorbed 1--20\,keV luminosity,
	and the outburst decay was fitted with eq. (\ref{eq:Omori}). 
}
\label{fig:luminosity_decay}
\end{center}
\end{figure}

Figure \ref{fig:luminosity_decay} 
	shows the 0.1--20 keV luminosities of these outbursts 
	after correcting for 
	the photo-absorption (\S\ref{subsec:swift_rxte_observations}) 
	and distances $d$ (Appendix \S\ref{section:location}).
Such fading light curves in Figure~\ref{fig:outburst_flux_decay}
	from different objects 
	have so far been fitted by some empirical formulae: 
	an exponential \citep{Rea2009MNRAS.396.2419R}, % SGR~0501+4516 
	double-exponential \citep{Scholz2012ApJ}, % Swift~J1822.3$-$1606
	power-law \citep{Lyubarsky2002ApJ...580L..69L,Kouveliotou2003ApJ...596L..79K}
	% SGR~1900$+$14, SGR~1627$-$41
	or broken power-law functions. 
Since there is no commonly established function applicable to all the outbursts,
	nor consensus on physical understanding of the decay,
	it is still meaningful to search for an empirical formula to uniformly investigate the outbursts.
	
The light curves imply that 
	some transients exhibit
	a plateau-like constant period at the earliest phase, 
	followed by a power-law decay,
	such as three transients: 
	SGR~0501$+$4516,
	SGR~0418$+$5729,
	and Swift~J1822.3-1606.
In order to represent this behaviour, we employed a mathematical form as 
\begin{equation}	
L(t) = \frac{L_0}{(1+t/\tau_0)^p}.
\label{eq:Omori}
\end{equation}	
Hereafter we call this Eq.(\ref{eq:Omori}) the plateau-decaying (PD) function 
	since 
	it becomes constant $L_0$ at $t\ll \tau_0$, %\,erg\,s$^{-1}$\,cm$^{-2}$
	while it becomes a power-law with an index of $p$ at $t\gg \tau_0$.
For sources which plateau is not clear, we just used the power-law (PL) model,
	 $L_{\rm x}=L_0 (t/\tau_0)^{-p}$ which is normalized at $\tau_0=1$\,d. 
The resultant parameters are summarized in Table~\ref{tab:list_of_outbursts},
	and further comparison among difference fittings 
	are described in Appendix~\ref{appendix:decay_function}.
As described there, 
	the plateau is not an artefact of the definition of the time origin.
The plateau period are $\tau_{\rm 0}$$\sim$11--43\,d for the above three sources, 
	and 
	subsequent decay indices are $p$$\sim$0.7--2.
Other sources (e.g., 1E~1547.0$-$5408 and Swift~J1834.9$-$0846) 
	faded with a flatter index $p$$\sim$0.1--0.3 
	without such a clear plateau duration.
		
The total energy $E_{\rm total}$ released 
	during a single outburst, an integration of Eq. (\ref{eq:Omori}),
	becomes $E_{\rm total}=\tau_0L_0/(p-1)$, 
	in a steep decay case ($p>1$),
	while that in the flatter decay 
	is calculated with assuming 
	an integration upper bound at 100\,d
	as a typical decaying duration. 
Even under present uncertainties of distance measurements,
	as listed in Table~\ref{tab:list_of_distance}, 
	a typical $E_{\rm tot}$ is a few of $10^{41}$ erg. 
	
% ----------------------------------------------------
% ----------------------------------------------------	

% ==========================================
\section{DISCUSSION}
\label{sec:Discussion}
% ==========================================

We compiled  
	all archival \suzaku spectra 
	of fifteen AXP and SGR sources acquired in 2006--2013,
adding 10 new observations to Paper~I.
We also incorporated 
	the {\it Swift}/XRT, {\it RXTE}/PCA, and {\it NuSTAR} public data.
These data sets reinforced 
	the magnetar hypothesis for SGRs and AXPs.

% ==========================================
\subsection{Uniqueness of the two spectral components}
% ==========================================

% ----------------------------------------------------
\begin{figure}
\begin{center}
\includegraphics[width=88mm]{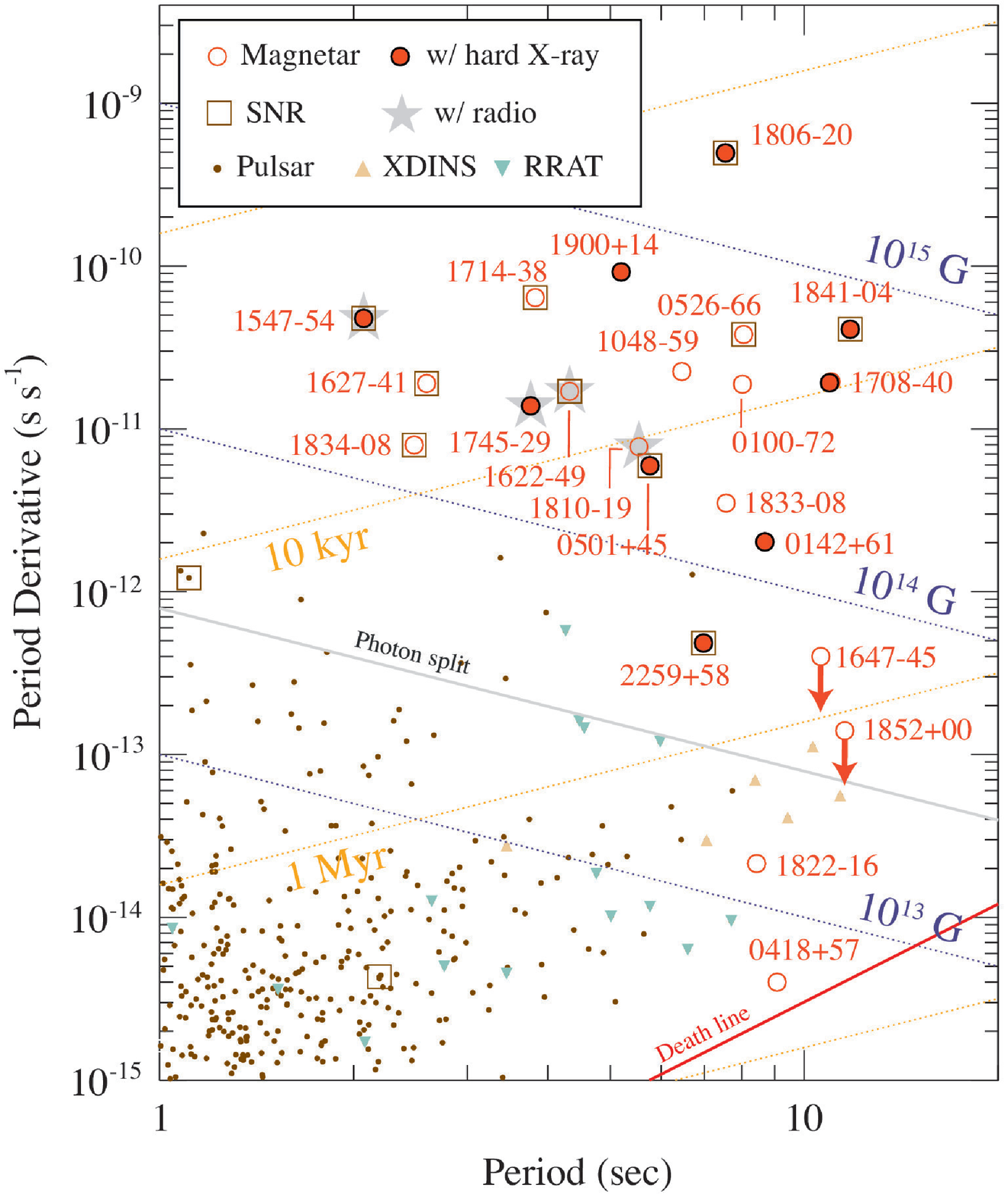}
\caption{
Magnetars (circle symbols) on the $P$-$\dot{P}$ diagram 
	(zoom-up from Figure~\ref{fig:p_pdot_diagram}).
Sources are shown in the filled symbols 
	if the HXC 
	are detected either by {\it Suzaku}, {\it INTEGRAL}, or {\it NuSTAR}.
Association with an SNR and detection in the radio band 
	are shown in square and star marks, respectively. 
Canonical pulsars (dot symbols), X-ray Dim Isolated Neutron Stars (XDINS; triangles),
	and Rotating RAdio Transient (RRAT; inverted triangles)
	are shown for comparison. 
}
\label{fig:ppdt_zoom}
\end{center}
\end{figure}
% ----------------------------------------------------

The characteristic spectral shapes, 
	composed of the SXC ($\lesssim$10 keV) and HXC ($\gtrsim$10 keV),
	have been revealed to be a common feature of this class.
Figure~\ref{fig:ppdt_zoom},
	an expanded $P$-$\dot{P}$ plot, 
	summarizes the currently available HXC information of all AXPs and SGRs. 
Combining {\it INTEGRAL}, {\it RXTE}, {\it Suzaku}
	(with 7 detections), and {\it NuSTAR} 
	which afforded the HXC detections from 
	1E~2259$+$586 \citep{Vogel2014ApJ...789...75V}
	and SGR~J1745$-$29 near Sgr A$^{*}$
	\citep{Mori2013ApJ...770L..23M,Kaspi2014ApJ...786...84K}, 
	the HXC has been confirmed from 9 objects
	among $\sim$23 confirmed sources,
	and the detected 15--60 keV HXC flux level is in the range of 
	$\sim 0.3$--$11\times 10^{-11}$\,erg\,s$^{-1}$\,cm$^{-2}$.

One important property of the HXC is that it exhibits the clear scalings,
	in terms of its luminosity ratio $\xi$ 
	(\S\ref{SpectralModeling}; Fig.\ref{fig:correlations}; Eq.\ref{eq:xi_vs_P}, \ref{eq:xi_vs_B}, and \ref{eq:xi_vs_tau})
	and its spectral slope $\Gamma_{\rm h}$
	(\S\ref{PhotonIndex}; Fig.\ref{fig:Gh_vs_B}c; Eq.\ref{eq:Gh_vs_B}).
Since the scalings apply both to AXPs and SGRs,
	they can be collectively called magnetars 
	at least from the view point of wide-band spectrum. 
These relations also allow us to predict the HXC intensity of a source,
	when its $P$, $\dot{P}$, and the SXC intensity are given.
At \S\ref{Ratio of HXC to SXC vs. magnetic field}, 
	we pointed out the narrow $P$ clustering of magnetars,
	and showed the $\eta$ and $\xi$ correlations to $\dot{P}$. 
In future, if young and fast rotating magnetars
(e.g., $P\sim 0.1$\,s at $\sim$10--100\,yr, as an origin of fast radio bursts, e.g., see \citealt{2017arXiv170208644B}), 
or hypothetical slowly rotating magnetars (e.g., $P \sim 100$\,s,
	as related with gamma-ray bursts \citealt{2015ApJ...813...92R}), 
	would be discovered, 
	the wider range of $P$ will provide another hint to judge a main control parameter of the spectral shape.	
	
Do we observe such two-component spectra from other classes of magnetized neutron stars?
Some rotation-powered pulsars, such as Vela pulsar or Geminga, 
	exhibit two-component spectrum, consisting of thermal and non-thermal components. 
However, 	they have lower $kT_{\rm s}$ and steeper	 $\Gamma_{\rm h}\sim 1.5$--2.5 
	than those of the SGRs and AXPs,
	and show no clear scaling of $\xi$-$B_{\rm d}$.
Accretion-powered X-ray pulsars 
emit predominantly in a single hard component (at least in $>$2\,keV),
	regardless of their luminosity (e.g., \citealt{2006ApJ...648L.139T}).
Thus, the two-component spectral composition is considered to be specific to magnetars. 	

% ==========================================
\subsection{The reality of $B_{\rm d}$} 
% ==========================================
Different from the magnetar model, 
	an alternative scenario to explain SGRs and AXPs
	assumes that 
	$B_{\rm d}$ calculated from $P$ and $\dot{P}$ does not represent 
	the true dipole field strength near a stellar surface
	nor exceed the critical field $B_{\rm QED}$.
In this case, 
	X-ray radiation is not powered from the magnetic energy 
	but from accretion of a fallback disk 
	which is left over from a supernova explosion of the progenitor
	\citep{Chatterjee2000ApJ...534..373C,Alpar2001ApJ...554.1245A,2013ApJ...778..119B}.
The bulk-motion Comptonization of the accretion column is attributed 
	to the HXC radiation 
	\citep{Trumper2010A&A...518A..46T, Trumper2013ApJ...764...49T, Kylafis2014A&A...562A..62K}.	

The accretion scenario has still difficulties to explain the broad-band X-ray observations:
1) Firstly, as already stated above, the accretion-powered pulsars do not show 
	distinct two spectral components even in the low X-ray luminosity. 
2) The power-law HXC extend up to $\sim$100\,keV with the hard $\Gamma_{\rm h}$
	without any absorption nor cutoff features 
	which are usually expected from the electron cyclotron resonance
	of the ordinary accretion-powered neutron stars.
3) Finally, the spectral scalings implies that the $B_{\rm d}$ values 
	play an important role for the emission mechanism, especially to the HXC.
The nominal value $B_{\rm d}$ derived from $P$ and $\dot{P}$
	is, therefore, considered to be a true poloidal component 
	near the stellar surface. 
	
The present work suggested that 
	$B_{\rm d}$ is one main control parameter of the HXC radiation.
Some theoretically motivated emission models 
	have been already developed so far 
	in the magnetar scheme;
	these include 
	thermal bremsstrahlung \citep{2005ApJ...634..565T},	
	synchrotron radiation  \citep{2005ApJ...634..565T,2005MNRAS.362..777H},
	resonant scattering \citep{2007Ap&SS.308..109B,2008MNRAS.389..989N,2012JPhCS.342a2013V,2013ApJ...762...13B},
	and down-cascade due to the photon splitting (Paper~I).
Although the mechanism is still observationally poorly understood, 
	the appropriate radiation scenario would include the physics in the strong magnetic field
	to explain the scaling. 
For example,
	the hard photon index $\Gamma_{\rm h}$ in our sample is correlated with $B_{\rm d}$.
This can not be explained only by the difference of viewing angles,
	and should be further compared with models (e.g., \citealt{2013ApJ...762...13B}).
	
One potential clue is how $\xi$ or $\Gamma_{\rm h}$ 
	behave on the scaling plots during the outburst.
For example,
	there is a signature of a faster decay of the HXC than the SXC
	in some sources: 
	e.g., SGR~0501+4516, \citep{Rea2009MNRAS.396.2419R,Enoto2010ApJ...715..665E},
	and 1E~1547.0$-$5408 (Enoto et al., in prep.).
This question has not yet to be clearly answered in the present {\it Suzaku} data.
In addition, 
	our current observations of the HXC are limited up to $\sim$60\,keV.
The HXC cutoff is expected at $\sim$400\,keV by the fossil disk model 
	\citep{Kylafis2014A&A...562A..62K},
	while the annihilation line is suggested in some magnetar models \citep{2013ApJ...762...13B}.
Thus, soft gamma-ray observation is expected to provide a smoking gun. 

% ==========================================
\subsection{Implications of the $L_{\rm x}$-$L_{\rm sd}$ diagram}
%: \\ Comparison to rotation-powered pulsars} 
\label{subsec:discussion_rotation_powered_pulsars}
% ==========================================		

In Figure~\ref{fig:Lx_vs_Lsd} and Table~\ref{tab:suzaku_obs_Lx}, 
	we compare 
	the total X-ray luminosity $L_{\rm x}=L_{\rm h}+L_{\rm s}$ (\S\ref{sec:Analysis_and_Results})
	with the spin-down powers $L_{\rm sd}$. 
The 2--10 keV luminosities of 41 ordinary rotation-powered pulsars 
	follow 
	a linear empirical relation, $\log L_{\rm x}=1.34 \log L_{\rm sd}-15.34$
	\citep{Possenti2002A&A...387..993P}.
They are well below the critical luminosity
	$\log L_{\rm crit}=1.48 \log L_{\rm sd} -18.5$,
	which is thought to be the maximum efficiency of conversion of
	the spin-down luminosity to X-ray emission.
Our inclusion of $L_{\rm h}$ 
	improved $L_{\rm x}$ measurements,
	and more clearly show the luminosity excess ($L_{\rm x}\gg L_{\rm sd}$).
Furthermore,
	$L_{\rm x}$ values exhibit little dependence on $L_{\rm sd}$. 
These properties have indeed been providing strong support to the magnetar hypothesis. 		

Highly variable transients mostly reach $\sim 10^{35}$\,erg\,s$^{-1}$ at an early stage of their outbursts
	and gradually decay back to quiescence below $\lesssim 10^{33}$\,erg\,s$^{-1}$.
Some of them (e.g., 1E~1547.0$-$5408 and Swift~J1834.9$-$0846) 
	decay back to $L_{\rm x}<L_{\rm sd}$, 
	to a region on Figure 14 which is adjacent to  
	radio-loud normal rotation-powered pulsars 
	\citep{Rea2012ApJ...748L..12R}.	
However, 	
	it is still unclear whether 
	AXPs and SGRs in their deep quiescence,
	can become dimmer than $L_{\rm crit}$. 

% ==========================================
\subsection{Maximum luminosity and decay law of transients}
\begin{figure*}[h]
\begin{center}
\includegraphics[width=150mm]{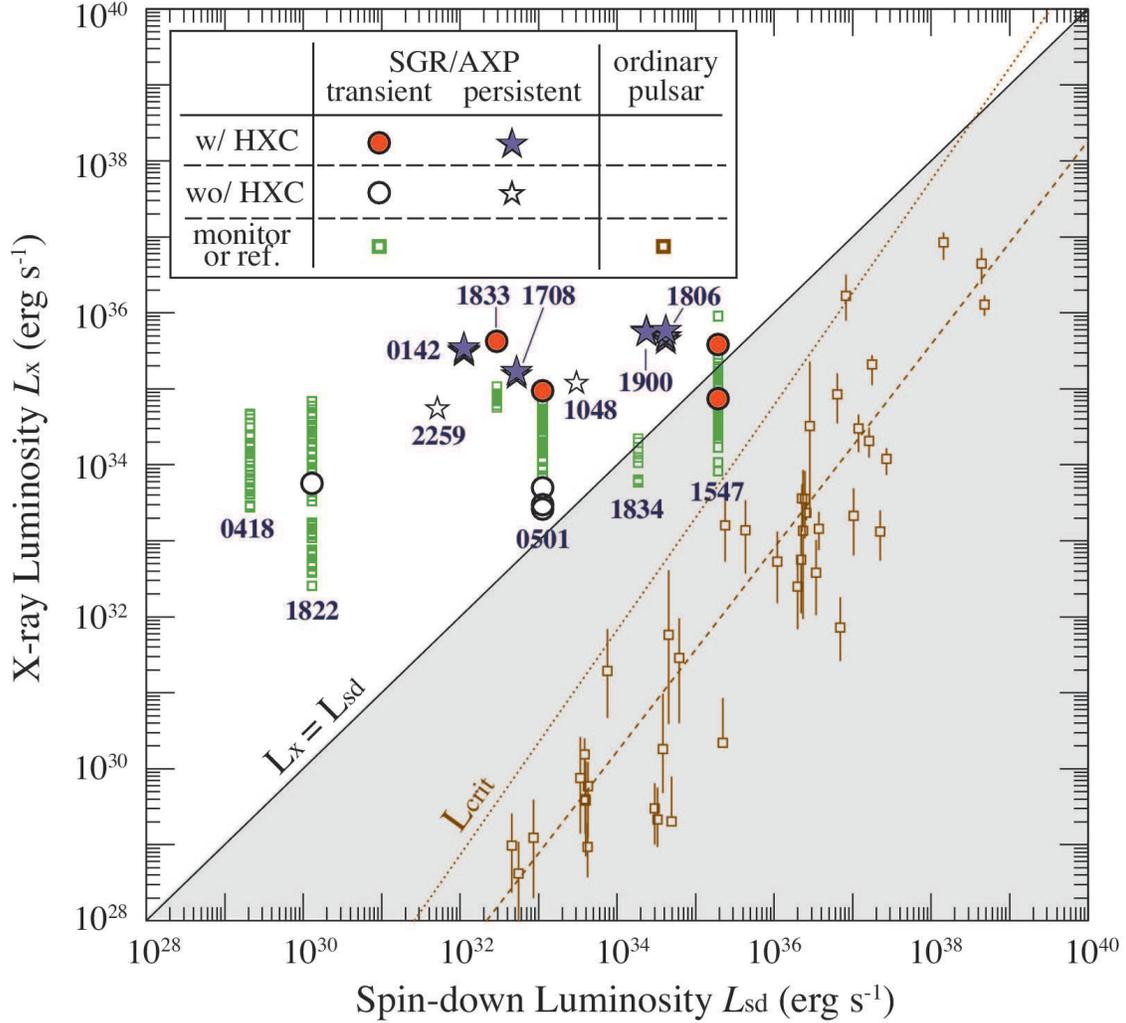}
\caption{
Observed 1--60\,keV X-ray luminosity $L_{\rm x}=L_{\rm s}+L_{\rm h}$ (SXC+HXC)
	of the objects listed in Table~\ref{tab:suzaku_obs_Lx}, 
	compared with their spin-down luminosity $L_{\rm sd}$.
Filled stars and circles indicate 
	the persistently bright objects 
	and 	
	transient ones, respectively.
The sources are shown in filled symbols
	if the HXC are detected with {\it Suzaku}, 
	while the others are open ones.
The decaying X-ray luminosity monitored with {\it Swift} and {\it RXTE}
	are shown with small green squares.
Ordinary rotation powered pulsars are shown in brown squares 
	from Table~1 in \cite{Possenti2002A&A...387..993P}.
The employed distances and periods refer to 
	Appendix \ref{section:location}
	and the McGill catalog \citep{Olausen2013arXiv1309.4167O}.	
}
\label{fig:Lx_vs_Lsd}
\end{center}
\end{figure*}

%Comparison to accretion-powered pulsars}
\label{subsec:Discussion_decay_law}
% ==========================================
On the $L_{\rm x}$-$L_{\rm sd}$ diagram in Figure~\ref{fig:Lx_vs_Lsd},
	the persistent X-ray emission do not exceed $\sim$$10^{36}\,$erg\,s$^{-1}$
	either in persistent sources or in transients during an early phase of outbursts.
In the light curves (\S\ref{section:outburst_lightcurve}),
	transients sometimes show an initial plateau phase for  $\sim$10--40 d 
	keeping $L_{\rm s}\sim 10^{35}$--$10^{36}$\,erg\,s$^{-1}$,
	in the same $L_{\rm s}$ range as the persistently sources.	
If the emission is a pure blackbody radiation,	
	this typically corresponds to 
$L_{\rm max}=4\pi R_{\rm spot}^2 \sigma T^4 = 1.3 \times 10^{35} (R_{\rm spot}/1\,{\rm km})^2
(kT/1\,{\rm keV})^4$\,ergs\,s$^{-1}$. 
Such a ceiling X-ray luminosity 
	could be interpreted as regulation via a temperature-sensitive neutrino cooling, 
	if the magnetic energy is converted to thermal energy in the stellar crust 
	\citep{2007Ap&SS.308..353P, Rea2013ApJ,Pons2012ApJ,Esposito2013MNRAS}.

% ----------------------------------------------------
% ----------------------------------------------------

% ----------------------------------------------------
\begin{figure}[!t]
\begin{center}
\includegraphics[width=80mm]{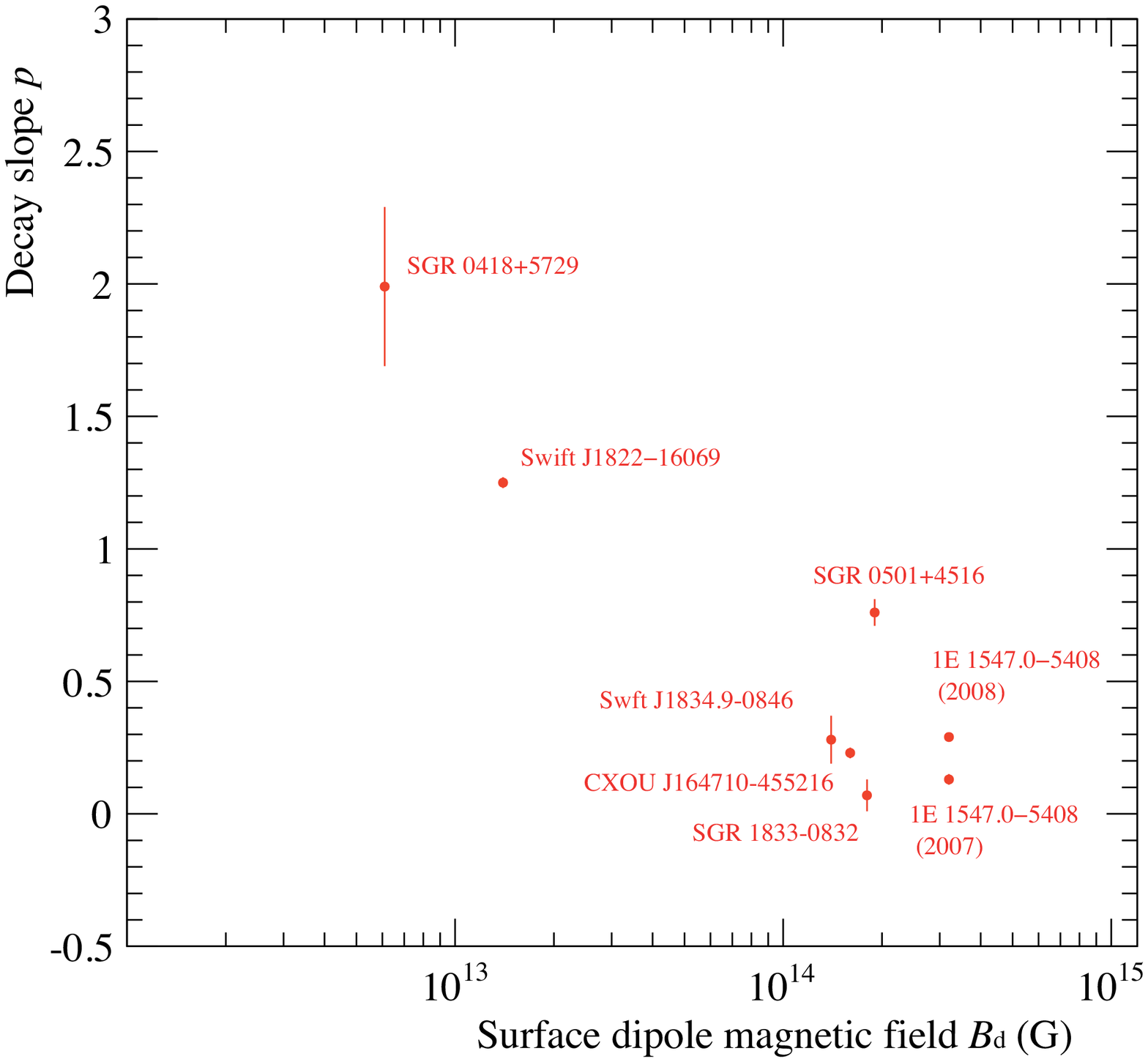}
\caption{
The decay slope $p$ (Eq. \ref{eq:Omori}) 
	shown as a function of $B_{\rm d}$.
}
\label{fig:B_vs_p}
\end{center}
%\end{figure*}
\end{figure}
% ----------------------------------------------------	

As shown in Figure~\ref{fig:B_vs_p},
	outbursts of higher-$B_{\rm d}$ magnetars show more prolonged decay than 
	those of lower-$B_{\rm d}$ sources. 
This supports the basic concept that outbursts from magnetars are 
	powered by sudden dissipation of the magnetic energy,
	although we do not yet know whether the release occurs 
	in the stellar interior \citep{Pons2012ApJ,2015ApJ...815...25L}
	or in the magnetosphere \citep{2007ApJ...657..967B,2014PTEP.2014b3E01K,2015MNRAS.447.2282B},
	or both.

The total emitted energy during a single X-ray outburst 
	is typically $\sim 10^{41}$--$10^{42}$\,erg.
If this outburst can be attributed to the magnetic energy
	stored in the magnetar crust 
	with a thickness of $\triangle R$ 
	and a hot spot radius of $R_{\rm spot}\sim 1$\,km,
	i.e., 
	$E_{\rm mag}= \pi R_{\rm spot}^2 \triangle R \cdot B^2/8\pi 
	=2.4\times 10^{41} 
	(B/B_{\rm QED})^2
	(R_{\rm spot}/1\,{\rm km})^2
	(\triangle R/1\,{\rm km})\,{\rm erg}$,
	the required $\triangle R \sim 1$\,km is comparable to the crust thickness. 
Since low-$B_{\rm d}$ sources below $B_{\rm QED}$ 
	(SGR~0418+5729 and Swift~J1822.3$-$1606)
	also radiate a similar amount of energy 
	to those of other high-$B_{\rm d}$ sources, 
	the actual surface field of  low-$B_{\rm d}$ transients
	would be much stronger than estimated from $P$-$\dot{P}$ method.

Incidentally, 	
	the fall-back disk model tries to explain the outbursts in terms of sudden mass accretion
	\citep{2001ApJ...554.1245A,2013ApJ...778..119B,2015MNRAS.447.2282B}. 
However,
	in such cases, 
	the luminosity would hit a ceiling at the Eddington limit ($\sim 10^{38}$\,erg\,s$^{-1}$)
	rather than at $\sim 10^{36}$\,erg\,s$^{-1}$;
	the outburst would decay exponentially \citep{1996ARA&A..34..607T},
	or sometimes falls abruptly via the propeller effect \citep{2013ApJ...773..117A}
	rather than in power-law form;
	and the spectrum around the outburst peak would exhibit
	clear fluorescence Fe-K line (e.g., EW$>$100\,eV),
	if the fast initial rise of X-ray outbursts indicates 
	a mass reservoir close to the neutron star.
Therefore, the scenario falls to account for several important properties of the actually observed 
	outburst of magnetars. 	

% ==========================================
\subsection{Speculative relation to seismology}
% ==========================================
In the magnetar scenario, 
	fading X-ray radiation of outbursts are usually theoretically modeled 
	as a transient thermal response to a sudden energy release 
	inside the stars. 
Originally proposed  
	 for post-glitch afterglows of ordinary neutron stars 
	\citep{Eichler1989ApJ, Hirano1997ApJ},
	this idea is recently investigated 
	in the magnetar context  \citep{Lyubarsky2002ApJ...580L..69L}
	and applied to X-ray observations of transient sources 
	\citep{Rea2012ApJ...754...27R, Scholz2012ApJ}.

Instead of a large delta-function energy release assumed in previous models, 
	we may alternatively assume that many small subsequent energy deposits 
	take place during the X-ray afterglow
	as small starquakes or reconnections. 
Such a scheme remind us of seismology,
	where the occurrence frequency of aftershock sequence $n(t)$
	after a large earthquake
	is known to follow an empirical formula,
	called the modified Omori's law, 
	as $n(t) \propto (t+c)^{-p}$ \citep{Utsu1961}.
In seismology,
	$t$ is the elapsed time after the mainshock,
	$p$$\sim$0.7--1.5 is the slope index of decay,
	and $c\lesssim 1$\,d is an empirical plateau phase duration \citep{Utsu1995}. 
This formula, thought to represent 
	an underlying relaxation process after a mainshock,
	has the same mathematical form as 
	the plateau-decaying function we used for X-ray outbursts 
	(Eq. \ref{eq:omori} in \S\ref{subsec:swift_rxte_observations}).
		
We already know another statistical similarity 
	of seismology to magnetars, 
	i.e., Gutenberg-Richter law \citep{Ishimoto1938,Gutenberg1941} seen in SGR short bursts:
	the number-intensity relations of their 
	short burst fluence 
	follows a power-law distribution
	\citep{Gotz2006A&A...449L..31G, Nakagawa2007PASJ...59..653N}
	as commonly seen in earthquakes and solar flares.
This statistical relation is believed to represent  
	self-organized criticality of the system
	\citep{Aschwanden2011soca.book.....A}.
If, furthermore, the magnetar persistent emission (or a part of it)
	is composed of small energy releases, 
	the plateau-decaying light curve 
	is understood as changes in the normalization 
	of the Gutenberg-Richter law during the outbursts \citep{Nakagawa2009PASJ...61..109N},
	providing another similarity 
	of underlying physical process to seismology,
	where subsequent energy dissipation triggered by a previous event.

% ==========================================
\section{CONCLUSION}
\label{sec:Summary}
% ==========================================
Our \suzaku observations of the SGR and AXP classes,
	combined with {\it NuSTAR}, {\it Swift}, and {\it RXTE} 
	 data, have yielded the following results. 
The magnetar interpretation have been 
strengthened from several aspects.
	
\begin{enumerate}
\item
The HXC, $\sim$0.3--11$\times 10^{-11}$\,erg\,s$^{-1}$\,cm$^{-2}$
in the 15--60\,keV band,  is thought to be a common feature of the this class.
\item 
The  scalings of $\xi$ and $\Gamma_{\rm h}$ found in Paper I 
(Eq. \ref{eq:xi_vs_P}, \ref{eq:xi_vs_B}, and \ref{eq:xi_vs_tau}), 
have been reinforced. $B_{\rm d}$ derived from $P$ and $\dot{P}$ 
is a key control parameter of the spectral evolution. 
\item
After including the HXC,
	differences from the ordinary rotation- or accretion-powered pulsars 
	become more prominent: 
	e.g., $L_{\rm x}>L_{\rm sd}$
	and the ceiling luminosity at $10^{35-36}$\,erg\,s$^{-1}$.
\item 
The fading $L_{\rm s}$ during outbursts is approximated by 
	a mathematical form of Eq.~\ref{eq:Omori},
	implying a similarity to seismology.
\end{enumerate}

\acknowledgments
The authors would like to express their thanks to the \suzaku team 
	for their prompt ToO observations,
	and to Prof. Hiroyuki Nakanishi for his providing 
	the hydrogen column density map of our Galaxy.
This work was supported by 
	JSPS KAKENHI Grant Numbers,
	12J03320, % JSPS Fellow of TE
	15H00845,  % Shingakuzyutu NS / TE
	15H03653, % Kazuo Makishima / Kiban-B
	16H00869, % Suwa / Shingakuzyutu 
	16H02198, % Kiban-A / Tamagawa 
	16K17665, % Suwa / Wakate-B 
	16J06773, % Kisaka / 
	25105507, % Kazuo Makishima / Shingakuzyutu 
	25400221, % Shibata Shinpei / Kiban-C
	and Hakubi project at Kyoto University. 
%T.E. was supported by JSPS KAKENHI, 
%	Grant-in-Aid for JSPS Fellows, 24-3320,
%	15H00845, and Hakubi project at Kyoto University. 
%Y. S was supported by JSPS KAKENHI, Grant-id-Aid for	
%	16H00869 and 16K17665.
%K.M. acknowledges the JSPS KAKENHI support,
%	Grant-in-Aid for Scientific Research (B). 
This research was also supported by the Munich Institute for Astro- and Particle Physics (MIAPP) 
	of the DFG cluster of excellence ``Origin and Structure of the Universe."
\bibliographystyle{apj}
\bibliography{suzaku,reference,decay,nustar,distance}

% =========================
%\input{report/report20140131}
% =========================

\appendix
\clearpage %\newpage
% ==========================================
\section{Note on individual sources updated from Paper~I}
\label{section:suzaku:individual_sources}
% ==========================================
%Some of \suzaku data were published in previous work 
%	listed in the note of Table~\ref{tab:list_suzaku_log}.
Here we summarize updated information of individual observations from Paper I.

% -------------------------------------------------------
\subsection{SGR~1833$-$0832}
% -------------------------------------------------------
On 2010 March 19,
	SGR 1833$-$0832 was discovered from short bursts.
A \suzaku ToO was performed on March 27
	for an effective 40 ks exposure, 
	$\sim$8.4 days after the discovery.
%During the $\sim$90\, ks gross observation,
The 0.5--10 keV XIS light curve is stable without any clear short burst detections. 
%This \suzaku data has not yet to be reported elsewhere. 
%The XIS0 data acquired with 1/8 window mode (1-s timing resolution)
%	was utilized for the pulse search 
%	with the barycentric correction via {\tt aebarycen} \citep{Terada2008PASJ...60S..25T}.
The pulsation was detected with XIS0 (1/8 window mode) 
	at $P=7.565\pm0.001$\,s at the MJD epoch of 55272.
The pulsed fraction is 
	${\rm PF}=(F_{\rm max}-F_{\rm min})/(F_{\rm max}+F_{\rm min})
	\sim$60--80\% (2--10 keV), 
	where $F_{\rm max}$ and $F_{\rm min}$ are the background-subtracted maximum 
	and minimum count rates of the pulse profile.
%Our timing result is consistent with the measurement by 
%	{\it Swift} and {\it RXTE} \citep{Gogus2010ApJ...718..331G,Esposito2011MNRAS.416..205E}.
When fitting the 1--10 keV XIS data alone,
	the SXC spectrum is fitted 
	by a single blackbody 
	of $kT$=1.22\,keV with
	$N_{\rm H}$=$9.0\times 10^{22}$\,cm$^{-2}$
	or 
	by a single power-law of $\Gamma$=$3.1$ 
	with $N_{\rm H}$=$1.6\times 10^{23}$\,cm$^{-2}$.
The HXC detection is marginal (Table~\ref{tab:suzaku_detections}).
	%at $\sim$2.9$\sigma$ level in Table~\ref{tab:suzaku_detections}.
%During the pouting toward SGR~1833-0832,
%	an Earth occultation data was acquired with the HXD 
%	which can be utilized to study the accuracy 
%	of the simulated HXD-PIN NXB \citep{Fukazawa2009PASJ...61S..17F}.
Comparing with Earth occultation data,
	the simulated HXD-PIN NXB
	was found to be slightly under-estimated by $\sim$2\% than other observations,
	and if we assigned this additional 2\% to the NXB uncertainty,
	the HXC detection significance reduced to 2.4$\sigma$ level. 
Thus, 
	we just regard the HXC of SGR~1833$-$0832 as a marginal signal.
%When including the possible HXC power-law of $\Gamma_{\rm h}=0.75^{+0.63}_{\-0.55}$,
%	the BB temperature is reduced to $kT=0.93\pm0.14$\,keV from $\sim$1.2\,keV.

% -------------------------------------------------------
\subsubsection{4U~0142+61} 
% -------------------------------------------------------
First and second \suzaku observations were performed in 2007 and 2009
	and already reported 
	focusing the spectral feature and pulse modulation, respectively 
	\citep{Enoto2011PASJ...63..387E,Makishima2014PhRvL.112q1102M}.	
A $\sim$37\,ks ToO was performed in 2011 September after 
	short burst activities on July 29, 2011
	\citep{Oates2011GCN..12209...1O}
	which states showed a slight brightening than previous observations.
Another $\sim$80\,ks AO8 observation in 2013 July
	is planned to report elsewhere.
%Both the SXC and HXC were confirmed from all the observations,
%	and they have been rather stable 
%	(see also, \citealt{Gonzalez2010ApJ...716.1345G}).

% -------------------------------------------------------
\subsubsection{AX~J1818.8$-$1559}
% -------------------------------------------------------
A magnetar candidate AX~J1818.8$-$1559 was discovered 
	from the ASCA Galactic survey \citep{Sugizaki2001ApJS..134...77S}.
\cite{Mereghetti2012A&A...546A..30M} performed a {\it Suzaku} follow-up
	observation, combined with {\it Chandra} and {\it XMM-Newton} spectral studies, 
	after a detection of a magnetar-like short burst with {\it INTEGRAL} in 2007.
%The similarity of the X-ray spectral and a burst behavior indicate
%	that AX~J1818.8$-$1559  is a new magnetar candidate.	
Due to low photon counts of the XIS spectra,
	we fixed $N_{\rm H}$ value 
	within the error range of \citep{Mereghetti2012A&A...546A..30M},
	and the derived flux and $kT$ values are consistent with their report.
%Although the source is a promising magnetar candidate,
%	we did not include this source as a magnetar class in the following discussion. 	
	
% -------------------------------------------------------
%\subsubsection{SGR~0501+4516} 
% -------------------------------------------------------
%After an early ToO \citep{Enoto2009ApJ...693L.122E,Enoto2010ApJ...715..665E},
%	four follow-up observations are performed in 2008, 2009, 2010, and 2013.
%The soft thermal emission
%	is gradually fading at least by one order of magnitude.
%The BB temperature of the SXC decreases from $\sim$0.5\,keV to $\sim$0.Y\,keV
%	with shrinking the BB radius, 0.3\,km to 0.0Y\,km.
%The HXC was only detected during the first observation. 

% -------------------------------------------------------
\subsubsection{1E~1841$-$045 (Kes~73)}
% -------------------------------------------------------
An angular radius $\sim$2$'$ of the shell-type SNR Kes 73
	is comparable with the HPD of the XRT. % (HPD of 2').
To take into account the SNR contamination into the XIS data,
%	Thus, the SNR component is highly contaminated 
%	into the XIS data.	
\cite{Morii2010PASJ...62.1249M} 
	used the {\it Chandra} SNR observation 
	\citep{Morii2003PASJ...55L..45M}
	to fit the {\it Suzaku} spectrum, 
	and we previously used the {\it XMM} result
	\citep{Vink2006MNRAS.370L..14V}
	in Paper I.
In this paper, we replaced it to the latest study 
	based on 
	the combined {\it Chandra} and {\it XMM} analyses 
	described in Table 3 by \cite{Kumar2014ApJ...781...41K},
	where two soft and hard components,
	corresponding to the swept-up interstellar medium 
	and ejecta,
	are represented by 
	\texttt{VPSHOCK}+\texttt{VPSHOCK} models in XSPEC.
Since there is spatial variation of the emission lines,
	we let the relative normalization of the SNR 
	to the point source and intensities of emission lines 
	free within errors described in \citep{Kumar2014ApJ...781...41K}, 
	while fixed the normalization between two SNR component.

%The difference of the SNR modeling affects by $\sim$10\% level 
%	to the AXP flux level.

% -------------------------------------------------------
\subsubsection{Swift~J1822.3$-$1606} 
% -------------------------------------------------------
The 8.4-s X-ray pulsar Swift~J1822.3$-$1606
	was 
	discovered on 2011 July by {\it Swift}/BAT 
	and 
	observed by different X-ray satellites 
	\citep{Cummings2011ATel.3488....1C,Rea2012ApJ...754...27R,Scholz2012ApJ}.
The derived dipole field $B_{\rm d}=1.4\times 10^{13}$\,G 
	is weaker than the critical field $B_{\rm QED}$,
	and called a low-field magnetar. 
Suzaku observation was performed on September 13,
	nearly two month later after the discovery.
The soft X-ray data was already reported in \cite{Rea2012ApJ...754...27R},
	and we report here 
	the 15--60\,keV absorbed flux upper limit 
	at $1.2\times 10^{-11}$\,erg\,s$^{-1}$\,cm$^{-2}$.

% -------------------------------------------------------
\subsubsection{CXOU~J171405.7$-$381031 (CTB37B)} 
% -------------------------------------------------------
We reanalyzed the same {\it Suzaku} data in 2006 (ObsID 501007010) 
	which \cite{Nakamura2009PASJ...61S.197N} already published. 
With a characterization of a supernova remnant CTB~37B, 
	they suggested a central compact source CXOU~J171405.7$-$381031
	is an AXP from their {\it Chandra} and {\it Suzaku} spectroscopies.
The source was soon identified as a magnetar
	from {\it XMM-Newton} detections of slow pulsation and its large derivative
	\citep{Sato2010PASJ...62L..33S}.
Since our XIS spectra is contaminated from 
	a thermal component of CTB~37B,
	we added a ``vnei" model with parameters fixed at values 
	described in \cite{Nakamura2009PASJ...61S.197N}.
Despite the difference of employed spectral models
	(i.e, Comptonized blackbody with the power-law 
	vs. a pure power-law),
	our derived parameters of $\Gamma_{\rm s}\sim 3.3$ and
	$N_{\rm H}\sim 3.5\times 10^{22}$\,cm$^{-2}$ are 
	consistent within errors with \cite{Nakamura2009PASJ...61S.197N}.
The hard X-ray data is not analyzed in detail due to a potential contamination from 
	a surrounding CTB~37B.
	
% -------------------------------------------------------
%\subsubsection{1E~1547.0$-$5408} 
% -------------------------------------------------------
%The second follow-up observation of 1E~1547.0$-$5408
%	was performed nearly 1.5 year after the X-ray outburst in 2009 January
%	\citep{Enoto2010PASJ...62..475E}.
%The detailed analyses was already reported in \cite{Iwahashi2013PASJ...65...52I}.

% -------------------------------------------------------
\subsubsection{1E~1048.1$-$5937} 
% -------------------------------------------------------
The X-ray source 1E~1048.1$-$5937 was observed with {\it Suzaku} on 2008 November 30
	with $\sim$85\,ks exposure.
The SXC was clearly detected, 
	but the HXC was not able to detected 
	with its 3$\sigma$ upper limit at $1.3 \times 10^{-11}$\,erg\,s$^{-1}$cm$^{-2}$
	in the 15--60 keV band. 
The HXC was not detected with {\it NuSTAR} either in 2013 July,
	while the SXC was detected up to $\sim$20 keV
	\citep{2015arXiv150908115Y,2015ApJ...815...15W}.
\cite{2015arXiv150908115Y} reported the 20--79 keV flux upper limit 
	at $<4.15\times 10^{-12}$\,erg\,s$^{-1}$\,cm$^{-2}$ (3$\sigma$),
	which is converted to
	$F_{\rm 15-60}<6.7\times 10^{-12}$\,erg\,s$^{-1}$\,cm$^{-2}$
	in the 15--60\,keV band, assuming the spectral shape $\Gamma_{\rm s}=3.64$ in \cite{2015arXiv150908115Y}.
Combining with $F_{1-10}=8.9\times 10^{-12}$\,erg\,s$^{-1}$\,cm$^{-2}$,
	$\eta<0.75$.
Since the true $\Gamma_{\rm h}$ is not known, we did not use $F_{\rm h}$ and $\xi$ values. 
%	 and $\xi<0.34$ both at 3$\sigma$ upper-limits. 

% -------------------------------------------------------
%\subsubsection{1RXS J170849.0$-$400910} 
% -------------------------------------------------------	
%The second observation towards 1RXS J170849.0$-$400910
%	was performed on 2010 September.
%The SXC and HXC spectral fluxes and spectral parameters are 
%	close to the first observation on 2009 August. 
	
% -------------------------------------------------------
%\subsubsection{Swift J1834.9$-$0846 } 
% -------------------------------------------------------
%Swift J1834.9$-$0846 is a new transient source 
%	discovered by the outburst occurred in 2011 August. 
%\suzaku observed it on 2013 October 17
%	to search for a surrounding supernova remnant,
%	but we did not detect a compact source
%	even in the soft X-rays below 10 keV.
%	and we 
%	assgined
%	**-** keV upper limit is derived at $**\times 10^{-12}$\,erg\,s$^{-1}$cm$^{-2}$.
%Thus, we did not use this source in the following discussion. 

\clearpage % ==========================================
\section{Distances}
\label{section:location}
\begin{figure}[h]
\begin{center}
\includegraphics[width=80mm]{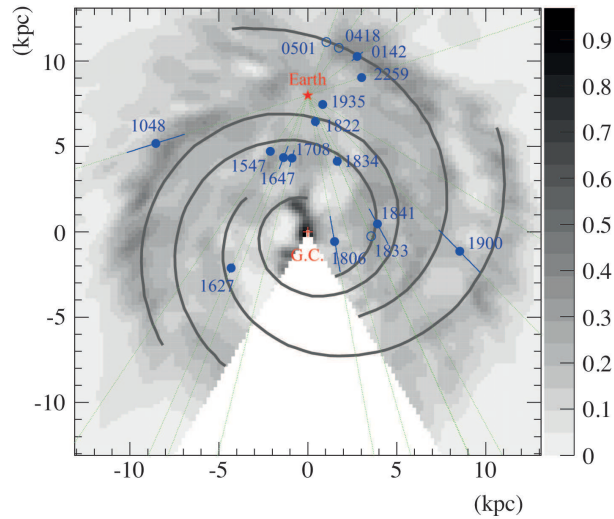}
\caption{
Galactic locations of SGRs and AXPs used in the present paper.	
The background plot indicates 
	the atomic hydrogen ($H_{\rm I}$) density map 
	at the Galactic plane
	measured from a radio 21-cm line survey 
	 \citep{Nakanishi2003PASJ}.
The filled circles are objects 
	with distance measurements (Table~\ref{tab:list_of_distance}),
	while open symbols are sources without known distances
	and to be assumed on the Galactic spiral arms
	\citep{Nakanishi2006PASJ}.
}
\label{fig:location}
\end{center}
\end{figure}

% ==========================================
In the present paper,
	we employed distances listed in Table~\ref{tab:list_of_distance},
	and illustrate their locations in Figure~\ref{fig:location}.	
%Referring to \cite{Olausen2014ApJS..212....6O} with some updates,	
%	we tabulated available recent literature, 
%	even though currently available methods are still limited
%		only during outbursts or available to sources associated with SNR; 
%	e.g.,
%	association with supernova remnants \citep{Kothes2002ApJ,Tian2010MNRAS},
%	transient radio emission during outbursts \citep{Camilo2007ApJ,Minter2008ApJ},
%	dust scattering X-ray halos \citep{Tiengo2010ApJ,Rivera-Ingraham2010ApJ},
%	or a red clump star method %(core helium-burning giant) method 
%	\citep{Durant2006ApJ}.
For three sources, 
	SGR~0501$+$4516, SGR~0419$+$5729, and SGR~1833$-$0832, 
	which do not have any reliable distance measurement,
	we assumed that sources are on the Galactic spiral arms.
%	since SGRs and AXPs are suggested to be young population ($\lesssim 10^{4-5}$ yr).
We attribute SGR~0501$+$4516 and SGR~0419$+$5729
	to the Perseus arm. %towards the anti-galactic center direction.
We assume SGR~1833$-$0832 on the Scutum-Crux arm,
	since SGR~1833$-$0832 shows 
	a similar high X-ray extinction $N_{\rm H}$ as that of SGR~1806$-$20
	toward the inner Galaxy.
It should be noted that
	the present distance are still highly uncertain,
	e.g., even for the well-studied object 1E~2259+586 hosted by SNR CTB~109,
	the distance has been revised several times within a range by a factor of a few
	(see note in Table~\ref{tab:list_of_distance}, $d$$\sim$3.0--7.5\,kpc for 1E~2259+586).

% ----------------------------------------------------
% ----------------------------------------------------

% ------------------------------------
% ~~~~~~~~~~~~~~~~~~~~~~~~~~~~~~~~~~~~~~~~~~~~~~~~~~~~
\begin{deluxetable*}{lrlcccccc}
\tabletypesize{\scriptsize}
%\tabletypesize{\small}
\tablecaption{Distances $d$ to known SGRs and AXPs used in this paper.
\label{tab:list_of_distance}}
\tablewidth{0pt}
\tablehead{
\colhead{Source name} &
\colhead{$d$} &
\colhead{Assumption or Method (Reference)} %& 
%\colhead{Lowest flux [Band]} 
\\
& (kpc)  &  % & ($10^{-12}$\,erg\,s$^{-1}$\,cm$^{-2}$) 
}
\startdata
SGR~1806$-$20 & $8.7_{-1.5}^{+1.8}$ & host cluster 1806$-$20 (G10.0$-$0.3) \citep{Bibby2008MNRAS.386L..23B} & %-- & 
\\
1E~1841$-$045 & $8.5_{-1.0}^{+1.3}$ & association to SNR Kes 73 \citep{Tian2008ApJ...677..292T} & % -- & 
\\
SGR~1900$+$14 & $12.5_{-1.7}^{+1.7}$ & host cluster C1 1900$+$14 \citep{Davies2009ApJ...707..844D} & %-- & 
\\
1RXS~J170849.0$-$400910 & $3.8_{-0.5}^{+0.5}$ & red clump star method \citep{Durant2006ApJ} & %-- & 
\\
1E~1048.1$-$5937 & $9.0_{-1.7}^{+1.7}$ & red clump star method \citep{Durant2006ApJ} & %-- & 
\\
4U~0142$+$61 & $3.6_{-0.4}^{+0.4}$ & red clump star method \citep{Durant2006ApJ} & %-- & 
\\
1E~2259$+$586 & $3.2_{-0.2}^{+0.2}$ & radio observation to SNR CTB 109 \citep{Kothes2002ApJ} & % -- & 
\\
\hline 
1E~1547.0$-$5408 & $3.91_{-0.07}^{+0.07}$ & dust scattering halo \citep{Tiengo2010ApJ...710..227T} & % -- & 
\\
SGR~0501$+$4516 & $3.3$ & assuming on the Perseus arm & % $1.3$ [1--10 keV] \citep{Rea2009MNRAS.396.2419R} & 
\\
SGR~1833$-$0832 & $9.0$ & assuming on the Scutum-Crux arm & % -- & 
\\
CXOU~J164710.2$-$455216 & $3.9_{-0.7}^{+0.7}$ & assocation to Westerlund I \citep{Kothes2007} & % -- & 
\\
Swift~J1834.9$-$0846 & $4.2_{-0.3}^{+0.3}$ & association to SNR W41 \citep{Leahy2008AJ....135..167L} & % -- & 
\\
Swift~J1822.3$-$1606 & $1.6_{-0.3}^{+0.3}$ & association to H$_{\rm II}$ region M17 \citep{Nielbock2001} & % $0.04$ [1--10 keV] \citep{Rea2012ApJ...754...27R} & 
%\\
%SGR~0418+5729 & $3.3$ & assuming on the Perseus arm & % -- & 
%\\
%SGR~1627$-$41 & $11.0_{-0.3}^{+0.3}$ & association with SNR G337.0-0.1 \citep{Corbel1999} & %0.06 [2--10 keV] \citep{Esposito2008MNRAS.390L..34E} & 
%\\
%SGR~1935$+$2154 & $1.0_{-0.001}^{+0.001}$ & --- & % -- & 
%\\
%1E~1547.0$-$5408 & $3.91_{-0.07}^{+0.07}$ & dust scattering halo \citep{Tiengo2010ApJ...710..227T} & % -- & 
\enddata
%\tablenotetext{a}{note-a}
%\tablenotetext{b}{note-b}
\tablecomments{
Referring to up-to-date data from \cite{Olausen2014ApJS..212....6O} 
	with previous distance studies: 
1E~1841$-$045, $>$5\,kpc (a red clump star method, \citealt{Durant2006ApJ}); 
1RXS~J170849.0$-$400910, 3.2--4.0\,kpc (dust scattering halo, \citealt{Rivera-Ingraham2010ApJ...710..797R}); 
1E~1048.1$-$5937, 5.7--6.2\,kpc (dust scattering halo \citealt{Rivera-Ingraham2010ApJ...710..797R}); 
4U~0142$+$61, 3.5--6.8\,kpc (dust scattering halo, \citealt{Rivera-Ingraham2010ApJ...710..797R}); 
1E~2259$+$586, 3.0$\pm$0.5\,kpc (radio observation to SNR CTB~109, \citealt{Kothes2002ApJ}), 7.5$\pm$1.0 (Red clump star method, \citealt{Durant2006ApJ}), 4.0$\pm$0.8 (radio observation to SNR CTB~109, \citealt{Tian2010MNRAS}); 
1E~1547.0$-$5408, $\sim$9\,kpc (diparsion measure, \citealt{Camilo2007ApJ...666L..93C}), $\sim$4\,kpc (possible association with SNR G327.24$-$0.13, \citealt{Gelfand2007ApJ...667.1111G}); 
SGR~0501$+$4516 0.8$\pm$0.4 kpc (possible association to SNR HB9, \citealt{Leahy2007A&A...461.1013L}), $\sim 1.5$\,kpc (possible association to SNR HB9, \citealt{Gaensler2008GCN..8149....1G}), $\sim 2$\,kpc (\citealt{Lin2011ApJ...739...87L}); 
Swift~J1834.9$-$0846, $\sim$5.4\,kpc (dust scattering halo, \citet{Esposito2013MNRAS.429.3123E});  
%SGR~0418+5729, $\sim 2$\,kpc (the Perseus arm \citealt{Horst2010ApJ...711L...1V}); 
%1E~1547.0$-$5408, $\sim$9\,kpc (diparsion measure, \citealt{Camilo2007ApJ...666L..93C}), $\sim$4\,kpc (possible association with SNR G327.24$-$0.13, \citealt{Gelfand2007ApJ...667.1111G}); 
}
\end{deluxetable*}
% ~~~~~~~~~~~~~~~~~~~~~~~~~~~~~~~~~~~~~~~~~~~~~~~~~~~~

% ------------------------------------

\clearpage \section{Decay Function}
\label{appendix:decay_function}

\begin{figure}
\begin{center}
\includegraphics[width=180mm]{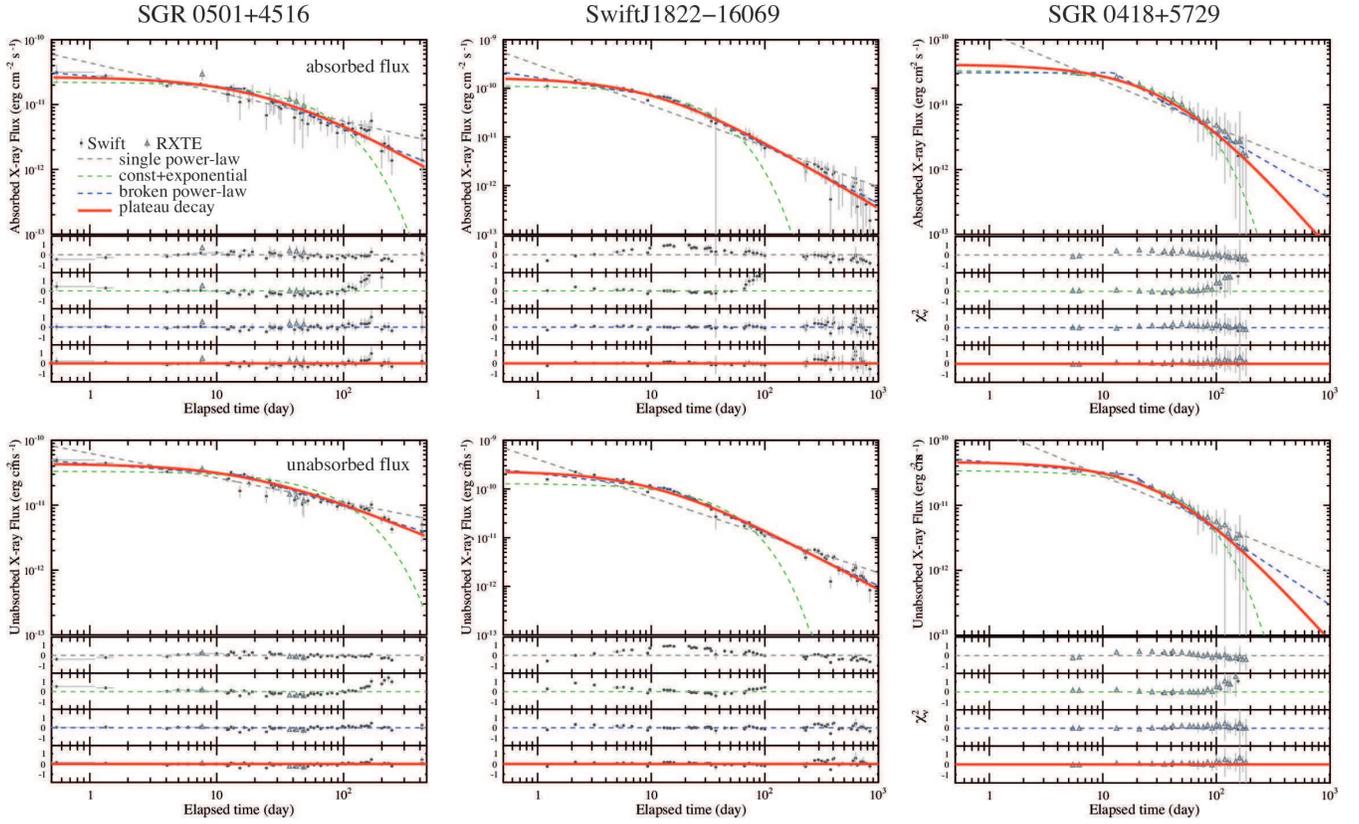}
\caption{
Examples of the SXC decays of
	SGR~0501+4516 (left),
	Swift~J1822$-$16069 (center),
	and SGR~0418$+$5729 (right).
Top and bottom figures are 
	the absorbed and unabsorbed 2--10 keV X-ray fluxes.
Fittings are performed with 
	a single power-law (PL) model (gray dashed lines),
	a constant with an exponential mode (green dashed lines),
	a broken power-law model (blue dashed lines),
	and 
	a plateau decay models (red thick lines).
Ratio of data to models are shown in bottom panels of individual figures. 
}
\label{fig:decay_fit}
\end{center}
\end{figure}

Figure~\ref{fig:decay_fit} is fitting examples of X-ray light curves 
	during three outbursts which showed the plateau-like feature at their early phase. 	
We tested four empirical formulae generally used in literatures:
	absorbed or unabsorbed fluxes $F(t)$
	as a function of an elapsed time $t$ from onsets of outbursts are represented by
	1) a single power-law $F(t) = K_{\rm pl} \cdot t^{-\alpha}$
	where $\alpha$ and $K_{\rm pl}$\,(erg\,s$^{-1}$\,cm$^{-2}$) 
	are a slope and normalization,
	2) an exponential shape 
	$F(t) = F_{\rm const} + K_{\rm exp} \cdot \exp (-t / \tau_{\rm exp})$ 
	where $\tau_{\rm exp}$ (days), $K_{\rm exp}$\,(erg\,s$^{-1}$\,cm$^{-2}$), 
	and $F_{\rm const}$\,(erg\,s$^{-1}$\,cm$^{-2}$) are
	a decay time-scale, normalization, and constant,
	3) a broken power-law $F(t) = K_{\rm brkn} \cdot t^{-\beta_1}$ for $t<t_{\rm bread}$,
	while $F(t) =K_{\rm brkn} t_{\rm break}^{-(\beta_1-\beta_2)}t^{-\beta_2}$ 
	for $t\ge t_{\rm bread}$,
	where $K_{\rm brkn}$, $t_{\rm break}$, $\beta_1$, and $\beta_2$ 
        are a normalization,
        breaking time of the curve,
        and slopes before/after this break, 
	and 4)
	the plateau decay (FD) function defined in Eq.~(\ref{eq:Omori}).
The PD model (4) generally gives good approximation of the light curves
	even though $\chi^2$ is close to that of the broken power-law case.
For example, the $\chi^2$ (dof) of the absorbed flux of SGR~0418+5729
	are 44.4 (27), 24.7 (27), 8.0 (25), and 8.3 (26)
	for the above 1), 2), 3) and 4) models, respectively. 
%Even though $\chi^2$ is close to the broken power-law case,
%	the PD model generally gives good approximation of the light curves:
%	e.g., $\chi^2$ (dof) of 44.4 (27), 24.7 (27), 8.0 (25), and 8.3 (26)
%	for the above 1), 2), 3) and 4) models, respectively 
%	in the case of the absorbed flux of SGR~0418+5729.
%Since we only use fewer free parameters (3) in the PD model 
%	than the broken power-law (4) 
%	and 
%	do not need to include a differential discontinuity 
%	at the break, 
%	we employed the PD model in the main text. 

% ----------------------------------------------------
% ----------------------------------------------------

% === MAXI upper limits ===
The onset of light curves is fixed 
	at the first short burst detected by {\it Swift}/BAT.
%	but there is an uncertainty of this onset time. 
Since we do not know the exact onset time of the X-ray outbursts,
	this might artificially produce the plateau like structure.
Such a possibility was already discussed 
	to explain a plateau phase of the X-ray afterglow of gamma-ray bursts 
	\citep{Yamazaki2009ApJ...690L.118Y}.
If $\tau_{\rm 0}$ is an artificial lag of the onset time,
	the single power-law is expanded to past
	typically around $t\sim -\tau_0$.
Predicted X-ray flux at $t\sim -\tau_0$
	becomes enough bright 
	for 
	the Monitor of All-sky X-ray Image (MAXI) \citep{Matsuoka2009PASJ...61..999M}
	to detect it. 
However the MAXI did not show strong enhancement 
	(e.g., 7$\sigma$ upper-limit at $\sim$5 mCrab level)
	for 
	SGR~0418$+$5729,
	SGR~1833$-$0832,
	and 
	Swift~J1834.9$-$0846
	(private communication with T. Mihara).
Therefore, 
	the plateau-decaying shape is not the artifact of the choice of the time origin. 
%	i the true onset is much earlier, we would see a fake plateau like structure. 
%	it is not likely to miss  MAXI did no found,
%	and we XXX that 
%	the plateau phase can not explain only by the misalignment of the onset time.
%The can not be interpreted as the artifact of the choice of the time origin.
%\input{appendix_fitting.tex}

%\section{Spectral fitting}
%\label{appendix:spectral_fitting}
%\input{table/table_suzaku_fit_all.tex}

\end{document}